\documentclass[reprint,superscriptaddress,amsmath,amssymb,aps]{revtex4-2}

\usepackage{color}
\usepackage{soul}
\usepackage{graphicx}
\usepackage{dcolumn}
\usepackage{bm}
\usepackage{comment}
\usepackage{hyperref}
\usepackage{mathtools} 
\usepackage{subfiles} 

\usepackage[linesnumbered,ruled,vlined]{algorithm2e}

\SetCommentSty{mycommfont}

\DeclareMathOperator*{\argmin}{arg\,min}
\DeclareMathOperator*{\argmaxloc}{arg\,max\,loc}

\begin{document}

\title{
    Nonstationary Critical Phenomena: Expanding The Critical Point
}

\author{Richard E. Spinney}
\email{r.spinney@unsw.edu.au}
\affiliation{School of Physics, UNSW, Sydney, NSW 2052, Australia}
\affiliation{EMBL Australia Node in Single Molecule Science, School of Biomedical Sciences, UNSW, Sydney, NSW 2052, Australia}
\author{Richard G. Morris}
\email{r.g.morris@unsw.edu.au}
\affiliation{School of Physics, UNSW, Sydney, NSW 2052, Australia}
\affiliation{EMBL Australia Node in Single Molecule Science, School of Biomedical Sciences, UNSW, Sydney, NSW 2052, Australia}
\affiliation{ARC Centre of Excellence for the Mathematical Analysis of Cellular Systems, UNSW Node, Sydney, NSW 2052, Australia}

\date{June 19, 2025}

\begin{abstract}
    A prototypical model of symmetry-broken active matter--- biased quorum-sensing active particles (bQSAPs)--- is used to extend notions of dynamic critical phenomena to the paradigmatic setting of driven transport, where characteristic behaviours are nonstationary and involve persistent fluxes.
    To do so, we construct an effective field theory with a single order-parameter--- a nonstationary analogue of active Model B--- that reflects the fact that different properties of bQSAPs can only be interpreted in terms of passive thermodynamics in appropriately chosen inertial frames.
    This codifies the movement of phase boundaries due to nonequilibrium fluxes between coexisting bulk phases in terms of a difference in effective chemical potentials and therefore an {\it unequal} tangent construction on a bulk free energy density.
    The result is both an anomalous form of coarsening and, more generally, an exotic phase structure; binodals are permitted to cross spinodal lines so that criticality is no longer constrained to a single point.
    Instead, criticality, with exponents that are seemingly unchanged from symmetric QSAPs, is shown to exist along a line that marks the entry to an otherwise forbidden region of phase space.
    The interior of this region is not critical in the conventional sense, but retains certain features of criticality, which we term pseudo-critical. 
    Whilst an inability to satisfy a Ginzburg criterion implies that fluctuations remain relevant at macroscopic scales, finite-wavenumber fluctuations grow at finite rates and exhibit non-trivial dispersion relations.
    The interplay between the growth of fluctuations and the speed at which they move relative to the bulk results in distinct regimes of micro- and meso-phase separation.
\end{abstract}

\maketitle


\section{\label{sec:intro}Introduction}
%
Critical phenomena is a cornerstone of modern physics that involves the study of critical points: singular locations in phase-space that mark sites of continuous phase transitions and therefore the start (or end) of lines of discontinuous phase transitions \cite{cardy_scaling_1996,domb_phase_1976,stanley_scaling_1999,goldenfeld_lectures_2019}. At a critical point both correlations and susceptibilities to perturbations diverge according to power laws whose exponents can be used to group systems into universality classes \cite{cardy_scaling_1996,domb_phase_1976,stanley_scaling_1999,goldenfeld_lectures_2019}.
\par
However, despite the central notions being developed in the 1970s, the full extent to which critical phenomena apply far from equilibrium remains an open challenge. Although the original work concerning static, equilibrium states \cite{kadanoff_scaling_1966,fisher_renormalization_1974,wilson_renormalization_1974,wilson_renormalization_1975} was rapidly adapted to encompass the dynamical relaxation of passive, energy-minimizing systems under a stochastic forcing \cite{halperin_calculation_1972,bausch_renormalized_1976,hohenberg_theory_1977}, it has taken until more recently for the ideas to be applied to systems whose dynamics cannot be written in terms of the minimization of an energy functional \cite{toner_long-range_1995,wittkowski_scalar_2014}.
These systems are synonymous with particles that are `active'; they perform work endogenously--- {\it e.g.}, via self-propulsion--- thereby breaking microscopic energy conservation and driving the system far from equilibrium \cite{ramaswamy_mechanics_2010, marchetti_hydrodynamics_2013}.
Prominent examples of active critical phenomena include order-disorder transitions in flocking  \cite{toner_long-range_1995,chen_critical_2015,jentsch_critical_2023}, and motility-induced phase separation (MIPS) \cite{cates_motility-induced_2015,partridge_critical_2019,redner_structure_2013,solon_pressure_2015-1,speck_effective_2014, cates_when_2013,solon_active_2015,tailleur_statistical_2008}. Notably, MIPS can be captured by modifying the model B class of dynamic critical phenomena for conserved order parameters \cite{wittkowski_scalar_2014,solon_generalized_2018}.
So-called active model B (AMB) describes phase-coexistence densities that can no longer be identified by constructing a common tangent on the free-energy density; two `uncommon' tangents are needed whose different intercepts reflect the fact that active pressure is not a state variable \cite{solon_pressure_2015}.
\par
Despite being active, AMB captures {\it stationary} nonequilibrium phenomena: limiting density profiles are time-independent. By contrast, the paradigmatic far-from equilibrium setting is that of driven transport, where characteristic behaviours are {\it nonstationary} and protoypical models involve persistent fluxes \cite{derrida_exactly_1998,katz_nonequilibrium_1984,antal_asymmetric_2000,van_beijeren_excess_1985,dierl_classical_2012}.
Contact can, however, be made between active phase separation and driven transport.
This has been shown explicitly in the context of biased quorum sensing active particles (bQSAPs) \cite{bertin_biased_2024}, as well as phenomenologically via the convective Cahn Hilliard equation \cite{golovin_convective_2001}.
These models exhibit highly novel behaviours--- {\it e.g.} travelling fronts, `chaotic' regimes, and `binodals inside spinodals'--- although such notions are broadly considered exotic and have, we argue, remained poorly understood.
\par
Here, using a combination of fluctuating hydrodynamics, numerical schemes, and particulate simulations we shed new light on these and other topics at the intersection of critical phenomena and nonstationarity.
In particular, we construct an effective field theory for bQSAPs which leads to several notions that subvert traditional expectations.
These include: equilibriumlike thermodynamic interpretations that depend on the choice of inertial-frame; anomalous coarsening; critical behaviour that is {\it not} constrained to a single point, and; criticallike behaviour that bears some, but not all, of the hallmarks of criticality.
\par
The remainder of the paper is organised as follows.
In Section \ref{sec:model}, we describe bQSAPS and derive their 
fluctuating hydrodynamics.
In Section \ref{sec:singleorder} we use the deterministic limit of the latter to motivate and construct a single order-parameter field theory--- a nonstationary analogue of AMB. 
This captures the effective thermodynamics of bQSAPs in different inertial frames, leading to certain subtleties, which we discuss. 
Of note, moving phase boundaries, and the corresponding fluxes between coexistence densities, are codified by an {\it unequal} tangent construction: an extension of the uncommon tangents of AMB to include different chemical potentials as well as different pressures. 
In Section \ref{sec:anomalous}, the implications of this on the wider phase structure are demonstrated by using a novel numerical scheme to identify binodal lines, which are now permitted to penetrate spinodals.
The result is the expansion of the critical point to a line, and access to an otherwise forbidden region of phase space.
Although the latter is not critical in the conventional sense, there is, nevertheless, no way of satisfying a Ginzburg criterion; understanding behaviour in this region {\it necessarily} requires an appreciation of fluctuations, which we provide
in Section \ref{sec:fluctuations}.
This is then combined with particulate simulations of the underlying bQSAPs in Section \ref{sec:characteristic} to establish a full picture of the phenomenology associated with different areas of phase space.
These include a novel form of coarsening, as well as types of meso- and micro-phase separation in which some but not all aspects of criticality are present.
We term this pseudo-criticality.
We conclude by recapitulating the salient points of the study, highlighting outstanding questions, and discussing how our results fit into the wider landscape of criticality far from equilibrium, including biology.
%
\begin{figure}[!t]
   \centering
   \includegraphics[width=0.48\textwidth]{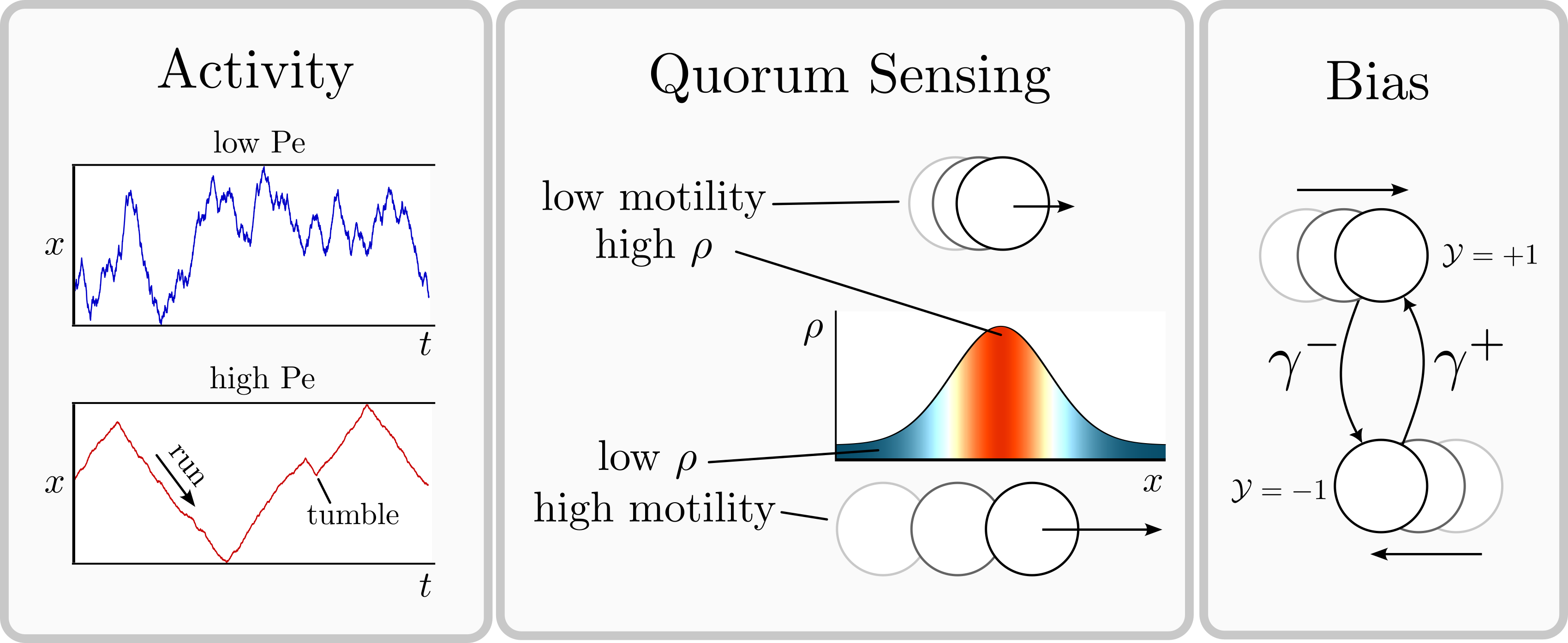} 
   \caption{
        {\bf Biased Quorum Sensing Active Particles (bQSAPs).} 
        bQSAPs are a form of over-damped self-propelled particle. In the absence of other particles, they move under a constant propulsion, subject to small, exogenous Gaussian fluctuations ({\it i.e.}, diffusion). 
        The direction of the self-propulsion changes according to a Poisson process, resulting in a `run-and-tumble' behaviour that is captured by the P\'eclet number ($\text{Pe}$), a dimensionless ratio of advective and diffusive rates of transport.
        A quorum-sensing interaction then serves to adjust the motile strength of the particles according to the local density, such that they slow down in the presence of other particles.
        Finally, particles are \emph{biased} when their tumbling rate is not symmetric, such that there is a preferred direction \emph{i.e.}, when $\gamma^+\neq\gamma^-$.
   }
   \label{fig:fig0}
\end{figure}
%

\section{bQSAPs: active phase separation meets transport}\label{sec:model}
%
In order to connect notions of active criticality and driven transport, we examine an explicit model: bQSAPs.
This provides a convenient sandbox for investigating how active criticality connects with driven transport.
Specifically, we combine a particulate formulation that can readily be simulated with a 
description of their fluctuating hydrodynamics for analysis.

\subsection{Microscopic Behaviour}
%
Quorum sensing active particles (QSAPs) \cite{solon_generalized_2018,bauerle_self-organization_2018} are `run-and-tumble' particles that slow down proportionally with their proximity to others.
In the standard 1D incarnation, the tumbling rate is left-right symmetric.
However, by adopting different rates of switching--- {\it e.g.}, left-to-right vs.~right-to-left--- it was recently shown that contact can be made with prototypical models of driven transport~\cite{bertin_biased_2024}.
\par
Specifically, we consider an off-lattice system of $N$ particles on a ring of length $L$.
Each particle has a position $X_i\in [0,L)$ and orientation $\mathcal{Y}_i\in\{+1,-1\}$, $i\in\{1,\ldots,N\}$.
These evolve in time according to the stochastic differential equations (SDEs)
\begin{subequations}
\label{eq:SDEs}
    \begin{align}
    \label{eq:SDEsA}
    dX_i&=\nu\mathcal{Y}_i\left[1-\frac{\rho_0L}{N}\sum_{\substack{j=1\\j\neq i}}^Nk_\sigma(X_i-X_j)\right]dt+\sqrt{2D}dW_i,\\
    \label{eq:SDEsB}
    d\mathcal{Y}_i&=2\delta_{-1,\mathcal{Y}_i}dN^{(i)}_{-,+}-2\delta_{+1,\mathcal{Y}_i}dN^{(i)}_{+,-}.
\end{align}
\end{subequations}
In (\ref{eq:SDEsA}), $\nu$ is a motile strength, $D$ is a diffusion coefficient, $\rho_0\in[0,1]$ is a coupling strength, and the $dW_i(t)$ are increments of independent Wiener processes such that $\mathbb{E}[ dW_i(t)dW_j(s)]=\delta_{i,j}\delta(t-s)dt$, with $\delta_{\cdot,\cdot}$ the Kronecker delta, $\delta(\cdot)$ the Dirac delta, and $\mathbb{E}[\cdot]$ an expectation.
$k_\sigma$ is a positive, normalised, interaction kernel with some characteristic width, $\sigma$, for which we stipulate $\sigma\ll L$. In (\ref{eq:SDEsB}), $dN^{(i)}_{-,+}\in\{0,1\}$ is an increment of a Poisson process with intensity $\gamma^+$ for particle $i$ and $dN^{(i)}_{+,-}\in\{0,1\}$ is an increment in a Poisson process with intensity $\gamma^-$ for particle $i$.
We assume, without loss of generality, that $\gamma^+\geq\gamma^-$ such that net motion is in the positive direction.
The key behaviours encoded by Eqs.~(\ref{eq:SDEs}) are illustrated in Fig.~\ref{fig:fig0}.
\par
When $\gamma^+=\gamma^-$, standard QSAPs are recovered, which exhibit MIPS
\cite{solon_generalized_2018}.
By allowing $\gamma^+\neq\gamma^-$ we obtain a model with biased, or asymmetric, mean particle orientation.
In extremis, when $\gamma^-/\gamma^+=0$, all motion is left-to-right, and the model becomes an off-lattice variant of the totally asymmetric exclusion process (TASEP); the canonical toy model of driven transport phenomena \cite{derrida_exactly_1998}.

\subsection{Fluctuating Hydrodynamics}
%
The collective behaviour of bQSAPs can be captured by 
representing the $N$ interacting SDEs using a Dean-like procedure adapted to incorporate Poisson jumps \cite{spinney_deankawasaki_2025}. 
The result is two coupled stochastic partial differential equations (SPDEs) that describe the system's fluctuating hydrodynamics. In non-dimensionalised form, and with assumption $N\gg 1$, (See Appendix \ref{app:SPDEs}) these can be written 
\begin{widetext}
\begin{subequations}
\label{eq:SPDE}
    \begin{align}
    \partial_t \rho(x,t)&=-\text{Pe}\,\partial_x \left[\psi(x,t)\left(1-\rho(x,t)\right)\right] +\partial_x^2\rho(x,t)+\partial_x\left(\sqrt{\frac{2N_L\rho_0\rho(x,t)}{N}}\eta_{\rho}(x,t)\right),\\
    \partial_t \psi(x,t)&=-\text{Pe}\,\partial_x \left[\rho(x,t)\left(1-\rho(x,t)\right)\right] +\partial_x^2\psi(x,t)-\frac{(1+\beta)^2}{2\beta}\psi(x,t)+\frac{1-\beta^2}{2\beta}\rho(x,t)+\partial_x\left(\sqrt{\frac{2N_L\rho_0\rho(x,t)}{N}}\eta_{\psi}(x,t)\right)\nonumber\\
    &\qquad\qquad+\sqrt{\frac{2N_L\rho_0}{N}\left[\frac{(1+\beta)^2}{2\beta}\rho(x,t)-\frac{(1-\beta^2)}{2\beta}\psi(x,t)\right]}\eta_{\leftrightarrow}(x,t).
    \end{align}
\end{subequations}
\end{widetext}
Here, $\rho$ and $\psi$ are dimensionless particle and polarisation density fields, respectively.
They are constructed from the dimensionless densities for the individual (left/right oriented) species, $\rho=\rho^++\rho^-$, $\psi=\rho^+-\rho^-$, which can be equivalently described through Eq.~(\ref{eq:SPDEs2}) in Appendix \ref{app:SPDEs}.
$N_L=L/l$ is the (non-dimensional) size of the system where $l=\sqrt{D/\bar\gamma}$ is a characteristic length based on the mean tumbling rate $\bar{\gamma}=2\gamma^+\gamma^-/(\gamma^++\gamma^-)$. For the fluctuations, $\eta_\rho$, $\eta_\psi$, and $\eta_\leftrightarrow$ are unit space time white noises, which are mutually independent except for between $\eta_\rho$ and $\eta_\psi$ which obey $\mathbb{E}[\eta_\rho(x,t)\eta_\psi(y,s)]=[\psi(x,t)/\rho(x,t)]\delta(x-y)\delta(t-s)$.
\par
There are three dimensionless control parameters: a mean effective density, $\rho_0={N_L}^{-1}\int_0^{N_L}\rho(x)\,dx$, derived from the underlying interaction strength; the P\'{e}clet number, $\text{Pe}=\nu/\sqrt{\bar{\gamma}D}$, and; an asymmetry parameter $\beta=\gamma^-/\gamma^+\in[0,1]$. This final control parameter is an addition to those found in conventional QSAPs. Of note, when $\beta<1$, the resulting asymmetry is not a {\it spontaneous} symmetry breaking as seen in a flocking transition \cite{toner_long-range_1995}, with particles instead behaving in the manner of response to an external applied field in a (fast) ionic conductor \cite{schmittmann_driven_1998}.
\par
Due to the weak coupling of the underlying SDEs, the $N\to \infty$ limit of Eqs.~(\ref{eq:SPDE}) obtains partial differential equations (PDEs) in the manner of a McKean-Vlasov diffusion \cite{mckean_class_1966,jabin_mean_2017} through a propagation of chaos result \cite{chaintron_propagation_2022-1,chaintron_propagation_2022}. 
As might be expected, the effects of changing $\beta$ mirror those seen at the particulate level.
Setting $\beta=1$, the system can be put in direct correspondence with AMB, the field theory that captures MIPS \cite{solon_generalized_2018}. 
Setting $\beta=0$, by contrast, recovers the Lighthill-Whitham-Richards (LWR) model of traffic flow \cite{lighthill_kinematic_1997,richards_shock_1956}, which is closely related to the viscous Burgers' equation, a classical PDE for driven transport that has applications in fluid dynamics and acoustics \cite{burgers_mathematical_1948,bonkile_systematic_2018}, and is also the continuum limit of TASEP \cite{rost_non-equilibrium_1981}. 
%
\subsection{Deterministic Limit and Linear Stability}
\label{sec:linear_stability}
%
In the deterministic, $N\to\infty$ limit, Eqs.~(\ref{eq:SPDE}) reduce to
\begin{subequations}
\label{eq:deterministic}
\begin{align}
    \partial_t\rho=&-\partial_x J,\\
    J=&\text{Pe}[\psi(1-\rho)]-\partial_x\rho,\\
    \partial_t\psi=& -\text{Pe}\,\partial_x \left[\rho(1-\rho)\right]+\partial_x^2\psi\nonumber\\
        &\qquad -\frac{(1+\beta)^2}{2\beta}\psi+\frac{(1-\beta^2)}{2\beta}\rho,
\end{align}
\end{subequations}
which possess the trivial homogeneous solution $\rho(x,t)=\rho_0=N_L^{-1}\int_0^{N_L}\rho(x,0)\,dx$ and $\psi(x,t)=\psi_0\coloneq \rho_0\,(1-\beta)/(1+\beta)$. 
We characterise the stability around this state by probing the behaviour of $\delta \rho(x,t)\coloneq \rho(x,t)-\rho_0$ and $\delta \psi(x,t)\coloneq \psi(x,t)-\psi_0$. 
To do so, we represent Eq.~(\ref{eq:deterministic}) in reciprocal space and real time via Fourier transform $\tilde{f}(k,t)=\mathcal{F}_x\{f(x,t)\}=\int_{0}^{N_L} dx\;e^{-ikx}f(x,t)$, with $k=2\pi n/N_L$, $n\in\mathbb{N}$. Linearising the resulting dynamic yields $d/dt(\delta\tilde{\rho},\delta\tilde{\psi})=\mathbf{A}(k)(\delta\tilde{\rho},\delta\tilde{\psi})$ with matrix
\begin{align}
\label{eq:A}
\mathbf{A}(k)&=
\begin{bmatrix}
-k^2+ik\text{Pe}\rho_0\frac{1-\beta}{1+\beta}&-ik\text{Pe}(1-\rho_0)\\
(\frac{1-\beta^2}{2\beta}-ik\text{Pe}(1-2\rho_0)) & -(k^2+\frac{(1+\beta)^2}{2\beta})
\end{bmatrix}.
\end{align}
The stability of the homogeneous solution $\{\rho_0,\psi_0\}$ can be assessed through the real component of the largest eigenvalue of $\mathbf{A}(k)$, computed as $\lambda_{\pm}(k)=[\text{Tr}(\mathbf{A}(k))\pm\sqrt{\text{Tr}^2(\mathbf{A}(k))-4\text{Det}(\mathbf{A}(k))}]/2$. 
\par
For large system size, $N_L\to\infty$, we expand $\text{Re}[\lambda_+]$ around $k=0$  yielding
\begin{align}
  \text{Re}[ \lambda_+ ]&=- \left(\frac{8 \beta ^2 \text{Pe}^2 \left(2 \rho _0^2-3 \rho _0+1\right)}{(\beta +1)^4}+1\right)k^2 +\mathcal{O}(k^4).
\end{align}
Solving for $\text{Re}[\lambda_+]=0$ in this limit defines the surface where we observe cross-over from linear stability to instability, characterising the spinodals of the system
\begin{align}
\rho^{\rm spinodal}_0&=\frac{1}{4}\left(3\pm\frac{\sqrt{\text{Pe}^2-(1+\beta)^4/\beta^{2}}}{\text{Pe}}\right).
\label{eq:spinodal}
\end{align}
The stability of homogenous solutions, compared to the unbiased $\beta=1$ case, can therefore be recast in terms of a $\beta$-dependent re-scaling of the P\'{e}clet numer: $\text{Pe}\to\text{Pe}(2+\beta^{-1}+\beta)/4$, with the bifurcation point occurring at $\text{Pe}=2+\beta^{-1}+\beta$, such that the required $\text{Pe}$ to destabilise the system diverges as $\beta\to 0$.

\section{Field Theory: Effective Thermodynamics in inertial frames}
\label{sec:singleorder}
%
To capture the behaviours codified by Eqs.~(\ref{eq:deterministic}), we aim to extract an underlying equilibriumlike structure in the sense of passive and active Model B field theories. This leads to a consideration of effective thermodynamics in different inertial frames. 
%
\subsection{Passive and Active Model B Redux}
\label{sec:AMB}
%
In passive Model B, a conserved scalar order parameter evolves according to $\dot{\phi}=-\partial_x J^{\rm MB}$, with $J^{\rm MB}=-M\partial_x\mu^{\rm eq}+\Lambda$.
$\Lambda$ is white noise and the chemical potential, $\mu^{\rm eq}$, is \emph{integrable} in the sense that we may write $\mu^{\rm eq}=\delta \mathcal{F}[\phi]/\delta \phi$ with $\mathcal{F}[\phi]=\int dx\,(f_{\rm bulk}(\phi)+\kappa (\partial_x\phi)^2/2)$, with $f_{\rm bulk}$ a double-welled bulk free energy.
In the case of $\Lambda=0$ and $f_{\rm bulk}=(1-\phi^2)^2/4$ it is referred-to as the Cahn-Hilliard equation \cite{bray_coarsening_2003}.
Limiting, $t\to\infty$, solutions of this equation coincide with the minimisation of $\mathcal{F}[\phi]$, with approximately piece-wise constant profiles accordant with a common tangent construction on $f_{\rm bulk}$.
This ensures that bulk phases are in mutual equilibrium, sharing the intensive state functions of chemical potential (gradients of $f_{\rm bulk})$ and pressure (Legendre transforms, or intercepts of the tangent lines, of $f_{\rm bulk}$).
\par
\emph{Active} Model B makes a minimal modification to passive Model B, such that the chemical potential is no longer integrable.
Specifically, it casts the chemical potential as $\mu^{\rm AMB}=\mu^{\rm eq}+\lambda|\partial_x\phi|^2=\mu^{\rm eq}_{\rm bulk}-\kappa \partial_x^2\phi+\lambda|\partial_x\phi|^2$, introducing activity through its hallmark $\lambda$ term which causes the \emph{boundaries} to reside out of equilibrium.
The inclusion of such a term prevents solutions corresponding to the minimisation of $\mathcal{F}[\phi]$ and thus a common tangent construction on $f_{\rm bulk}$, despite $f_{\rm bulk}$ determining behaviour in the bulk phases. Instead, solutions correspond to an \emph{uncommon} tangent construction on $f_{\rm bulk}$ \cite{wittkowski_scalar_2014,solon_generalized_2018}, where chemical potentials (gradients) are equal between the bulk phases, but pressures (intercepts) differ.
This reflects {\it i}) the absence of flux in the emergent steady state ($\partial_x\mu^{\rm AMB}=0$) and {\it ii}) the failure of mechanical and thermodynamic pressures to be equivalent in the context of activity \cite{solon_pressure_2015}, respectively. 
\par
Crucially, both passive and active Model B are stationary and absent  a persistent current, properties that bQSAPs [Eqs.~(\ref{eq:deterministic})] explicitly break.
%
\subsection{An Implicit Field Theory in a Co-Moving Frame}
\label{sec:comoving}
For $\beta=1$, Eqs.~(\ref{eq:deterministic}) capture the $N\to\infty$ limit of conventional, symmetric QSAPs.
The resulting expression can be put in direct correspondence with AMB by eliminating the polarisation density in a stationary state \cite{kourbane-houssene_exact_2018}.
\par
However, for $\beta<1$, this strategy is not available, since solutions are nonstationary--- {\it i.e.}, in the limit $t\to\infty$, we have $\partial_t\rho\neq 0$ and $\partial_t\psi\neq 0$ on any time scale.
To proceed, we instead make the ansatz that limiting solutions uniformly travel at constant velocity, $v$. 
As such, we perform a Galilean transform, or boost, resulting in
\begin{subequations}
\label{eq:Gal}
\begin{align}
    \partial_t\rho_{v}&=-\partial_x J_v,\\
    \partial_t\psi_{v}&= -\text{Pe}\,\partial_x \left[\rho_{v}(1-\rho_{v})\right]+\partial_x^2\psi_{v}\nonumber\\
        &\quad -\frac{(1+\beta)^2}{2\beta}\psi_{v}+\frac{(1-\beta^2)}{2\beta}\rho_{v}+v\partial_x\psi_{v},
\end{align}
\end{subequations}
where $\rho_v(x,t)=\rho(x+vt,t)$, $\psi_v(x,t) = \psi(x+vt,t)$, and
\begin{equation}
\label{eq:j_v}
    J_v = \text{Pe}\,\left[ \psi_{v}(1-\rho_{v})\right]-\partial_x\rho_{v}-v\,\rho_{v},
\end{equation}
with $v$ the constant speed of the boost.
\par
Our motivation here is to ask: if an observer is walking alongside a travelling solution, do they see behaviour that is equilibirumlike, and how much is it captured by AMB?
To proceed, we notice that may treat $J^{\rm st}_{v}=\lim_{t\to\infty}J_v$ as a constant, which is equal to the spatially \emph{homogeneous} current observed in the co-moving frame.
This permits us to identify
\begin{align}
    \psi^{\rm st}_{v}&=\frac{(\rho^{\rm st}_{v})'+(v\rho^{\rm st}_{v}+J^{\rm st}_{v})}{\text{Pe}(1-\rho_v^{\rm st})},
    \label{eq:psistationary}
\end{align}
where $\rho^{\rm st}_{v}(x)$ and $\psi^{\rm st}_{v}(x)$ are the time-independent, limiting profiles in the co-moving frame, and the prime notation indicates a derivative.
Substituting (\ref{eq:psistationary}) into (\ref{eq:Gal}) with the $t\to\infty$ condition $\partial_t\psi_v^{\rm st}=0$ yields the condition for a stationary solution in the co-moving frame
\begin{align}
    \label{eq:L}
    0 
    &=\text{Pe}\,\partial_x\left[\rho^{\rm st}_{v}(1-\rho^{\rm st}_{v})\right] -\partial_x^2\left[\frac{\partial_x\rho^{\rm st}_{v}+(v\,\rho^{\rm st}_{v}+J^{\rm st}_{v})}{\text{Pe}(1-\rho^{\rm st}_{v})}\right] \nonumber\\
    &\quad
    - v\,\partial_x\left[\frac{\partial_x\rho^{\rm st}_{v}+(v\,\rho^{\rm st}_{v}+J^{\rm st}_{v})}{\text{Pe}\,(1-\rho^{\rm st}_{v})}\right]    -     \frac{(1+\beta)^2}{2\beta\text{Pe}}\partial_x\ln(1-\rho^{\rm st}_{v})\nonumber\\
    &\quad+\frac{(1+\beta)^2}{2\beta}\frac{(v\,\rho^{\rm st}_{v}+J^{\rm st}_{v})}{\text{Pe}(1-\rho^{\rm st}_{v})}   - \frac{1-\beta^2}{2\beta}\rho^{\rm st}_{v}.
\end{align}
This motivates the construction of a co-moving, effective field theory for a single conserved order-parameter of the form
\begin{align}
    \dot{\phi}
    =\partial_xM(\partial_x\mu - F),
    \label{eq:NSAMB}
\end{align}
such that it reduces to active model B when $\beta=1$ (\emph{cf.} Sec~\ref{sec:AMB}).
\par
Our field theory comprises three parts. First, a chemical potential
\begin{subequations}
\begin{align}
    \mu &= \mu_0 + \lambda(\phi)(\partial_x\phi)^2- \kappa(\phi)\partial^2_x\phi+\pi(\phi)\partial_x\phi,\\
    \lambda(\phi)&=-\frac{1}{\text{Pe}(1-\phi)^2},\\
    \kappa(\phi) &= \frac{1}{\text{Pe}(1-\phi)},\\
    \pi(\phi)&=-\frac{J^{\rm st}_v+v(2-\phi)}{\text{Pe}(1-\phi)^2},\\
    \mu_0&=\text{Pe}\phi(1-\phi)-\frac{(1+\beta)^2}{2\beta\text{Pe}}\ln(1-\phi)\nonumber\label{eq:mu0}\\
    &\qquad-\frac{v(v\phi+J^{\rm st}_v)}{\text{Pe}(1-\phi)}-\frac{1-\beta}{1+\beta}J^{\rm st}_v,
\end{align}
\end{subequations}
which differs from that appearing in AMB by the left-right symmetry-breaking term $\pi(\phi)\partial_x\phi$, as well as modifications to $\mu_0$. (Note: we have added an unimportant constant to $\mu_0$ for convenience). Second, a non-conservative force
\begin{align}
    F&=-\frac{(1+\beta)^2}{2\beta}\frac{(v\,\phi+J^{\rm st}_{v})}{\text{Pe}(1-\phi)}   + \frac{1-\beta^2}{2\beta}\phi,
\end{align}
that cannot be written in terms of the gradient of a chemical potential. Third, an (ostensibly) arbitrary mobility function, $M$. Since such a theory reduces to the form of AMB which describes QSAPs in the case $\beta=1$ and $v=J^{\rm st}_v=0$, we call this effective field theory \emph{co-moving} nonstationary active model B (nsAMB). 
\par
Crucially, Eq.~(\ref{eq:NSAMB}) shares the same solution profiles as Eq.~(\ref{eq:deterministic}), but its limiting, $t\to\infty$ state, $\phi^{\rm st}\coloneq\lim_{t\to\infty}\phi$, is both \emph{stationary} and possesses \emph{no flux}. 
Transforming the highly out-of-equilibrium behaviour of bQSAPs into a stationary and flux-free analogue has, however, come at a cost.
Such an equation is now \emph{implicit}, through the inclusion of the unknown constants $v$ and $J^{\rm st}_v$. 
In this sense, the stationary solutions to Eq.~(\ref{eq:NSAMB}) should be thought of as the \emph{triples} $v$, $J^{\rm st}_v$ and $\phi^{\rm st}$, which together satisfy $0=\partial_xM(\partial_x\mu - F)$. 
Once all three are known, we can reconstruct the limiting flux of Eq.~(\ref{eq:deterministic}) in the laboratory frame ($J_{\rm lim}=\lim_{t\to\infty}J$) using
\begin{align}
   \label{eq:linearJ}
   J_{\rm lim}&=J^{\rm st}_v+v\phi^{\rm st}.
\end{align}
Whilst a numerical method for obtaining such solution triples is described later in Sec.~\ref{sec:shooting}, we can characterize the relevant effective thermodynamics by examining two cases that are analytically tractable: solutions that are {\it i}) homogenous, or {\it ii}) phase separated.
%

\subsection{Homogeneous Solutions}
\label{sec:homogeneous}
%
We first consider homogeneous solutions of Eqs.~(\ref{eq:deterministic}), where all field gradients vanish, $\rho=\rho_0$, and $\psi=\psi_0=\rho_0(1-\beta)/(1+\beta)$. In this case the system experiences a uniform `bulk current' in the lab-frame  
\begin{align}
    \label{eq:Jbulk}
    J_{\rm bulk}(\rho)&\coloneq\text{Pe}\frac{1-\beta}{1+\beta}\rho(1-\rho),
\end{align}
which is otherwise absent in the case of symmetric QSAPs ($\beta=1$). 
The corresponding stationary condition for the flux-free effective theory in Eq.~(\ref{eq:NSAMB}) is $F=0$.
This recapitulates Eq.~(\ref{eq:linearJ}) but with $\phi^{\rm st}=\rho_0$ and $J_{\rm lim}=J_{\rm bulk}$--- {\it i.e.},
\begin{align}
    \label{eq:F=0}
    J_{\rm bulk}(\rho_0)&=J^{\rm st}_v+v\rho_0.
\end{align}
Therefore we see that the notion of the self-consistent solution triple is under-determined in the case of a constant field: we may choose any $v$, so long as $J^{\rm st}_v$ is chosen, correspondingly, so as to satisfy Eq.~(\ref{eq:F=0}).  In other words, because homogenous solutions to Eqs.~(\ref{eq:deterministic}) are flat, they are captured by our field theory, Eq.~(\ref{eq:NSAMB}), irrespective of the observer's frame. 

\subsubsection{The Bulk Chemical Potential is Invariant to the Inertial Frame}
\label{sec:mubulk}
Whilst not immediately apparent from the form of Eq.~(\ref{eq:mu0}), this reasoning extends to the {\it bulk} component of $\mu$: it is invariant to our choice of $v$ and $J^{\rm st}_v$ and therefore the observer's frame. To demonstrate this, one need only substitute the stationary condition for the bulk, $J^{\rm st}_v=J_{\rm bulk}(\phi)-v\phi$ from Eq.~(\ref{eq:F=0}), into Eq.~(\ref{eq:mu0}) to obtain the following quantity free of both $v$ and $J^{\rm st}_v$ 
\begin{align}
\label{eq:mubulk}
    {\mu}_{\rm bulk}&=\frac{4\beta\text{Pe}}{(1+\beta)^2}\,\left[\phi(1-\phi)\right]     -     \frac{(1+\beta)^2}{2\beta\text{Pe}}\ln(1-\phi).
\end{align}
Explicitly, a locally flat region which is part of any limiting solution with triple $\{v,J^{\rm st}_v,\phi^{\rm st}\}$, possesses $\mu=\mu_{\rm bulk}$ and $F=0$. 
This in turn implies an equally frame-invariant bulk free energy density (up to an unimportant linear term in $\phi$) of
\begin{align}
\label{eq:fbulk}
f_{\rm bulk}(\phi)&=\int_\phi \mu_{\rm bulk}(p)\,dp\nonumber\\
&=\frac{2\text{Pe}\beta}{3(1+\beta)^2}(3-2\phi)\phi^2\nonumber\\
&\qquad+\frac{(1+\beta)^2}{2\text{Pe}\beta}(1-\phi)\ln(1-\phi).
\end{align}
Moreover, this quantity is physically meaningful--- 
its curvature accurately predicts the stability of Eqs.~(\ref{eq:deterministic}). Explicitly, solving for $\phi$ in $f_{\rm bulk}''(\phi)=\mu_{\rm bulk}'(\phi)=0$ determines the spinodal line, in exact agreement with Eq.~(\ref{eq:spinodal}). Further, $\mu_{\rm bulk}$ coincides with $\mu_0$ [Eq.~(\ref{eq:mu0})] in the case of $\beta=1$, as required. 
\par
Through the identification of $f_{\rm bulk}$ the bulk states of bQSAPs can seemingly be captured by an effective, equilibriumlike structure. In the $\beta=1$ case of symmetric QSAPs and their associated AMB, this is a natural consequence of the coarse-graining of a stationary flux-free system. For $\beta<1$, however, the meaning of $f_{\rm bulk}$ is less clear, since our effective field theory does not, in general, correspond to gradient descent in $f_{\rm bulk}$. By examining the stability of Eq.~(\ref{eq:NSAMB}) we can, however, find an inertial frame in which it does. 

\subsubsection{Equilibriumlike Stability Requires a Specific Inertial Frame}
\label{sec:eq_inertial}
Whilst $f_{\rm bulk}$ accurately predicts the stability properties of Eq.~(\ref{eq:deterministic}), somewhat counter-intuitively, the stability of Eq.~(\ref{eq:NSAMB}) still depends on the choice of $v$ and $J^{\rm st}_v$. This is because the implicit dependence on $v$ and $J^{\rm st}_v$ in $\mu_0-\mu_{\rm bulk}$ contributes to even the smallest perturbations away from the bulk.  
However, by considering the linear response of Eq.~(\ref{eq:NSAMB}) (see Appendix \ref{app:NSAMB_stability}), we find that if we use the unique inertial frame characterised by
\begin{align}
    \label{eq:vJdegen}
    v&=\text{Pe}\frac{1-\beta}{1+\beta}(1-2\rho_0),\quad J^{\rm st}_v=\text{Pe}\frac{1-\beta}{1+\beta}\rho_0^2,
\end{align}
its large scale ($k\to 0$) stability properties are determined solely by $f_{\rm bulk}$--- {\it i.e.}, the dynamics around bulk states are captured in terms of gradient descent of $f_{\rm bulk}$ without any dependence on $\mu_0-\mu_{\rm bulk}$. 
This choice is not arbitrary: it is precisely the velocity at which perturbations around $\rho=\rho_0$ travel in the full description of bQSAPs in the same limit (see Sec.~\ref{sec:fluctuations}). As such, it is simultaneously the frame where the response of Eq.~(\ref{eq:NSAMB}) to perturbations is purely dispersive, and the field appears time reversible, on large scales.
\par
As a result, we see that an effective, equilibriumlike system can be associated with the bulk states of Eqs.~(\ref{eq:deterministic}) by coarse-graining in a frame where fluctuations do not travel, and where the bulk free energy $f_{\rm bulk}$ solely determines its behaviour. The notion that a system appears equilibriumlike in a specific frame is not especially strange; in the case of $\beta=1$ one must also choose a unique frame, but it is trivially $v=0$. With asymmetry ($\beta<1$) the difference is that we have $v=v(\rho_0)$--- {\it i.e.}, the frame in which bulk states appear equilibriumlike depends on the density of that state.

\subsection{Phase Separated Solutions}
\label{sec:phasesep}
%
Phase-separated solutions of Eq.~(\ref{eq:NSAMB}) comprise interfacial boundaries that separate bulk liquid and gaseous phases, at densities denoted $\rho_L$ and $\rho_G$, respectively. 
Whilst we expect these interfaces to travel in the laboratory frame, we have a description in Eq.~(\ref{eq:NSAMB}) where the boundary is stationary and there is vanishing net flux. 
We thus explore the extent to which we can construct a thermodynamic description in terms of two connected equilibriumlike systems. 

\begin{figure*}[!htp]
   \centering
   \includegraphics[width=\textwidth]{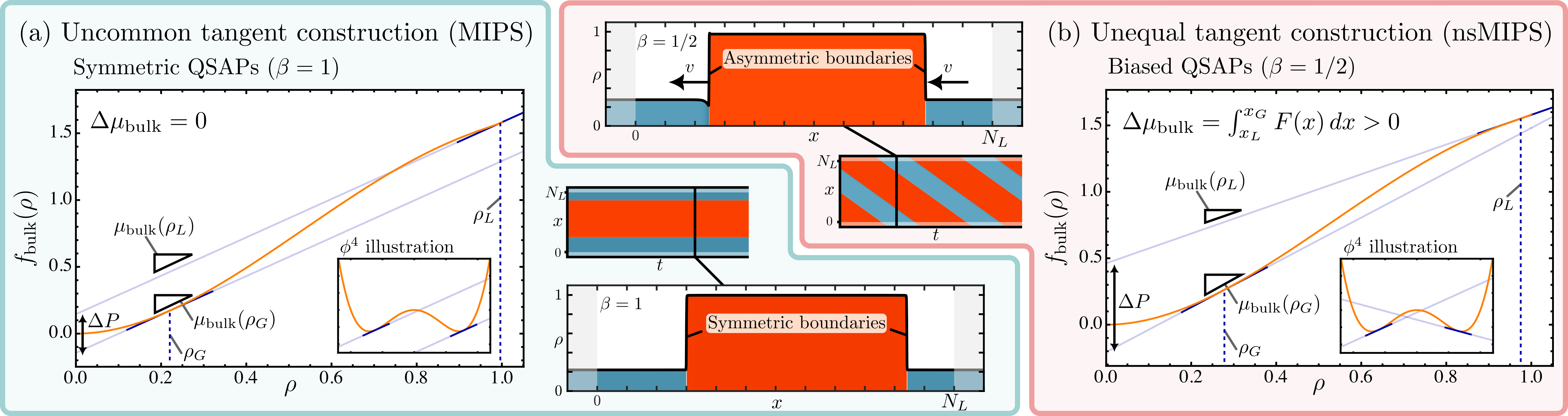} 
   \caption{
        {\bf Uncommon and Unequal Tangent Constructions: Stationary vs.~Nonstationary MIPSs.}
        When $\beta=1$ [Panel (a), left, teal bounding box] QSAP tumbling rates are symmetric and the chemical potentials of the bulk phases are equal.
        This corresponds to conventional AMB, which is characterised by an uncommon tangent construction on a bulk free energy density (same gradients, different intercepts) such that there is zero flux across the domain boundaries.
        Density profiles from numerical integration are symmetric and domain boundaries are stationary (\emph{i.e.}, conventional, stationary MIPS).
        When $\beta<1$ [Panel (b), right, pale pink bounding box], QSAP tumbling rates differ and bulk chemical potentials are unequal.
        This corresponds to nsAMB, which is characterised by an unequal tangent construction (different gradients, different intercepts) such that there {\it is} a flux across the boundaries.
        As a result, density profiles from numerical integration are {\it asymmetric} and domain boundaries move at characteristic speed $v$ (\emph{i.e.}, \emph{non}stationary MIPS).
        Examples shown use $(\beta=1,\text{Pe}=9,\rho_0=0.7)$ and $(\beta=1/2,\text{Pe}=8,\rho_0=0.7)$ with numerical integration of Eqs.~(\ref{eq:deterministic}) performed using finite elements (see Appendix~\ref{sec:simulation}, code available in \cite{spinney_rspinneyamips_2024}).
   }
   \label{fig:unequal}
\end{figure*}

\subsubsection{Coexistence Densities are Dual to the Co-Moving Frame}
%
A crucial property of our effective theory is a duality between the implicit frame variables and the coexistence densities $\{\rho_G,\rho_L\}$. 
This equivalence exists because each bulk state must separately satisfy the condition in Eq.~(\ref{eq:F=0}).
We can view this in graphical terms by considering solutions of Eq.~(\ref{eq:F=0}) as the intersection of the parabola, $J_{\rm bulk}(\rho)$, by the linear construction of the lab-frame current, $J^{\rm st}_v + v\phi^{\rm st}$; the two intersection points correspond to the coexistence densities (see Appendix \ref{sec:duality_figure}). 
Pairs of $v$ and $J^{\rm st}_v$ for which there are no intersections do not satisfy the bulk stationary condition, whilst pairs for which the line is tangent to the parabola correspond to homogeneous solutions in their effective equilibrium frame [\emph{i.e.} using Eqs.~(\ref{eq:vJdegen})].
\par
We can find expressions for $v$ and $J^{\rm st}_v$ in terms of $\{\rho_G,\rho_L\}$ by solving the simultaneous equations $J_{\rm bulk}(\rho_L)=J^{\rm st}_v+v\rho_L$ and $J_{\rm bulk}(\rho_G)=J^{\rm st}_v+v\rho_G$, giving
\begin{align}
    \label{eq:vdual}
    v&=\text{Pe}\frac{1-\beta}{1+\beta}(1-\rho_G-\rho_L),\quad J^{\rm st}_v=\text{Pe}\frac{1-\beta}{1+\beta}\rho_G\rho_L.
\end{align}
As such, if one knows $\{v,J^{\rm st}_v\}$ they automatically know $\{\rho_G,\rho_L\}$, and \emph{vice versa}. 
\par
Importantly, this duality allows us to express all elements of nsAMB in terms of the coexistence densities. Explicitly, we now have the following, alternative, forms for $\mu_0$, $\pi$, and $F$
\begin{align}
\label{eq:pi_dual}
\pi(\phi)=&\frac{1-\beta}{1+\beta}\frac{\rho _G \left(2-\rho _L-\phi \right)-(\phi -2) \left(\rho
   _L-1\right)}{(1-\phi)^2},\\
   \label{eq:mu0_dual}
    \mu_0=&\text{Pe}\,\phi(1-\phi)-\frac{(1+\beta)^2}{2\beta\text{Pe}}\ln(1-\phi)-\text{Pe}\frac{(1-\beta)^2}{(1+\beta)^2}\rho_G\rho_L\nonumber\\
    &+\frac{(1-\beta)^2}{(1+\beta)^2}\text{Pe} \left(1-\rho _G-\rho _L\right)\nonumber\\
    &\qquad\times\frac{  \rho _G \left(\phi
   -\rho _L\right)+\phi  \left(\rho _L-1\right)}{ 1-\phi},\\
    \label{eq:Ffict_dual}
    F=&-\frac{1-\beta^2}{2\beta}\frac{(\phi-\rho_G)(\phi-\rho_L)}{1-\phi}.
\end{align}
In these forms it is immediately apparent that all terms that deviate from AMB (\emph{i.e.}, $\mu_0-\mu_{\rm bulk}$, $\pi$, and $F$) vanish in the case of $\beta=1$. 

\subsubsection{Unequal Tangent Construction}
\label{sec:unequal_tangents}
%

%
In the case of phase separation, inspection of Eq.~(\ref{eq:NSAMB}) immediately implies
\begin{align}
    \Delta \mu=\int_{x_L}^{x_G}\partial_x\mu\,dx=\int_{x_L}^{x_G}F\,dx,
\end{align}
where the notation $\int_{x_L}^{x_G}$ indicates an integral over a liquid-to-gas phase boundary. 
\par
Crucially, even though the values of $v$ and $J^{\rm st}_v$  which characterise a phase-separated state [Eqs.~(\ref{eq:vdual})] are different from those where the bulk states appear equilibriumlike [Eq.~(\ref{eq:vJdegen})], the bulk chemical potential remains frame invariant (Sec.~\ref{sec:mubulk}). As such, $\mu_0(\rho_G)=\mu_{\rm bulk}(\rho_G)$ and $\mu_0(\rho_L)=\mu_{\rm bulk}(\rho_L)$, and thus
\begin{align}
    \Delta \mu=\Delta\mu_{\rm bulk}.
\end{align}
In other words, phase-separated solutions are characterised by the bulk, equilibriumlike properties of the phases. 
Consequently, we associate the stationary solutions of Eqs.~(\ref{eq:NSAMB}) [and thus the limiting, $t\to\infty$ solutions of Eqs.~(\ref{eq:deterministic})] with the integral equation
\begin{align}
    \Delta\mu_{\rm bulk}&\coloneq \mu_{\rm bulk}(\rho_G)-\mu_{\rm bulk}(\rho_L)\nonumber\\
    &=f_{\rm bulk}'(\rho_G)-f_{\rm bulk}'(\rho_L)=\int_{x_L}^{x_G} F\,dx.
\end{align}
As such, phase-separated solutions of bQSAPs manifest as \emph{unequal} tangents on $f_{\rm bulk}$, with distinct chemical potentials and thermodynamic pressures between the two bulk states, thereby generalising the uncommon tangent construction of AMB (Sec.~\ref{sec:AMB}) where pressures differ, but chemical potentials are equal. 
\par 
An example of a numerical solution to Eqs.~(\ref{eq:deterministic}), exhibiting nonstationary MIPS (nsMIPS), and its corresponding unequal tangent construction on $f_{\rm bulk}$ is shown in Fig.~\ref{fig:unequal}(b).
In particular, we note that the chemical potential is \emph{greater} in the gaseous phase, $\Delta\mu_{\rm bulk}>0$. This property is general and it can be inferred from our bulk free energy (\ref{eq:fbulk}).
In the limit $\beta\to 0$, there is no phase separation, with all homogeneous states stable [\emph{cf.} Eq.~(\ref{eq:spinodal})], but the surviving term in $f_{\rm bulk}$ is convex, supporting arbitrary differences $\Delta\mu_{\rm bulk}<0$, but no differences $\Delta\mu_{\rm bulk}>0$, implying that the latter is required for coexistence with $\partial_x\mu=F$.
We will also see that such a property
can be determined from the form of the solution profiles which are characteristically asymmetric (see Sec.~\ref{sec:asym_boundaries}). 

\subsubsection{Asymmetric Interfacial Profiles}
\label{sec:asym_boundaries}
%
An immediate consequence of the characterisation of $\Delta\mu_{\rm bulk}$ between homogeneous solutions, and the form of $F$ in Eq.~(\ref{eq:Ffict_dual}), concerns the nature of the boundary solutions between the liquid and gaseous phases.
\par
In the case of stationary phase seperation (\emph{e.g.} $\beta=1$ in our model), phase boundaries are symmetric (\emph{i.e.} a liquid-to-gaseous boundary solution is the mirror image of a gaseous-to-liquid solution). 
However, in the nonstationary case this property is lost. 
This may be understood in thermodynamic terms from the integral form of the stationary requirement for $\Delta\mu_{\rm bulk}$ 
\begin{align}
    \Delta\mu_{\rm bulk}&=\int_{x_L}^{x_G} F\,dx\nonumber\\
    &=-\frac{(1-\beta^2)}{2\beta}\int_{x_L}^{x_G}\frac{(\phi^{\rm st}(x)-\rho_G)(\phi^{\rm st}(x)-\rho_L)}{1-\phi^{\rm st}(x)}\,dx,
\end{align}
comprised of positive contributions where $\phi\in(\rho_G,\rho_L)$, and negative contributions where $\phi\in[0,\rho_G)$ or $\phi\in(\rho_L,1)$, strongly tying the solution profile to changes in $\mu$.
\par
For example, since we expect $\Delta\mu_{\rm bulk}> 0$ at the liquid-to-gas boundary, a monotonic decreasing $\phi^{\rm st}(x)$ satisfies this condition. However, at the opposing boundary, the chemical potential difference \emph{must} be of equal magnitude, but of the opposite sign. Consequently, because $F$ is invariant at each boundary, the gas-to-liquid phase boundary \emph{cannot} be monotonic, nor reside exclusively within $[\rho_G, \rho_L]$, such that the solution must have a characteristic `overshoot' of the coexistence densities--- a prediction that is verified from simulation [Fig.~\ref{fig:unequal}(b)], reaffirming $\Delta\mu_{\rm bulk}>0$. 

\subsection{Implicit Field Theory in the Lab Frame}

\subsubsection{Chemical Potential Gradients are Conjugate to a Boundary Current}

Until now, we have only considered the relationship in Eq.~(\ref{eq:linearJ}) in the context of bulk, or flat, states; equating the currents in two, different bulk states was how we identified the inertial frame that co-moves with phase boundaries. To further understand the behaviour of bQSAPs across phase boundaries, however, we must consider densities where spatial gradients are {\it not} zero. In this case, Eq.~(\ref{eq:linearJ}) becomes
\begin{align}
    J_{\rm lim}=J^{\rm st}_v+v\phi^{\rm st}=J_\partial+J_{\rm bulk},
\end{align} 
where $J_{\rm bulk}$ is defined by Eq.~(\ref{eq:Jbulk}), as before. The new term, $J_\partial$, represents a \emph{boundary} current, which vanishes in the bulk, and is associated with the maintenance of phase interfaces.
\par
Crucially, we may write the non-conservative force, appearing in our field theory, as
\begin{align}
    F&=-\frac{(1+\beta)^2}{2\beta\text{Pe}(1-\phi)}\left[(v\phi+J^{\rm st}_v)-\text{Pe}\frac{1-\beta}{1+\beta}\phi(1-\phi)\right],
\end{align}
such that in the limiting state, $\phi\to\phi^{\rm st}$, the term in brackets equals $J_\partial$. 
Thus, in the limiting state, where $\partial_x\mu=F$, we obtain the flux-force relationship
\begin{align}
    \label{eq:muJpartial}
    J_\partial &=-\frac{2\beta\text{Pe}(1-\phi^{\rm st})}{(1+\beta)^2}\partial_x\mu,
\end{align}
directly identifying a physical flux in the laboratory frame with the thermodynamic force associated with gradients in chemical potential in our effective theory. This relation also motivates the choice of mobility function to be $M(\phi)=2\beta\text{Pe}(1-\phi)/(1+\beta)^2$, such that Eq.~(\ref{eq:NSAMB}) can be put in correspondence with Eq.~(\ref{eq:deterministic}).

As a result, we may write a lab frame version of Eq.~(\ref{eq:NSAMB}) by undoing the initial Galilean transform--- {\it i.e.}, by subtracting a term in $v\partial_x\phi$. As $t\to\infty$, however, we may identify $v\partial_x\phi^{\rm st} = \partial_x(v\phi^{\rm st}+J_v^{\rm st})=\partial_x J_{\rm lim}=\partial_x(J_\partial+J_{\rm bulk})$, thus giving
\begin{align}
    \label{eq:NSAMB_lab}
    \dot{\phi}_{\rm lab}&=-\partial_x J_{\rm bulk}+\frac{2\beta\text{Pe}}{(1+\beta)^2}\partial_x[(1-\phi_{\rm lab})\partial_x\mu],
\end{align}
such that we can verify 
\begin{align}
    \lim_{t\to\infty}\dot{\phi}_{\rm lab}&=-\partial_xJ_{\rm bulk}-\partial_xJ_{\partial}=-\partial_xJ_{\rm lim}.
\end{align}
Notably, where Eq.~(\ref{eq:NSAMB_lab}) is self-consistent, such that the values of $v$ and $J^{\rm st}_v$ lead to solutions that travel at a constant, uniform, $v$, it then shares the exact limiting solutions (including flux) of Eq.~(\ref{eq:deterministic}). 
It also reduces exactly to Eq.~(\ref{eq:deterministic}) in the case $\beta\to 0$. 
Notwithstanding its implicit nature, we note that Eq.~(\ref{eq:NSAMB_lab}) has a form very similar to that of the convective Cahn-Halliard (cCH) equation \cite{golovin_convective_2001}, although both the steady current and form of $\mu$ have specific structure which differs from those conventionally used in cCH.

\subsubsection{Interfacial Thermodynamic Interpretation}
%

\begin{figure*}[!htp]
    \includegraphics[width=\textwidth]{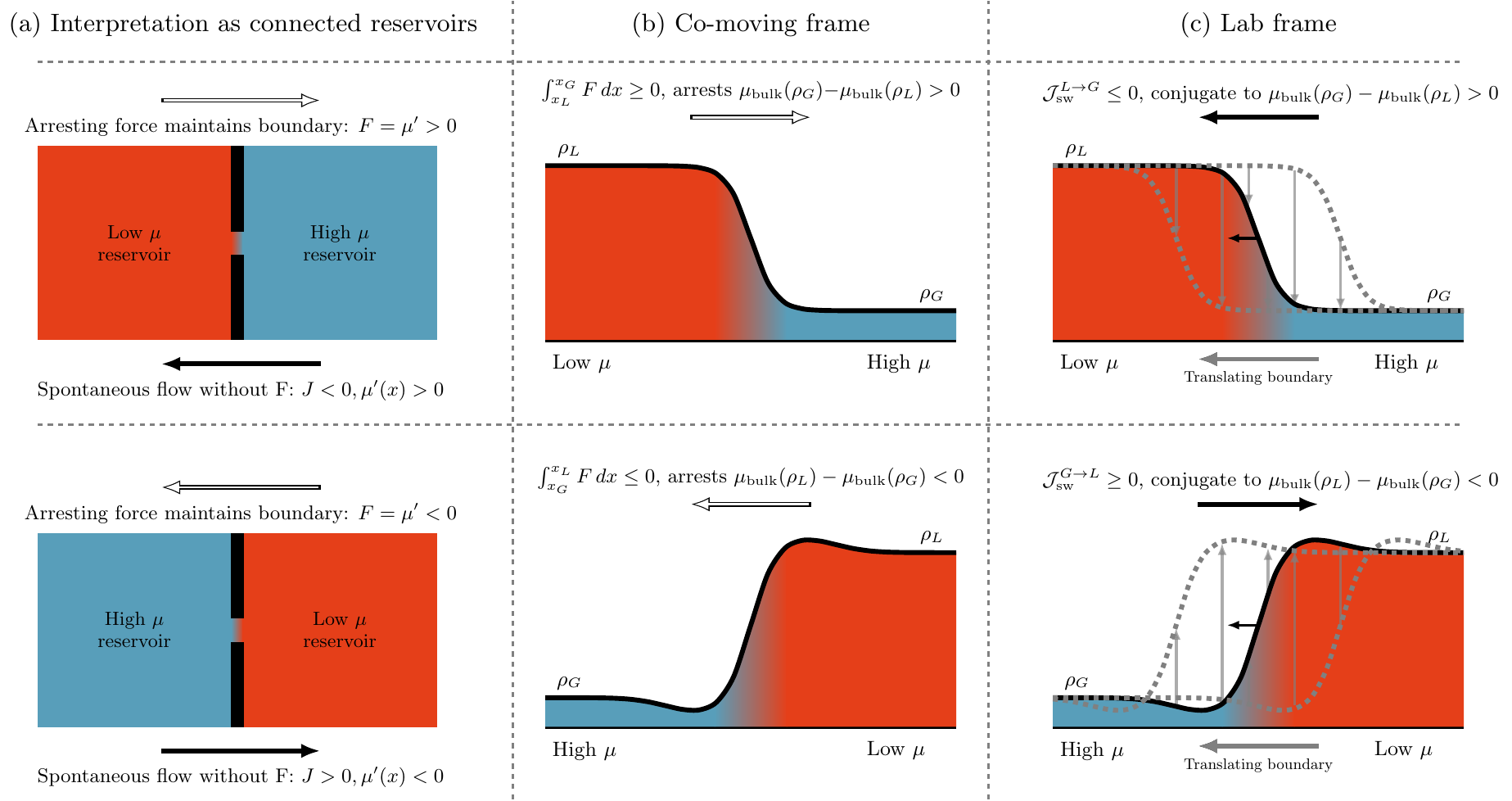} 
    \caption{
    {\bf Chemical Potential Differences in Different Frames: Arresting Forces and Swept Currents}.
    Since bulk phases appear equilibriumlike (see Sec.~\ref{sec:comoving}), we understand phase-separation by drawing analogy with two connected equilibrium reservoirs, each of which have different chemical potentials that are derived from their density [Panel (a), left].
    In the co-moving frame, the flux that would otherwise result from a difference in chemical potentials, $\Delta\mu_{\rm bulk}$, is arrested by an integrated force $\int_{x_L}^{x_G} F\, dx$ that maintains the interfacial boundary [Panel (b), middle].
    In the lab frame, $\Delta\mu_{\rm bulk}$ is conjugate to a flux that corresponds to the mass swept out by the moving boundary, $\mathcal{J}_{\rm sw}$ [Panel (c), right]. The top row illustrates the liquid-to-gas boundary, whilst the bottom row illustrates the gas-to-liquid boundary. 
    }
    \label{fig:boundary}
\end{figure*}
Phase separated solutions of Eq.~(\ref{eq:NSAMB}) (\emph{i.e.} in the co-moving frame) have a natural thermodynamic interpretation in terms of two reservoirs with distinct chemical potentials connected by a boundary solution. Ordinarily, one would expect a resultant flux. However, due to the non-conservative force, $F$, this flux is arrested leaving the system in a stationary, equilibriumlike, state, with such an interpretation illustrated in Fig.~\ref{fig:boundary}(a) and (b).
\par
By contrast, in the lab frame we have Eq.~(\ref{eq:muJpartial}), directly implying $\Delta\mu_{\rm bulk}\propto\int_{x_L}^{x_G}J_\partial/(1-\phi)\,dx$, 
however this is a difficult quantity to interpret. Instead, we consider the interpretation that the moving phase boundaries are associated with a flux which arises from the difference in chemical potentials between the effective reservoirs. 
To characterise this, we quantify the density that is `swept out' by the moving boundary per unit time across the liquid-to-gas interface through
\begin{align}
\label{eq:Jswept}
    \mathcal{J}_{\rm sw}^{L\to G}&\coloneq \int_{x_L}^{x_G} \partial_t\rho_{\rm lim}(x,t)\,dx=-\int_{x_L}^{x_G} \partial_xJ_{\rm lim}(x,t)\,dx\nonumber\\
    &=-v\int_{x_L}^{x_G} \partial_x\phi^{\rm st}(x)\,dx=-v\Delta\phi=-\Delta J_{\rm bulk}(\phi),
\end{align}
where $\Delta J_{\rm bulk}\coloneq J_{\rm bulk}(\rho_G)-J_{\rm bulk}(\rho_L)$. 
This quantity is strictly negative at the liquid-to-gas interface since $v<0$, and its equivalent, $\mathcal{J}_{\rm sw}^{G\to L}$, at the gas-to-liquid interface similarly must be positive. 
Meanwhile, we may characterise the difference in chemical potential across such a boundary solution using Eq.~(\ref{eq:mubulk}) as
\begin{align}
    \Delta\mu_{\rm bulk}&=\frac{4\beta\text{Pe}}{(1+\beta)^2}(1-\rho_G-\rho_L)(\rho_G-\rho_L)\nonumber\\
    &\quad+\frac{(1+\beta)^2}{2\beta\text{Pe}}[\ln(1-\rho_L)-\ln(1-\rho_G)],
\end{align}
which may immediately be written in terms of the above current 
\begin{align}
    \label{eq:deltamubulk}
    \Delta \mu_{\rm bulk} &= \frac{4\beta}{1-\beta^2}\Delta J_{\rm bulk}-\frac{(1+\beta)^2}{2\beta\text{Pe}}\Delta \ln(1-\phi)\nonumber\\
    &= -\frac{4\beta}{1-\beta^2}\mathcal{J}_{\rm sw}^{L\to G}-\frac{(1+\beta)^2}{2\beta\text{Pe}}\Delta \ln(1-\phi),
\end{align}
where $\Delta \ln(1-\phi)\coloneq \ln(1-\rho_G)-\ln(1-\rho_L)$, thus explicitly relating motion of the phase boundaries to the change in chemical potential. Moreover, we always have $\Delta\mu_{\rm bulk}>0$ and $\mathcal{J}_{\rm sw}^{L\to G}<0$, whilst both vanish in the symmetric case $\beta=1$, indicating that these may be thought of as a conjugate force-flux pair, with the information-theoretic like term only contributing quantitative differences due to dispersion along the boundary. This interpretation is illustrated in Fig.~\ref{fig:boundary}(a) and (c).
\par
In addition to this, by rearranging the above such that we have
\begin{align}
    \mathcal{J}^{L\to G}_{\rm sw}&=-\frac{1-\beta^2}{4\beta}\Delta\mu_{\rm bulk}-\frac{(1-\beta)^2(1+\beta)^2}{8\beta^2\text{Pe}}\Delta \ln (1-\phi),
\end{align}
we can relate the chemical potential difference to the inter-species current that has been coarse-grained out of the full model [Eq.~(\ref{eq:deterministic})]. Explicitly, we may write
\begin{align}
    &\mathcal{J}^{L\to G}_{\rm sw}\nonumber\\
    &=-\frac{1-\beta^2}{4\beta}\int_{x_L}^{x_G} \partial_x\mu\,dx-\frac{(1-\beta)^2(1+\beta)^2}{8\beta^2\text{Pe}}\Delta \ln (1-\phi)\nonumber\\
    &=\frac{(1-\beta^2)(1+\beta)^2}{8\beta^2\text{Pe}}\int_{x_L}^{x_G} \frac{J_\partial+\partial_x\phi}{1-\phi}\,dx\nonumber\\
    &=\frac{(1-\beta^2)(1+\beta)^2}{8\beta^2}\int_{x_L}^{x_G} \left( \psi - \frac{1-\beta}{1+\beta}\rho\right)\,dx\nonumber\\
    &=\frac{1-\beta^2}{2\bar{\gamma}\beta}\int_{x_L}^{x_G} (\gamma^-\rho^+-\gamma^+\rho^-)\,dx\nonumber\\
    &\eqqcolon\frac{1}{\bar{\gamma}}\frac{1-\beta^2}{2\beta}\mathcal{J}_{+\to -}^{L\to G},
\end{align}
where $\mathcal{J}_{+\to -}^{L\to G}$ is defined as the integral in the penultimate line which captures the net flux--- and, correspondingly, the breakage of detailed balance--- between the right and left moving particles across the interface. In the above we have used $\Delta\mu_{\rm bulk}=\int_{x_L}^{x_G}\partial_x\mu\,dx$ on the first line,  then Eq.~(\ref{eq:muJpartial}) on the second line. To reach the third line we used $J_{\rm lim}=J_\partial+J_{\rm bulk}=\text{Pe}\psi(1-\rho)-\partial_x\rho$, alongside the equivalence $\phi^{\rm st}=\lim_{t\to\infty}\rho$. The penultimate line is then reached by using relations $\psi=\rho^+-\rho^-$, $\rho=\rho^++\rho^-$, $\beta=\gamma^-/\gamma^+$, and $\gamma^+=\bar{\gamma}(1+\beta)/\beta$.
\par
Finally, by comparison with Eq.~(\ref{eq:deltamubulk}) we can cast $\Delta\mu_{\rm bulk}$ as an integrated hidden flux and a difference in a modified entropy density
 \begin{align}
 \Delta\mu_{\rm bulk}&=2\bar{\gamma}^{-1}\mathcal{J}^{L\to G}_{+\to -}-\frac{(1+\beta)^2}{2\beta\text{Pe}}\Delta \ln(1-\phi).
  \end{align}
  In the case of $\beta=1$, $\Delta\mu_{\rm bulk}=0$--- the integrated flux exactly matches a change in the entropy density term. Where $\beta<1$ these quantities differ, leading to a chemical potential difference, which is in turn associated with the procession of the phase boundaries.

\subsection{Implicit vs. Explicit Theories}
\label{sec:implicit_explicit}
A notable peculiarity of our effective field theory, be it in the co-moving [Eq.~(\ref{eq:NSAMB})] or lab [Eq.~(\ref{eq:NSAMB_lab})] frame, is its implicit nature. Specifically, its form depends on properties of its own solution--- namely its constant procession velocity, $v$, and limiting current in the frame where it appears stationary, $J^{\rm st}_v$. As such, whilst its formulation has been useful to obtain an effective thermodynamics and construct solutions that match the underlying bQSAPs, it cannot be expected to embody a truly physical theory. Moreover, whilst Eq.~(\ref{eq:NSAMB_lab}) shares solutions and limiting flux with the underlying system of bQSAPs, it cannot capture the exact stability properties of phase separated states because different bulk states appear equilibriumlike in distinct frames. Specifically, phase-separated states use implicit variables from Eq.~(\ref{eq:vdual}) whereas bulk states require Eqs.~(\ref{eq:vJdegen}). The ramifications of this are explored in the Supplemental Material \cite{SM}, where it is shown that self-consistent solutions to Eq.~(\ref{eq:NSAMB_lab}) nevertheless possess the same qualitative phase structure as Eq.~(\ref{eq:deterministic}). As we will go on to discuss in more detail in Secs.~\ref{sec:fluctuations} and \ref{sec:characteristic}, this type of frame-dependence ultimately derives from the property that bQSAPs support multiple characteristic velocities simultaneously.
\par
It is natural, therefore, to ask if an \emph{explicit} single order parameter theory could be constructed which captures the essential properties of bQSAPs. Indeed, a plausible form is motivated by Eq.~(\ref{eq:NSAMB_lab}), which, after ignoring any implicit dependencies, becomes
\begin{subequations}
\label{eq:NSAMBexplicit}
\begin{align}
\dot{\phi}&=-\partial_xJ,\quad J=J_{\rm bulk}-\frac{2\beta\text{Pe}}{(1+\beta)^2}(1-\phi)\partial_x\mu,\\
\mu&=\mu_{\rm bulk}(\phi)-\kappa(\phi)\partial_x^2\phi\nonumber\\
&\qquad\qquad+\lambda(\phi) (\partial_x\phi)^2+\frac{\pi}{\text{Pe}(1-\phi)}\frac{1-\beta}{1+\beta}\partial_x\phi,
\end{align}
\end{subequations}
in this case leaving $\pi$ as a constant or some other explicit function of $\phi$ (or simply even $\pi=0$). As written, Eq.~(\ref{eq:NSAMBexplicit}) trivially possesses the same $k\to 0$ stability properties and bulk current as Eq.~(\ref{eq:deterministic}), and reduces to the appropriate form of AMB for symmetric QSAPs in the case $\beta\to 1$. Moreover, it shares, in principle, the same thermodynamic structure (\emph{i.e.}, unequal tangent construction, bulk chemical potential differences \emph{etc.}) as our implicit theory.
\par
Indeed, such a model can be further simplified so as to place it alongside the traditional suite of canonical field theories (model B, AMB, AMB+, active model H \emph{etc.} \cite{halperin_calculation_1972,cates_active_2022,tjhung_cluster_2018}). In arbitrary dimension, this leads to an \emph{explicit} nsAMB (in the lab frame)
\begin{subequations}
\label{eq:NSAMBgeneral}
\begin{align}
    \partial_t\phi &= -\nabla\cdot J,\quad J = J_0(\phi)\,\hat{n} - M\nabla\mu + \Lambda,\\
    \mu &= \frac{\delta {\mathcal F}}{\delta\phi} + \lambda \left\vert \nabla \phi\right\vert^2 + \pi\nabla \phi\cdot\hat{n},\\
    \mathcal{F} &= \int dV \left[ -\phi^2/2 + \phi^4/4 +\kappa\left(\nabla\phi\right)^2/2\right],
\end{align}
\end{subequations}
where $\Lambda$ is a (vector-valued) spatio-temporal white noise, $J_0(\phi)$ is a steady driving flux, $\hat{n}$ is a unit-vector specifying the direction of the underlying bias, and with all other parameters treated as constants.
\par
 However, we urge caution; use of such heuristically constructed models could be misleading. Since solutions of Eqs.~(\ref{eq:NSAMBexplicit}) and (\ref{eq:NSAMBgeneral}) cannot be put in {\it quantitative} correspondence with the underlying system of bQSAPs, they will not share coexistence densities or resultant thermodynamic properties, $\Delta\mu_{\rm bulk}$ \emph{etc.}. 
 Here, we consider correspondence with substantive microscopic models paramount, and therefore we do not consider (\ref{eq:NSAMBexplicit}) or (\ref{eq:NSAMBgeneral}) further. 
We nevertheless propose such explicit theories as an interesting avenue for future study. In particular, knowing the extent to which choices, for $\mu$, $J_0$ \emph{etc.}, can reproduce qualitative behaviours of bQSAPs, including phase structures, limiting solutions, and transient properties would provide much needed clarity.
\par
That an explicit single order-parameter theory seemingly cannot be put in complete correspondence with bQSAPs, whilst one can in the case of symmetric QSAPs ($\beta=1$), prompts consideration about the inherent limits of capturing increasingly non-equilibrium behaviour with a single order-parameter. Active field theories (AMB, AMB+, \emph{etc.}) all perturb an essentially equilibrium structure (\emph{e.g.}, Model B), where bulk states are taken to be in some effective equilibria, implicitly in a single frame of reference ($v=0$). By contrast, bulk states of bQSAPs are manifestly out-of-equilibrium, with persistent flux, only appearing equilibriumlike in distinct inertial frames. 
We thus conjecture that nonstationary models will, in general, cease to be captured quantitatively by single parameter theories, with their description thus requiring additional fields, such as the inclusion of an effective momenta [\emph{e.g.}, Eq.~(\ref{eq:deterministic})].

%
%
%
%
%
%
%
%
%
%
%
%
\section{Anomalous phase structure: decoupling spinodals from binodals}
\label{sec:anomalous}
%
The most striking consequence of our unequal tangent construction is that the binodal lines will not appear in their usual position on a phase diagram. In particular, the otherwise strong constraint of binodals lying outside the spinodal line, as in passive and active model B, is lost.

\subsection{Symmetric QSAPs Possesses a Conventional Phase Structure}
In conventional phase separation, binodal solutions always lie outside the spinodal region and are thus always stable, both locally and globally.
Moreover, they can be calculated directly from the common tangent construction, which provides the two necessary constraints for obtaining $\rho_L$ and $\rho_G$: $\Delta\mu^{\rm eq}=\Delta P=0$.
\par
On first inspection, such a procedure appears under-determined for AMB owing to the uncommon tangent characterisation and \emph{a priori} unknown $\Delta P\neq 0$.
This suggests that coexistence densities must be recovered from numerical solution to, for example, Eq.~(\ref{eq:deterministic}) (with $\beta=1$).
However, a striking property of AMB holds if a transform of the order parameter, $R(\phi)$, exists such that $\kappa(\phi)R''(\phi)=-(2\lambda(\phi)+\kappa'(\phi))R'(\phi)$.
If this is true [as it is in the case of $\beta=1$ through $R(\phi)=\ln(1-\phi)$] the entirety of the deterministic component of the AMB dynamic may be written in terms of a functional derivative \emph{viz.}
\begin{align}
\dot{\phi}&=\partial_x\left[M(\phi)\partial_x\frac{\delta \mathcal{G}[R]}{\delta R}\right],
\label{eq:AMBR}
\end{align}
where $\mathcal{G}[R]=\int dx\,g(R,|\partial_xR|^2)$, and $g(R,(\partial_xR)^2)=g_{\rm bulk}(R)+(\kappa/2R')|\partial_xR|^2=g_{\rm bulk}(R)-|\partial_xR|^2/2\text{Pe}$ \cite{solon_generalized_2018}.
Consequently, AMB has phase boundary solutions and thus coexistence densities (but not phase volumes) that coincide with those of a fictitious \emph{equilibrium} dynamic in the non-physical variable $R$, $\dot{R}=\partial_xM\partial_x\delta\mathcal{G}[R]/\delta R$.
Since these fictitious dynamics are integrable in $R$, they extremise $\mathcal{G}[R]$ and thus, in our $\beta=1$ case, correspond to a \emph{common-tangent} construction on $g_{\rm bulk}(R)=\int_R \mu_{\rm bulk}(1-e^r)\,dr=\text{Pe}e^R(2-e^R)/2-R^2/\text{Pe}$, the result of which is shown in Fig.~\ref{fig:phase_diagrams}(a) and (b). 
Importantly, the resulting phase structure is qualitatively identical to passive model B since $R=\ln(1-\phi)$ is monotonic on $\phi\in[0,1]$ and  chemical potentials are equal across both descriptions, \emph{i.e.} $\mu_{\rm bulk}\equiv g'_{\rm bulk}(R)=f'_{\rm bulk}(\phi)$.
Consequently both formulations agree on the location of the spinodals where $f''_{\rm bulk}(\phi)=g''_{\rm bulk}(R)=0$. As such, spinodals are still rigorously within the binodal solutions, meeting at the critical point, with the effect of the active $\lambda$ component only serving to quantitatively change the location of the binodals within the stable region of phase space.
\par
Crucially, there is no transform that allows nsAMB to be written in such an integrable way. Consequently, nsAMB is equivalently characterised by an unequal tangent construction on $g_{\rm bulk}(R)$ and on $f_{\rm bulk}(\phi)$, with the difference between the two formulations manifesting only as a difference in the intercepts/pressures.

\subsection{Unequal Tangents Imply Offset Binodals}
Notably, the difference in bulk chemical potentials that is implied by the unequal tangent construction must persist even arbitrarily close to the bifurcation point that marks the splitting of binodal lines.
Explicitly, if we consider the limit of the bifurcation point, approached from above in $\text{Pe}$, such that $\rho_L,\rho_G\to \phi=\rho_0$ represented as $\rho_{L}=\rho_0+ d\phi/2$, $\rho_{G}=\rho_0- d\phi/2$, we have to first order in $d\phi=\rho_L-\rho_G$
\begin{align}
    \Delta\mu_{\rm bulk}&=\mu_{\rm bulk}(\rho_G)-\mu_{\rm bulk}(\rho_L)\nonumber\\
    &=-\mu'_{\rm bulk}(\rho_0)d\phi + \mathcal{O}(d\phi^2)\nonumber\\
    &=-f''_{\rm bulk}(\rho_0)d\phi + \mathcal{O}(d\phi^2).
\end{align}
Since we have $\Delta \mu_{\rm bulk}>0$, the above indicates that at the bifurcation point the curvature of the free energy density is \emph{negative}, in turn implying that it is located \emph{within} the spinodal region.
\par 
Further, if we consider the $\text{Pe}\to\infty$ limit of nsAMB it shares the same limiting form as AMB, such that all nonstationary terms are sub-dominant, with the system converging on the saturated solution $\{\rho_G,\rho_L\}=\{0,1\}$, with infinitely sharp boundaries and $\Delta\mu_{\rm bulk}=0$. Since AMB possesses a conventional phase structure with binodals always lying outside the spinodals, this implies that the binodal lines of nsAMB/bQSAPs will \emph{cross} the spinodals at intermediate values of $\text{Pe}$, with no requirement that the liquid and gaseous binodal lines cross the spinodals at the same value of $\text{Pe}$. 

\subsection{Self-Consistent Shooting Method for Finding Binodals Inside Spinodals}
\label{sec:shooting}
%

\begin{figure*}[!htp]
   \centering
   \includegraphics[width=\textwidth]{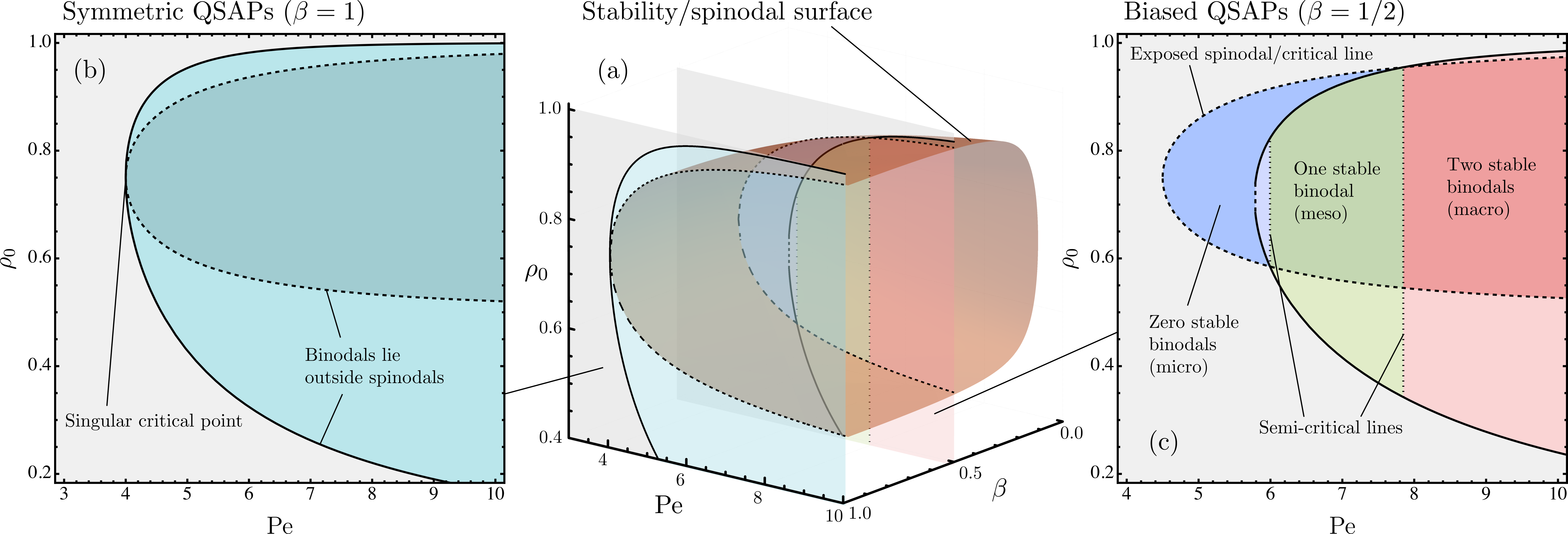} 
   \caption{
        {\bf Anomalous Phase Structure of bQSAPs}.
        Since symmetric QSAPs display conventional, stationary MIPS, the $\beta = 1$ `slice' of the phase diagram has the expected structure: 
        lines of stable phase coexistence (binodals, solid lines) encase those that mark the onset of spontaneous phase separation (spinodals, dashed lines), and the two meet at a single location, the critical point [panels (b) and (a), left and centre].
        Away from this limit, nonstationarity leads to an exotic phase structure, exemplified by the $\beta = 1/2$ `slice' [panels (c) and (a), right and centre].
        Here, the unequal tangent construction permits phase coexistence lines to enter the unstable region, which is otherwise forbidden in either passive phase separation or conventional, stationary, MIPS.
        The decoupling of binodals and spinodals permits the identification of three clear regions: (\textit{i}) where both binodal densities are stable, which we identify as macro-phase separation [panel (c), right, red]; (\textit{ii}) where a single binodal density is stable, which we identify with meso-phase separation [panel (c), right, green], and; (\textit{iii}) where there are no stable binodal densities, which we identify with micro-phase separation [panel (c), right, blue].
        Notably, there is no longer a single critical point, but instead an entire section of the spinodal line which is exposed, acting as a critical \emph{line}. Thus, where one of the coexistence densities lie precisely on the spinodal we identify that only one of the phases experiences criticality and thus call lines in phase space where this occurs semi-critical lines. These occur at $\text{Pe}\simeq 5.994$ and $\text{Pe}\simeq 7.848$. Data is absent in this vicinity of the binodal bifurcation point, but we estimate it to occur at around $\text{Pe}\simeq 5.78$ and $\rho_0\simeq 0.71$ for the $\beta=1/2$ case in panel (c). Binodals for the $\beta=1/2$ case are deduced using the methodology described in Sec.~\ref{sec:shooting}.
    }
   \label{fig:phase_diagrams}
\end{figure*}

Since solutions that aren't dynamically stable cannot be expected to be converged-on in simulation, we are, in general, unable to use numerical approximation of Eq.~(\ref{eq:deterministic}) to find binodal solutions inside the spinodal region.
Instead, we recognise that such solutions, by definition, coincide with the condition $\partial_x\mu-F=0$ from Eq.~(\ref{eq:L}), where they are defined without reference to any time evolution. In other words, we can use our effective thermodynamic treatment to sidestep the issue of dynamical stability.
However, Eq.~(\ref{eq:L}) is defined in terms of $v$ and $J^{\rm st}_v$, which are unknown.
As such, we exploit the duality in Eqs.~(\ref{eq:vdual}), valid for the infinite, bulk phase separated solutions that we seek, and construct an operator ${\mathcal{L}}_{\rho_L,\rho_G}[\phi]=\partial_x\mu -F$ by exchanging all of $\mu_0$, $\pi$, and $F$ for those that appear in Eqs.~(\ref{eq:pi_dual}-\ref{eq:Ffict_dual}).
\par
This operator then defines an ODE (${\mathcal{L}}_{\rho_L, \rho_G}[\phi]=0$) whose solution specifies our weak-sense coexistence densities according to the self-consistency requirement
\begin{align}    
    \label{eq:BVP}
    \{\rho_L,\rho_G\}&=\Big\{\left\{a,b\right\}|\,{\mathcal{L}}_{a,b}[\phi]=0,\nonumber\\
    &\qquad\,\text{s.t.}\,\lim_{x\to -\infty}\phi(x)=a,\,\lim_{x\to +\infty}\phi(x)=b\Big\},
\end{align}
such that the limiting, $t\to\infty$ bulk densities of a liquid-to-gas solution profile match the densities appearing in the operator itself \footnote{Note, we explicitly consider the liquid-to-gas boundary solution as it is the boundary that sets the coexistence densities for the case $\gamma^+\geq\gamma^-$ that we consider (See Supplemental Material \cite{SM}).}.
That is, the coexistence densities $\{\rho_G,\rho_L\}$ are those which, when used as constants in ${\mathcal{L}}_{\rho_L,\rho_G}[\phi]=0$, lead to a phase separated solution at densities $\rho_L$ and $\rho_G$. Finding the $\rho_L$ and $\rho_G$ that satisfy Eq.~(\ref{eq:BVP}) thus amounts to solving an implicit boundary value problem, where the derivatives of the function at the boundaries are known (they tend to zero in the bulk), but the values of the function at the boundary appear as unknown parameters in the ODE.
\par
We convert this into an initial value problem by way of a shooting method. This involves the repeated generation of trial solutions, through integration of ${\mathcal{L}}_{\rho_L,\rho_G}[\phi]=0$, from an initial, approximately bulk, solution at $\phi(0)\simeq\rho_L$, using different values of $\rho_L$, $\rho_G$ and initial conditions $\phi'(0)$ and $\phi''(0)$ until Eq.~(\ref{eq:BVP}) is (approximately) satisfied.  Performing a full parameter sweep is not tractable, so instead we treat the task as an optimisation problem through the construction of a suitable objective function which vanishes when the trial solution is self-consistent. By performing this procedure for different values of $\text{Pe}$ we are able to compute the binodal lines shown in Fig.~\ref{fig:phase_diagrams}(c). Full details are given in Appendix \ref{app:shooting}. (Pseudo-code for the algorithms is given in the Supplemental Material \cite{SM}).

\subsection{Structure of the Phase Diagram}
For AMB (\emph{i.e.}, symmetric QSAPs) and equilibrium phase separation, a point in phase space being within/outside the spinodals indicates the stability of a homogeneous solution, whilst being within/outside the binodals indicates the existence of a phase separated solution.
Conventionally, this leads to the identification of three distinct regions of phase space: a one-phase mixed state; an unstable region with phase separated solutions (phase separation via spinodal decomposition), and; a meta-stable region with phase separated solutions (phase separation via nucleation and growth), [\emph{cf.} Fig.~\ref{fig:phase_diagrams}(b)].
\par
By contrast, in the case of bQSAPs, the phase structure can consist of up to \emph{seven} regions based simply on the geometric intersections of the now offset spinodal and binodal lines. Specifically, whereas with passive or active Model B, binodals must always lie outside the spinodals, here there can now be: regions which are unstable, but where there are no binodal solutions at all; regions where both binodals lie inside the spinodal; regions where only one lies inside the spinodal, and; regions where both lie outside the spinodal. These are shown in distinct colours in Fig.~\ref{fig:phase_diagrams}(c).
\par
The different regions of the phase diagram correspond to different underlying behaviours. This can be seen from study of the deterministic equations of motion [Eq.~(\ref{eq:deterministic})], 
which is described in detail in the Supplemental Material \cite{SM}. 
In the red regions, where both coexistence densities are stable, we observe (nonstationary) phase separation on arbitrary length-scales, with stability of initial, approximately homogeneous, states governed by their relation to the spinodal line.
In the blue region, where there are no coexistence solutions, or both coexistence densities are unstable, we observe evolution towards short, oscillatory, travelling peaks that we term `micro-phase separation'.
These states can be reached by crossing the exposed spinodal line from the one phase region.
The exposed spinodal acts as a critical line (see Sec.~\ref{sec:critical_spinodal}), which, when crossed, incurs a Hopf bifurcation as the real part of the complex eigenvalues of Eq.~(\ref{eq:A}) change sign.
As such, the system transitions from one where there is no apparent procession ($\rho=\rho_0$) to one with small perturbations that travel with velocity $v$, in the manner of a continuous phase transition.
Then, as the system moves further into the spinodal region, these small perturbations grow into larger peaks.
Finally, in the green regions, coexistence solutions exist where the gas density is stable, but the liquid density is unstable. Here, we do not see a disintegration of the liquid phase into micro-phase separation in the sense observed in the blue regions. Instead, at high $\text{Pe}$ we observe phase-separation that is effectively the same as that observed in the red regions, with the liquid coexistence densities indistinguishable from the coexistence lines that make up the binodals in Fig.~\ref{fig:phase_diagrams}(c).
Then, as $\text{Pe}$ is reduced, the maximal length of the liquid phase becomes truncated, with the liquid domains becoming shorter as the liquid coexistence density moves into the spinodal region.
Because, in this region, gas phases can persist on arbitrary length-scales, whilst liquid phases appear bulk-like with tuneable length-scales, we term such behaviour `meso-phase separation'. We note that between the `micro' and `meso' regions, and between the `meso' and `macro' regions, one of the binodal lines crosses the (critical) spinodal line--- this motivates the description of the `semi-critical' lines on the phase diagram in Fig.~\ref{fig:phase_diagrams}(c).
\par
To understand the full implications of the anomalous phase structure, however, requires a quantitative understanding of the role of fluctuations.
%
\section{Fluctuations and criticality}
\label{sec:fluctuations}
%
In passive or active Model B (and indeed almost all other field theories) the use of deterministic equations of motion are commonly justified by satisfying a Ginzburg criterion \cite{chaikin_principles_1995}.
Via sufficient coarse-graining and/or reduction in noise strength, this is achievable everywhere except for the critical point.
At the critical point, coexistence densities converge on the spinodal line and fluctuations are manifestly important to understanding behaviour.
\par
For our system of bQSAPs, the region where such a strategy fails has now been dramatically expanded.
Wherever part or all of the system lies on or within the spinodal region [the blue and green regions in Fig.~\ref{fig:phase_diagrams}(c)], fluctuations are now amplified; a Ginzburg criterion cannot be satisfied and fluctuations, however small, will grow and become important on macroscopic scales.

\subsection{Stochastic Dispersion Relations}
%

\begin{figure*}[!htp]
   \centering
   \includegraphics[width=\textwidth]{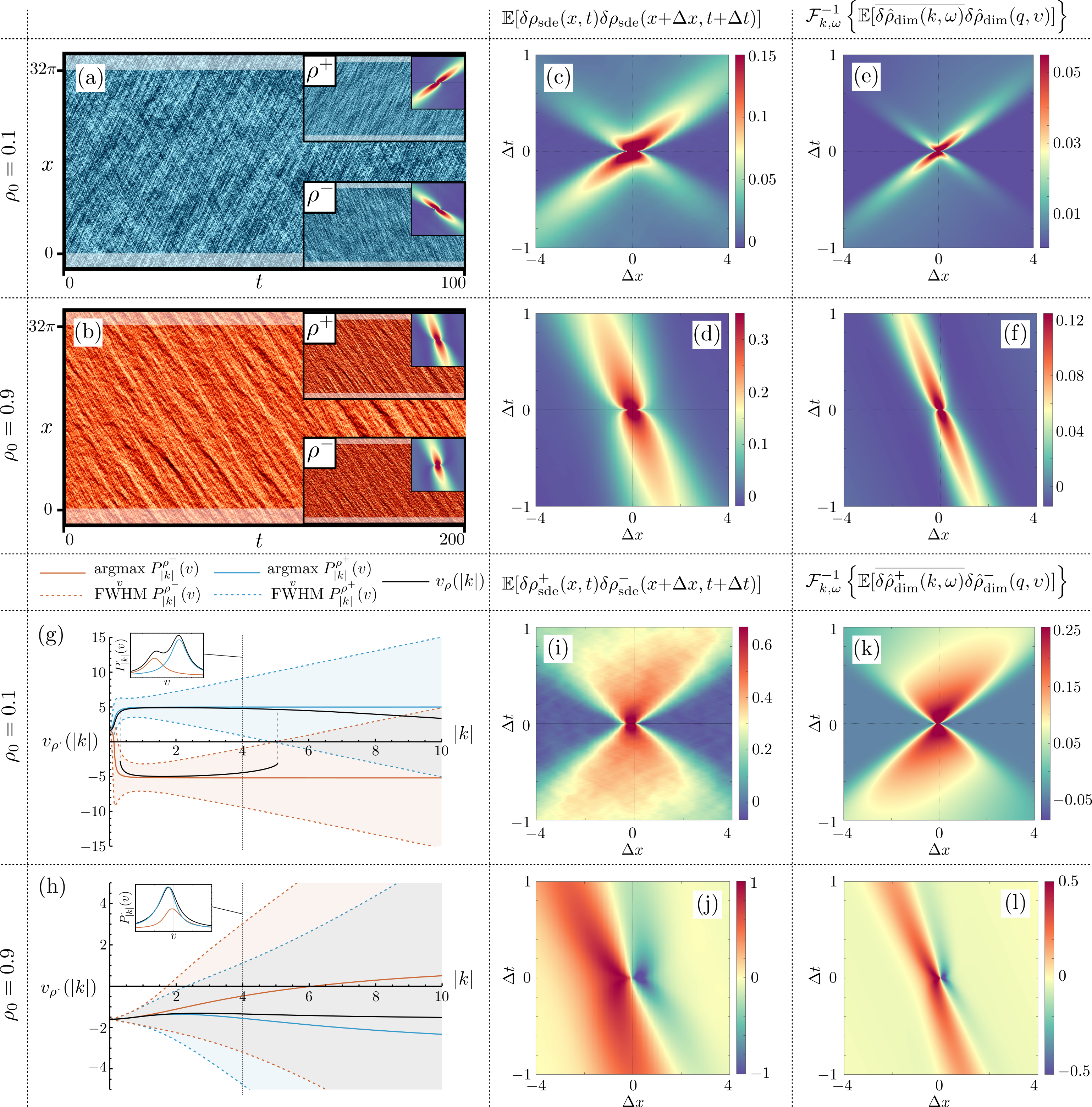} 
   \caption{
        {\bf Fluctuating Behaviour of bQSAPs ($\beta=1/2$)}. 
       Top left panels: Total density ($\rho$) kymographs for low [panel (a), $\rho_0=0.1$, $\text{Pe}=9$] and high [panel (b), $\rho_0=0.9$, $\text{Pe}=4.5$] densities/interaction strengths. Inset are density kymographs and empirical correlation functions for individual sub-species ($\rho^\pm$). Top right panels: Empirical density-density space-time correlation functions for the low and high density simulations [panels (c) and (d)] alongside inverse Fourier transforms of the redimensionalised analytical power spectra [panels (e) and (f)]. Bottom right panels: Empirical sub-species cross correlation functions [panels (i) and (j)] alongside inverse Fourier transforms of the associated power spectra [panels (k) and (l)]. Bottom left panels: Properties of the velocity power spectra against wavenumber for low [panel (g)] and high [panel(h)] densities. The black lines show the stochastic dispersion relation, $v_\rho(|k|)$, [Eq.~(\ref{eq:stoch_dispersion})]. Blue lines and regions show the peak and width of the velocity power spectra for $\rho^+$, respectively. Red lines and regions show the same quantities for $\rho^-$. Inset are velocity power spectra at an illustrative $|k|=4$. At low density the predominant fluctuation velocity is positive [\emph{cf.} Eq.~(\ref{eq:vfluct})], but the species are weakly coupled such that both species are visible in the density kymographs, the correlation functions, and dispersion relations. At high density the predominant fluctuation velocity is negative [\emph{cf.} Eq.~(\ref{eq:vfluct})], with the species highly coupled such that below $|k|\sim 6$ both species' fluctuations travel in the negative direction, with features in $\rho^-$ following those in $\rho^+$ with a characteristic lag, as can be deduced from the cross-correlation function.
    }
   \label{fig:fluctuations}
\end{figure*}

To characterise the dispersive, nonstationary nature of the full, fluctuating dynamics [Eq.~(\ref{eq:SPDE})], we start by analysing the space-time Fourier transforms of small fluctuations around the homogeneous state, $\delta\rho=\rho-\rho_0$, $\delta\psi=\psi-\psi_0$.
Defining $\delta\hat{\rho}(k,\omega)=\mathcal{F}_{x,t}\{\delta\rho(x,t)\}=\int_{0}^{N_L} dx\;\int_{-\infty}^{+\infty} dt\;e^{-i(kx- \omega t)}\delta\rho(x,t)$ and $\delta\hat{\psi}(k,\omega)=\mathcal{F}_{x,t}\{\delta\psi(x,t)\}=\int_{0}^{N_L} dx\;\int_{-\infty}^{+\infty} dt\;e^{-i(kx-\omega t)}\delta\psi(x,t)$, this gives, to linear order 
\begin{align}
   \label{eq:linearSPDEkomega}
\begin{bmatrix}
\delta\hat{\rho}(k,\omega)\\
\delta\hat{\psi}(k,\omega)
\end{bmatrix}=-\sqrt{\varepsilon}\left(i\omega\mathbb{I}+\mathbf{A}(k)\right)^{-1}\hat{\boldsymbol{\xi}}(k,\omega),
\end{align}
with $\mathbf{A}(k)$ given by Eq.~(\ref{eq:A}). Here we have a noise strength $\varepsilon=\rho_0^2 N_L/N$ and a noise vector 
\begin{align}
\hat{\boldsymbol{\xi}}(k,\omega)&=
\begin{bmatrix}
\sqrt{2}ik\hat{\eta}_\rho(k,\omega)\\
\sqrt{2}ik\hat{\eta}_\psi(k,\omega)+2\hat{\eta}_\leftrightarrow(k,\omega)
\end{bmatrix},
\end{align}
comprised of Fourier transformed space-time white noises, $\hat{\eta}_\cdot(k,\omega)=\mathcal{F}_{x,t}\{\eta_{\cdot}(x,t)\}$, 
with constituent fields possessing correlations (at first order in $\{\delta\rho,\delta\psi\}$)
\begin{subequations}
\begin{align}
\mathbb{E}[\overline{\hat{\eta}_\leftrightarrow(k,\omega)}\hat{\eta}_\rho(q,\upsilon)]&=\mathbb{E}[\overline{\hat{\eta}_\leftrightarrow(k,\omega)}\hat{\eta}_\psi(q,\upsilon)]=0,\\
\mathbb{E}[\overline{\hat{\eta}_\rho(k,\omega)}\hat{\eta}_\psi(q,\upsilon)]&=\frac{1-\beta}{1+\beta}\delta({\omega-\upsilon})\delta_{k,q},\\
\mathbb{E}[\overline{\hat{\eta}_\cdot(k,\omega)}\hat{\eta}_\cdot(q,\upsilon)]&=\delta({\omega-\upsilon})\delta_{k,q},
\end{align}
\end{subequations}
where the overline notation indicates the complex conjugate, $\overline{x+iy}=x-iy$, $x,y\in\mathbb{R}$.
\par
This allows us to obtain the power spectral density of the total density fluctuations $C_{\rho,\rho}(k,\omega)=\mathbb{E}[\overline{\hat{\rho}(k,\omega})\hat{\rho}(k,\omega)]$, given in full by Eq.~(\ref{eq:app_psd}) in Appendix \ref{app:fluctuations}. A similar calculation can then be performed with dynamics for the individual (left/right oriented) species' fields, using Eq.~(\ref{eq:SPDEs2}) from Appendix \ref{app:SPDEs}. From these, we calculate the cross power spectral density $C_{\rho^+,\rho^-}(k,\omega)=\mathbb{E}[\overline{\hat{\rho}^+(k,\omega})\hat{\rho}^-(k,\omega)]$, in addition to the power spectral density for the individual species, the full result of which is also given in Appendix \ref{app:fluctuations}.
\par
Next, using these spectral densities we characterise the power spectral density over the \emph{velocity} of the fluctuations through 
\begin{align}
\label{eq:psd}
    P^\rho_{|k|}(v)&\coloneq \mathbb{E}[\delta\hat{\rho}(-|k|,-v|k|)\delta\hat{\rho}(|k|,v|k|)]\nonumber\\
    &=C_{\rho,\rho}(|k|,v|k|),
\end{align}
and then define a stochastic dispersion relation as the (possibly) multi-valued function
\begin{align}
\label{eq:stoch_dispersion}
&v_\rho(|k|)=\argmaxloc_{v} P^\rho_{|k|}(v)\nonumber\\
&\coloneq\{x\,|\,\exists \varepsilon >0 \,\text{s.t.}\, P^\rho_{|k|}(x)\geq P^\rho_{|k|}(z) \,\forall z\in \mathbb{R},\,||x-z||\leq\varepsilon\},
\end{align}
equal to the velocities that correspond to local maxima in the power spectra $P^{\rho}_{|k|}(v)$. Similar quantities can then be defined for the individual species fields $\rho^\pm$ and are illustrated in Fig.~\ref{fig:fluctuations}(g) and (h), for the case of $\beta=1/2$.
\subsubsection{$\beta\to 0$: Density Dependent Fluctuation Velocities}
The $\beta\to 0$ limit of our dynamics (where $\rho=\rho^+$) serves as a simple illustrative example of the above quantities. In this case the correlation function reduces to that found in \cite{worsfold_density_2023} yielding
\begin{align}
\label{eq:dispersion_beta0}
&\mathbb{E}[\overline{\delta\hat{\rho}(k,\omega)}\delta\hat{\rho}(q,\upsilon)]=\delta_{k,q}\delta(\omega-\upsilon)\nonumber\\
&\times\frac{2\varepsilon k^2}{k^2 \text{Pe}^2 \left(1-2 \rho _0\right)^2+k^4+2 k \text{Pe} \left(2 \rho _0-1\right) \omega +\omega ^2}.
\end{align}
This function is asymmetric in $k$ and $\omega$, indicative of nonstationary `density waves' which process at characteristic velocities relative to the laboratory frame. We can corroborate this behaviour through the stochastic dispersion relation, obtaining the simple result
\begin{align}
\label{eq:vk_beta0}
    v_\rho(|k|)&=\text{Pe}(1-2\rho),
\end{align}
indicating that the most powerful fluctuations at all length-scales (all $|k|$), travel at a \emph{density dependent} velocity $\text{Pe}(1-2\rho_0)$, with the power spectra unimodal. Note, at large ($\rho_0>1/2$, \emph{i.e.} liquid-like) densities this procession velocity is negative, whilst at small ($\rho_0<1/2$, \emph{i.e.} gas-like) densities this procession velocity is positive.
\par 
For completeness we note that in this simple case we can, in the $L\to\infty$ thermodynamic limit, inverse Fourier transform Eq.~(\ref{eq:dispersion_beta0}) obtaining 
\begin{align}
    &\mathbb{E}[\delta{\rho}(x,t)\delta{\rho}(x',t')]\nonumber\\
    &=\frac{\varepsilon}{2 \sqrt{\pi } \sqrt{| t-t'| }}e^{-\frac{\left(\text{Pe} \left(2 \rho _0-1\right) (t-t')-(x-x')\right)^2}{4 | t-t'| }},
\end{align}
where the density dependent procession is self-evident. A key corollary of this phenomenon is that, in the laboratory frame, the evolution of the field is \emph{not} time reversible due to the processive behaviour of these fluctuations. Note, however, they would look time reversible  (\emph{i.e.}, possess symmetric correlation functions) in a frame moving at $v=v_{\rho}$, \emph{cf.} Sec.~\ref{sec:eq_inertial}.  Density kymographs, space time correlation functions, and inverse Fourier transforms of the power spectral densities for this $\beta\to 0$ case are given in the Supplemental Material \cite{SM}.
\subsubsection{$\beta=1$: Symmetric Fluctuations and Species Decoupling}
In stark contrast to the $\beta\to 0$ case is the $\beta=1$ scenario, which we associate with symmetric QSAPs. Here the power spectral density is given by
\begin{align}
    &\mathbb{E}[\overline{\delta\hat{\rho}(k,\omega)}\delta\hat{\rho}(q,\upsilon)]=\delta_{k,q}\delta(\omega-\upsilon)\nonumber\\
    &\times\frac{2\varepsilon k^2 ((k^2+2) (k^2+\text{Pe}^2 (\rho _0-1)^2+2)+\omega
   ^2)}{(k^2 (\text{Pe}^2 (\rho _0-1) (2 \rho _0-1)+2)+k^4-\omega
   ^2)^2+4 (k^2+1)^2 \omega ^2},
\end{align}
which contains only even powers of $k$ and $\omega$, indicative of symmetric correlation functions in space/time. Such a function is very similar to that of the on-lattice model reported in \cite{agranov_exact_2021}. Here $v_{\rho}(|k|)$ is strictly either \emph{i}) uni-valued with a single value $v_\rho(|k|)=0$, or \emph{ii}) double valued at some $\pm u$. The behaviour of the $\rho$ field here appears \emph{time reversible} since fluctuations do not process in any favoured direction. Where $v_{\rho}=0$ the fluctuations are purely dispersive. In contrast where $v_{\rho}=\pm u$, procession of the individual $\rho^\pm$ species becomes apparent. However, the latter behaviour is only observed in stable configurations when $\rho_0<1/2$, at high values of $\text{Pe}$, and never in the $k\to 0$ limit. The exact form of $v_\rho(|k|)$ for $\beta=1$ alongside density kymographs, space time correlation functions and inverse Fourier transforms of the dispersion relations, is given in the Supplemental Material \cite{SM}.
\subsubsection{$\beta\in(0,1)$: Density Dependent Velocity and Species Asymmetry}
\label{sec:fluct_beta01}
Finally, at intermediate values of $\beta\in(0,1)$, relevant to bQSAPs, we interpolate between these two behaviours. Most crucial is that we still observe characteristic travelling behaviour in the fluctuations, but now in a system comprised of two distinct species such that it can phase separate. In particular, the stochastic dispersion relation is uni-valued in the micro and macroscopic limits, possessing the property
\begin{align}
    \label{eq:vfluct}
    v_{\rm fluct}(\rho)&\coloneq\lim_{|k|\to 0}v_\rho(|k|)=\lim_{|k|\to \infty}v_\rho(|k|)\nonumber\\
    &=\text{Pe}\frac{1-\beta}{1+\beta}(1-2\rho).
\end{align}
Note, as specified in Sec.~\ref{sec:eq_inertial}, this is equal to the frame velocity where bulk states appear time symmetric and are described by $f_{\rm bulk}$. 
Again, at liquid-like densities ($\rho>1/2$) the fluctuation procession speed is negative, whilst at gas-like densities this procession speed is positive. Moreover, in the context of phase separation, where each phase will behave as a bulk at $\rho\in\{\rho_G,\rho_L\}$ we have, by Eq.~(\ref{eq:vdual}), the following inequalities
\begin{align}
    \label{eq:vdiverge}
    v_{\rm fluct}(\rho_L)&=\text{Pe}\frac{1-\beta}{1+\beta}(1-2\rho_L)\nonumber\\
    &< v=\text{Pe}\frac{1-\beta}{1+\beta}(1-\rho_G-\rho_L)\nonumber\\
    &<\text{Pe}\frac{1-\beta}{1+\beta}(1-2\rho_G)=v_{\rm fluct}(\rho_G).    
\end{align}
Explicitly, fluctuations in the liquid phase will be moving with, and faster than, the procession of the phase boundaries, whilst fluctuations in the gas phase will be moving slower than the procession of the phase boundaries or in the opposite direction. From this, we can deduce that fluctuations will \emph{converge towards} the leading gas-to-liquid phase boundary, and \emph{diverge away from} the trailing liquid-to-gas boundary.
\par
The full behaviour of the fluctuations is illustrated for $\beta=1/2$ in Fig.~\ref{fig:fluctuations} where density kymographs show evolution of the empirical density ($\rho_{\rm sde}$) over time with observable fluctuations. The empirical correlation functions of these systems [specifically density density  correlations  ({\it i.e.}, $\rho$---$\rho$) and interspecies cross correlations ({\it i.e.}, $\rho^+$---$\rho^-$)] are shown and contrasted with the inverse Fourier transforms of the relevant redimensionalised analytical spectra \footnote{To compare with simulation, \emph{cf.} Appendix.~\ref{sec:simulation}, we can transform the spectra through $\mathbb{E}[\overline{\delta\hat{\rho}_{\rm dim}(k,\omega)}\delta\hat{\rho}_{\rm dim}(q,\upsilon)]=l\tau\mathbb{E}[\overline{\delta\hat{\rho}(lk,\tau\omega)}\delta\hat{\rho}(lq,\tau\upsilon)]$, with $l=\sqrt{D/\bar{\gamma}}$, $\tau=1/\bar{\gamma}$}, where we see excellent qualitative agreement. In addition, the typical fluctuation velocity in $\rho$, $v_\rho(|k|)$, at varying wave number is shown along with the behaviour of the power spectra in the individual species $\rho^\pm$. In broad terms, we validate Eq.~(\ref{eq:vfluct}), with fluctuations at low density predominantly moving forwards, whilst those at high density move backwards. However, at low (gas-like) densities the individual species becomes decoupled over a range of length-scales, observable as `criss-crossed' patterns in the kymographs, `cross-shaped' patterns in the density-density correlations, broadly symmetric, non-negative, cross correlation spectra, and non-overlapping velocity power spectra. In contrast, at high (liquid-like) densities, the species' fluctuations are much more tightly coupled, again observable in the kymographs, correlation functions, and overlapping velocity power spectra. In particular, the cross-correlation spectra indicate that features in $\rho^-$ lag behind, or `follow', those in $\rho^+$. 
\par
The fact that all fluctuations (be they in $\rho$, $\rho^+$ or $\rho^-$) travel in the same (negative) direction (and at the same speed), as $k\to 0$, at large (liquid-like) densities is highly important in the context of our anomalous phase diagram [\emph{cf.} Fig.~\ref{fig:phase_diagrams}(c)]. This is because, in states where only the liquid phase is unstable, it ensures that all fluctuations travel away from the liquid-to-gas phase boundary on the liquid side of the interface.

\subsection{Critical Behaviour of the Exposed Spinodal Line}
\label{sec:critical_spinodal}

One of the striking features of the phase structure in  Fig.~\ref{fig:phase_diagrams}(c) is that there is a whole section of the spinodal line that is now exposed outside of the binodal solutions. Classically, the only point at which the spinodal line can be reached is at a critical point, as shown in Fig.~\ref{fig:phase_diagrams}(b). We argue that all of this exposed section can be approached and that all of it has properties that are traditionally associated with a critical point--- in particular we show that at all points along the spinodal line both the correlation length diverges and the system undergoes a form of critical slowing down.
\subsubsection{Diverging Correlation Length}
\label{sec:fluct_corrlength}

\begin{figure}[!t]
   \centering
   \includegraphics[width=0.45\textwidth]{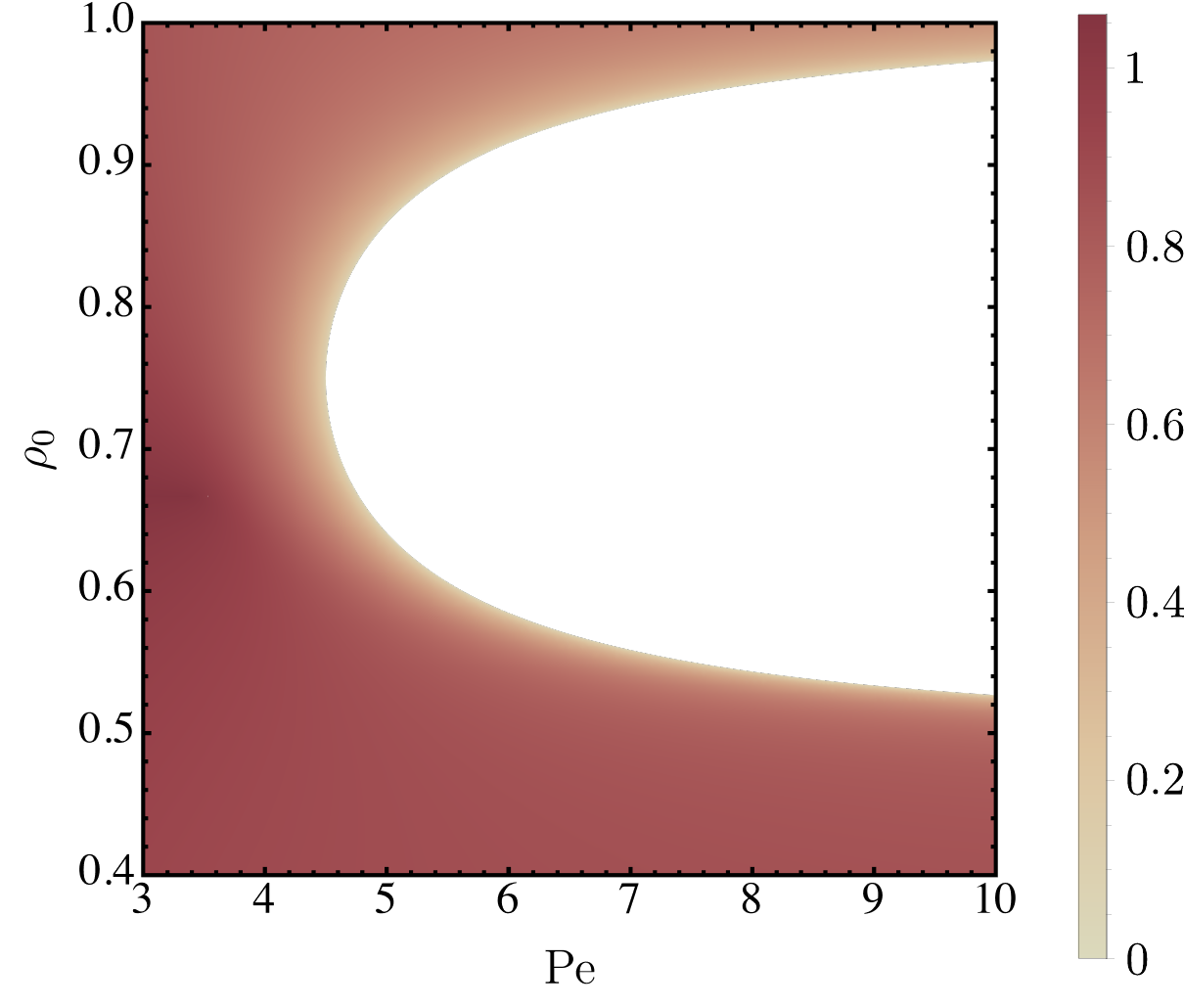} 
   \caption{
        {\bf Inverse Correlation Length for $\beta=1/2$}. The inverse correlation length vanishes on the spinodal line, indicating a diverging correlation length: a feature synonymous with criticality. The leading-order critical exponent is $\nu = 1/2$, the same as that found in a related, but symmetric lattice gas model similar to the $\beta=1$ case \cite{agranov_exact_2021}.       
    }
   \label{fig:corr_length}
\end{figure}

In the first case we investigate the correlation length by calculating the equal time density-density correlation function. To obtain this function we start with the linearised dynamics of Eq.~(\ref{eq:SPDE}) in reciprocal space, real time such that we consider the linear SDEs
\begin{align}
\label{eq:SDEsk}
&\frac{d}{dt}\begin{bmatrix}
\delta\tilde{\rho}(k,t)\\
\delta\tilde{\psi}(k,t)
\end{bmatrix}
=\mathbf{A}(k)\begin{bmatrix}
\delta\tilde{\rho}(k,t)\\
\delta\tilde{\psi}(k,t)
\end{bmatrix}+{\sqrt{\varepsilon}}\tilde{\boldsymbol{\xi}}(k,t),
\end{align}
where
\begin{align}
\tilde{\boldsymbol{\xi}}(k,t)&=
\begin{bmatrix}
\sqrt{2}ik\tilde{\eta}_\rho(k,t)\\
\sqrt{2}ik\tilde{\eta}_\psi(k,t)+2\tilde{\eta}_\leftrightarrow(k,t)
\end{bmatrix},
\end{align}
with Fourier transformed space-time white noises $\tilde{\eta}_{\cdot}(k,t)=\mathcal{F}_x\{{\eta}_{\cdot}(x,t)\}$ obeying (at linear order around $\{\rho_0,\psi_0\}$)
\begin{subequations}
\begin{align}
\mathbb{E}[\overline{\tilde{\eta}_\leftrightarrow(k,t)}\tilde{\eta}_\rho(q,s)]&=\mathbb{E}[\overline{\tilde{\eta}_\leftrightarrow(k,t)}\tilde{\eta}_\psi(q,s)]=0,\\
\mathbb{E}[\overline{\tilde{\eta}_\rho(k,t)}\tilde{\eta}_\psi(q,s)]&=\frac{1-\beta}{1+\beta}\delta(t-s)\delta_{k,q},\\
\mathbb{E}[\overline{\tilde{\eta}_\cdot(k,t)}\tilde{\eta}_\cdot(q,s)]&=\delta(t-s)\delta_{k,q}.
\end{align}
\end{subequations}
We then consider the integrating factor solution to the above SDEs
\begin{align}
   \begin{bmatrix}
      \delta\tilde{\rho}(k,t)\\
      \delta\tilde{\psi}(k,t)
      \end{bmatrix}&=\sqrt{\varepsilon}\int_{-\infty}^t dt'\;e^{-\mathbf{A}(k)(t-t')}\tilde{\boldsymbol{\xi}}(k,t'),
\end{align}
allowing us to calculate $\mathbb{E}[\overline{\delta\tilde{\rho}(k,t)}{\delta\tilde{\rho}(q,t)}]$, the full form of which is given in Appendix \ref{app:fluctuations}. 
\par
Performing the inverse Fourier transform yields a large, unwieldy, expression, but with identifiable factors in $e^{-|x|/l_{\rm corr}}$, with $l_{\rm corr}$ the dimensionless correlation length, such that it is given by the positive roots of the polynomial
\begin{subequations}
\begin{align}
   0&=\alpha_0-\alpha_1l^{-2}_{\rm corr}+\alpha_2l^{-4}_{\rm corr}-\alpha_3l^{-6}_{\rm corr},\\
   \alpha_0&=  c_5^3 + c_2 (c_1 c_4 c_5 + c_3 c_5^2)-c_2^2 c_4^2,\\
   \alpha_1&=c_1^2c_5+4c_2c_3c_5+5c_5^2,\\
   \alpha_2&=c_1^2+4c_2c_3+8c_5,\\
   \alpha_3&=4,
\end{align}
\end{subequations}
equal to the positive square roots of the solution to the equivalent cubic equation, where  $c_1=\text{Pe}\rho_0(1-\beta)/(1+\beta)$, $c_2=\text{Pe}(1-\rho_0)$, $c_3=\text{Pe}(1-2\rho_0)$, $c_4=(1-\beta^2)/2\beta$, and $c_5=(1+\beta)^2/2\beta$.
\par
In the case of $\beta=1$ (such that $c_1=c_4=0$) the only non-trivial root is $\sqrt{c_2c_3+c_5}$ such that
\begin{align}
   l_{\rm corr}^{\beta=1}&=\frac{1}{\sqrt{2+(1-\rho_0)(1-2\rho_0)\text{Pe}^2}},
\end{align}
with critical scaling exponent, $\nu$, of $1/2$ in the asymptotic behaviour around the critical point $\{\rho^*_0,\text{Pe}^*\}=\{3/4,4\}$, approached from below in $\text{Pe}$ at constant $\rho_0=\rho_0^*$
\begin{align}
   l_{\rm corr}^{\beta=1}&=\frac{1}{\sqrt{\text{Pe}^*-\text{Pe}}}+\mathcal{O}(\sqrt{\text{Pe}^*-\text{Pe}}),
\end{align}
placing the model in the mean-field (4D) Ising class, all in agreement with the model in \cite{agranov_exact_2021}.  Such a property can be contrasted with higher dimensional active particles which are generally, albeit with some debate, associated with the relevant Ising class \cite{dittrich_critical_2021, maggi_universality_2021, gnan_critical_2022}.
\par
Characterising the critical behaviour of the general ($\beta\in(0,1]$) case is less straight-forward, but it can be heuristically deduced using specific examples. 
Choosing $\beta=1/2$ and $\rho_0=\rho_0^*=3/4$, such that we similarly approach the critical/spinodal bifurcation point, $\{\rho^*_0,\text{Pe}^*\}=\{3/4,9/2\}$, from below in $\text{Pe}$, reduces the polynomial to
\begin{align}
   0=&(36\text{Pe}^2-729)+(1620-63\text{Pe}^2)l_{\rm corr}^{-2}\nonumber\\
   &+(28\text{Pe}^2-1152)l_{\rm corr}^{-4}+256l_{\rm corr}^{-6},
\end{align}
whose relevant root can be asymptotically expanded as
\begin{align}
   l_{\rm corr}^{-1}&=\frac{4}{\sqrt{17}}\sqrt{\text{Pe}^*-\text{Pe}}+\frac{826}{2601\sqrt{17}}\left(\text{Pe}^*-\text{Pe}\right)^{\frac{3}{2}}\nonumber\\
   &\quad+\mathcal{O}\left(\left(\text{Pe}^*-\text{Pe}\right)^{\frac{5}{2}}\right),
\end{align}
thus exhibiting the same critical scaling. 
The same scaling is then exhibited along the approach to the whole spinodal line, as can be intuited from the plot of the correlation length in Fig.~\ref{fig:corr_length}. We can make this claim explicit by considering the form of the correlation length at another point on the spinodal. Specifically, we choose the point on the spinodal line $\text{Pe}=8$, $\rho_0=(48+5\sqrt{7})/64$. In this case the polynomial reduces to 
\begin{align}
  0=& (1679616-26244\text{Pe}^2)      \nonumber\\
  &+(-3732480+30177\text{Pe}^2-4320\sqrt{7}\text{Pe}^2)l_{\rm corr}^{-2}    \nonumber\\
  &\quad +(2654208-13412\text{Pe}^2+1920\sqrt{7}\text{Pe}^2)l_{\rm corr}^{-4}\nonumber\\
  &\qquad-589824l_{\rm corr}^{-6},
\end{align}
the solution of which can be expanded as
\begin{align}
   l_{\rm corr}^{-1}&=\frac{648}{\sqrt{1801152+276480\sqrt{7}}}\sqrt{8-\text{Pe}}\nonumber\\
   &\qquad+\mathcal{O}\left((8-\text{Pe})^{\frac{3}{2}}\right),
\end{align}
indicating that $l_{\rm corr}$ indeed diverges at other points on the spinodal, and exhibits the same scaling with $\text{Pe}$. As such, when there are no binodal solutions, such that the spinodal line can be approached [as in Fig.~\ref{fig:phase_diagrams}(c)], the system will approach criticality.
\subsubsection{Critical Slowing}
Next, by turning back to the linearised dynamic in reciprocal space and time [Eq.~(\ref{eq:linearSPDEkomega})], we can extract the linear dynamical matrix $\mathbf{M}(k,\omega)=-(i\omega\mathbb{I}+\mathbf{A}(k))$. 
From this matrix we can deduce relaxation properties of the system from its spectra in terms of its eigenvalues. 
Expanding these eigenvalues around $k\to 0$, the upper has an $\mathcal{O}(1)$ leading term with value $\text{Pe}^*/2$, leaving the lower to determine dynamical time scales. 
Explicitly, the real component of such an eigenvalue, $\lambda^-$, reads
\begin{widetext}
\begin{align}
   \text{Re}[\lambda^-]&=\frac{\left(\beta  \left(\beta  \left(\beta  (\beta +4)+8 \text{Pe}^2 \left(\rho _0-1\right) \left(2 \rho _0-1\right)+6\right)+4\right)+1\right)}{(\beta +1)^4}k^2\nonumber\\
   &\quad-\frac{32 \beta ^4  \text{Pe}^4 \left(\rho _0-1\right) \left(2 \rho _0-1\right) \left((\beta  (9 \beta -26)+9) \rho _0^2-12 ((\beta -3) \beta +1) \rho _0+4 (\beta -3) \beta +4\right)}{(\beta +1)^{10}}k^4+\mathcal{O}(k^6).
\end{align}
\end{widetext}
The term in $k^2$ explicitly vanishes along the spinodals [\emph{cf.} Eq.~(\ref{eq:spinodal})], where the correlation length diverges, materially changing the speed of relaxation of the long, slow, modes, in these regimes, characterising a critical slowing down of the system on this line.
\subsection{Spectral Response and Growth Within the Spinodal Region}

\begin{figure*}[!htp]
   \centering
   \includegraphics[width=\textwidth]{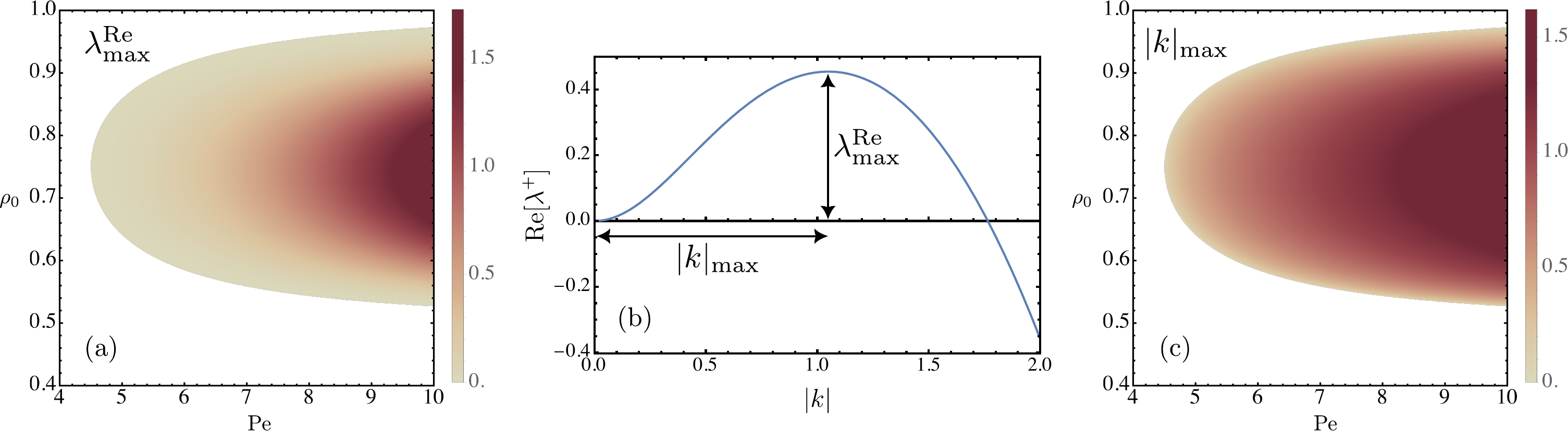} 
   \caption{
        {\bf Growth Dispersion Relation Within the Spinodal Region for $\beta=1/2$}. Within the spinodal line not all modes are dynamically stable, with the system exhibiting a type II instability [panel (b)]. Those with positive $\text{Re}[\lambda^+]$ grow exponentially with rate $\text{Re}[\lambda^+]$. This instability starts with the longest modes around $|k|=0$ as the spinodal line is crossed [panel (c)] with small growth rates [panel (a), \emph{cf.} critical slowing], with shorter modes destablising, and with a faster growth rate, as we consider points further into the spinodal region. The relation in panel (b) is created using control parameters $\text{Pe}=7$, $\rho_0=0.7$, and $\beta=1/2$.
    }
   \label{fig:inside_spinodal}
\end{figure*}

Finally, we numerically investigate the growth rate of small fluctuations \emph{within} the spinodal region, in anticipation of states that form within this traditionally forbidden area, owing to the anomalous phase structure in Fig.~\ref{fig:phase_diagrams}(c). We do this by investigating the real part of the largest eigenvalue of the dynamical matrix, $\mathbf{A}(k)$ [Eq.~(\ref{eq:A})], of the linearised reciprocal space SDEs in Eq.~(\ref{eq:SDEsk}).
\par
Unlike the broad region outside the spinodals, where $\text{Re}[\lambda_+(|k|)]\leq 0$, indicating that all fluctuations have a negative `growth rate' and are thus damped away, inside the spinodal region there are a range of long modes for which $\text{Re}[\lambda_+(|k|)]> 0$ indicating that they are amplified by the system.
\par
Behaviour of this growth dispersion relation is illustrated in Fig.~\ref{fig:inside_spinodal}, revealing that the system has a type-II instability \cite{cross_pattern_2009}. The most important characteristics to appreciate are that \emph{i}) the maximum growth rate near the spinodal is small and this grows as we consider points further into the spinodal region. When the system approaches the spinodal itself the maximal growth rates tends to zero in accord with a critical slowing down that the system experiences along the spinodal line. \emph{ii}) The length-scales that are unstable start with only the $|k|=0$ mode when the spinodal line is crossed (in accord with the criticality experienced there, where there are no finite length-scales associated with the system), but increasingly shorter length-scales are destabilised as we move into the spinodal region.
\par
These features become important in the context of the anomalous phase structure in Fig.~\ref{fig:phase_diagrams}(c) because branches of the binodals (especially the liquid binodal) reside in the spinodal region. Thus, if approximately homogeneous densities form within the spinodals in these states we can expect them to exhibit fluctuations with these growth rate properties.

\section{Simulations: coarsening and pseudo-criticality}
\label{sec:characteristic}
%
Armed with our thermodynamic interpretation of unequal tangent constructions, subsequent anomalous phase diagram, and a basic understanding of the underlying system's fluctuations, we rationalise and classify the behaviour we observe in large scale particulate simulations. 

\subsection{Nonstationary Macro-Phase Separation}

\begin{figure*}[!htp]
   \centering
   \includegraphics[width=\textwidth]{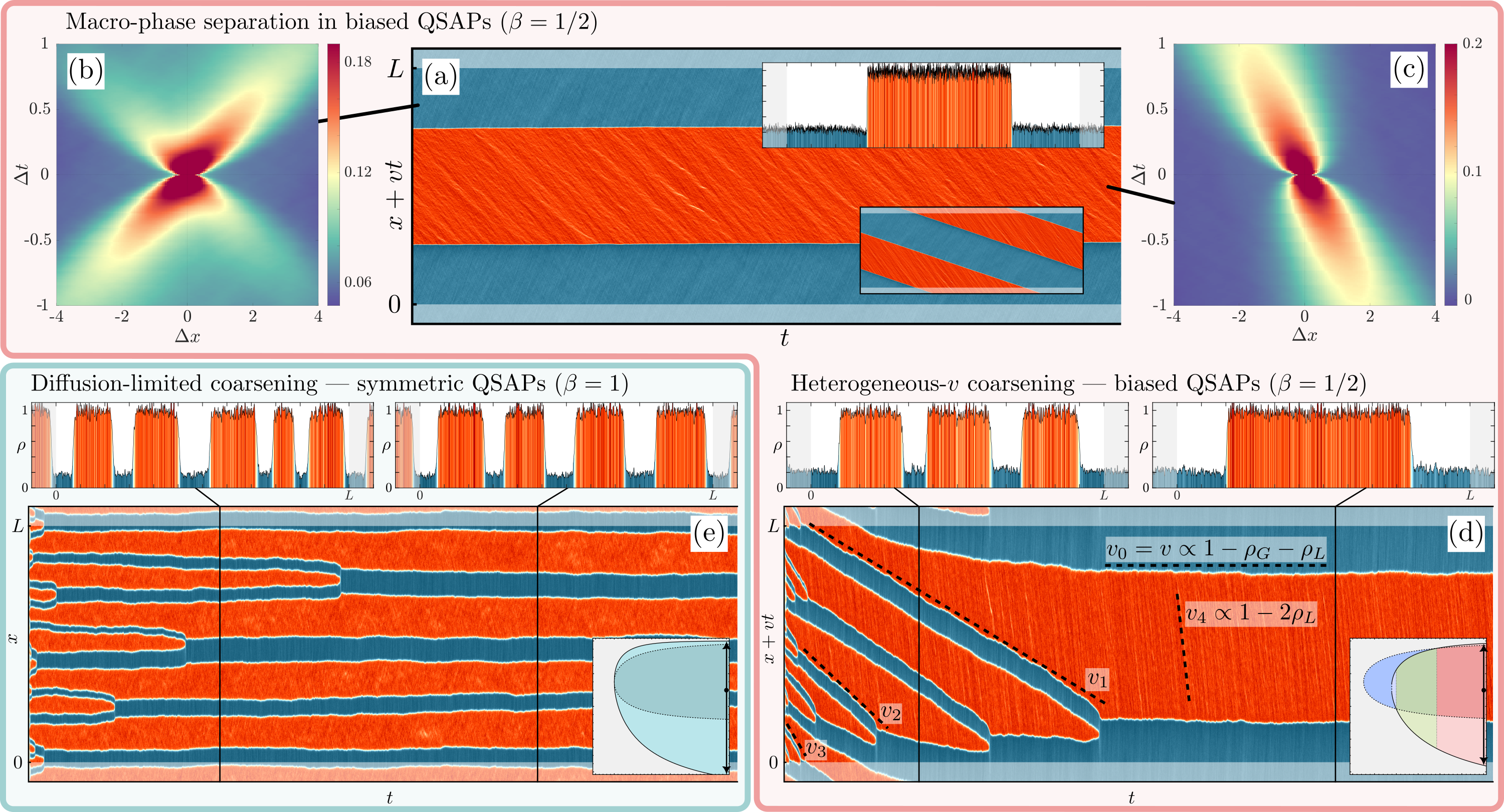} 
   \caption{
        {\bf Fluctuations, and Heterogeneous-$v$ Coarsening in Nonstationary Macro-Phase Separation.} 
        Top panels, pink bounding box: panel (a) is a density kymograph of biased QSAPs exhibiting macro-phase separation in the co-moving frame. Inset are the view in the laboratory frame and a field snapshot at the time horizon. Left [panel (b)] and right [panel(c)] show density-density correlations computed from the gas and liquid phases, respectively. Fluctuations behave differently in each phase, in aggregate diverging away from the liquid-to-gas boundary. 
        Bottom left, panel (e), teal bounding box: Density kymograph of symmetric QSAPs coarsening from a homogeneous initial state. Here since all relevant characteristic velocities are equal (\emph{i.e.} $v=v_{\rm fluct}=0$) and since there is no surface tension in 1D, coarsening is limited to diffusion of the phase boundaries.
        Bottom right, panel (d), pink bounding box: Density kymograph of biased QSAPs coarsening from a homogeneous initial state. Here, characteristic velocities of the system are unequal. This causes a spectrum of velocities to open, ranging from $v_{\rm fluct}\propto 1-2\rho_L$ for  small fluctuations/features in the liquid phase, to $v\propto 1-\rho_G-\rho_L$ for the largest features/phase boundaries.
        Consequently, we observe fluctations emerge in the liquid phase at $v_4=v_{\rm fluct}$, which bounds the speed of small bubble defects ($|v_3|<|v_4|$), which in turn bounds the speed of larger defects ($|v_2|<|v_3|$), \emph{etc}., until the bulk travels at the boost speed $|v|=|v_0|<|v_1|$.  
        This spectrum of velocities leads to smaller regions being ejected form larger ones, effecting a coarsening mechanism. 
        Example in panel (a) simulates the underlying particulate SDEs with parameters $(\beta=1/2, \text{Pe}=10, \rho_0=0.65)$ on a domain of length $L=128\pi$, with $N=320,000$ particles, and with a time horizon of $t_h=1400$. 
        Coarsening examples shown [panels (d) and (e)] use $(\beta=1,\text{Pe}=10,\rho_0=0.7)$ and $(\beta=1/2,\text{Pe}=10,\rho_0=0.7)$  with $L=32\pi$ and $N=80,000$. Time horizons for the coarsening density kymographs are $t_h=2000$ and $t_h=2800$ for $\beta=1$ and $\beta=1/2$, respectively. Details of the numerics can be found in Appendix~\ref{sec:simulation}, code available in \cite{spinney_rspinneyamips_2024}.
        }
   \label{fig:macro}
\end{figure*}

First, we examine the behaviour where the system most closely resembles AMB and conventional MIPS, namely the macro-phase (red) region described in Fig.~\ref{fig:phase_diagrams}(c), where both coexistence densities are outside the spinodals, and thus stable.
\subsubsection{Nonstationary Macro-Phase Separation and the Loss of Spatio-Temporal Symmetries}
In these regions the broad behaviour of the system manifests as a nonstationary analogue of conventional MIPS, as can be seen Fig.~\ref{fig:macro}(a), 
with stable phase boundaries that travel at the characteristic speed of the boost $v\propto 1-\rho_G-\rho_L$ [Eq.~(\ref{eq:vdual})], and with the asymmetric boundaries observed in the deterministic system faintly visible in the noisy density profile. In this regime, fluctuations are damped away on all length-scales, so that with sufficient coarse-graining and/or reduction in noise strength, the full fluctuating behaviour is in direct correspondence with the deterministic Eq.~(\ref{eq:deterministic}) and its  resulting behaviour [\emph{cf.} Fig.~\ref{fig:unequal}(b)].
\par
However, on closer inspection, we see additional structure in the fluctuations, absent from the deterministic dynamics. As predicted in Sec.~\ref{sec:fluct_beta01}, we see travelling density waves in both the liquid and gas phases, but the fluctuations in the liquid phase travel faster (more negative) than $v$, and fluctuations in the gas phase travel slower (more positive). This is corroborated by the empirical correlation functions calculated from each phase, also shown in Fig.~\ref{fig:macro}(b) and (c), and in broad agreement with analytics shown in Fig.~\ref{fig:fluctuations}.
\par
This phenomenon is in stark contrast to the case of conventional MIPS (AMB) or equilibrium phase separation where there is an implicit kinematic symmetry of the system: all characteristic, macroscopic ($k\to 0$), system velocities are equal--- namely, $v=v_{\rm fluct}(\rho_G)=v_{\rm fluct}(\rho_L)=0$--- and therefore all fluctuations are purely dispersive in the lab frame ($v=0$). In nonstationary MIPS, however, this symmetry is broken, and fluctuations travel at distinct velocities in each of the bulk states, both of which are distinct from the velocity at which the phase boundaries travel.

\subsubsection{Coarsening due to Heterogeneous Velocities}
\label{sec:coarsening}
Thus far, we have identified three characteristic speeds in our fluctuating system of bQSAPs, those of the ($k\to0$) fluctuations in the two bulk phases, and that of the interface between them.
However, as the system approaches its limiting, phase-separated state, we observe a coarsening behaviour that arises from a continuum of such velocities.
Specifically, we observe that, if tracking the behaviour of a gas-like region within a liquid bulk, small features on the scale of fluctuations emerge at the characteristic fluctuation velocity in the liquid bulk, $v_{\rm fluct}\propto 1-2\rho_L$. But as they grow, the features slow down, seemingly monotonically, until they become a macroscopic feature (\emph{i.e.} a gas bulk phase) and their procession velocity necessarily equals $v\propto 1-\rho_G-\rho_L$. The aggregate result of this continuum is that smaller bubble defects are ejected from larger liquid features, effecting a coarsening mechanism. We refer to this phenomenon as `heterogeneous-$v$ coarsening' and it is illustrated, and compared to symmetric MIPS ($\beta=1$) where no such coarsening mechanism is present, in  Fig.~\ref{fig:macro}(d) and (e). 
\par
Mechanistically, this continuum of speeds appears to derive from the effective truncation of the (in principle) infinite interfacial boundary solutions, when a bubble defect (for example) has a finite size. In thermodynamic terms, the change in chemical potential across the edge of the bubble differs from the theoretical coexistence value, and thus a different flux across the boundary is incurred, in turn leading to a distinct procession speed. The size of this effect then follows from the degree of truncation, and thus the size of the defect.
\par
We remark that such a phenomenon is surprising--- conventionally one does not expect such coarsening in 1D  owing to the lack of surface tension \cite{bray_coarsening_2003}, and indeed surface tension is absent here. Instead, this effect is strictly a consequence of active transport and the distinct emergent velocities that happen in the context of phase separation. Whilst the presented data is inconclusive in terms of empirically determining a scaling exponent for such coarsening, it would be interesting to ask whether the system exhibits typical $t^{1/3}$ behaviour associated with conserved dynamics (model B) \cite{bray_coarsening_2003}, or if the presence of persistent transport across domains allows for faster, $t^{1/2}$, scaling normally associated with model A, or otherwise.

\subsection{Pseudo-Criticality: Meso- and Micro-Phase Separation}

\begin{figure*}[!htp]
   \centering
   \includegraphics[width=\textwidth]{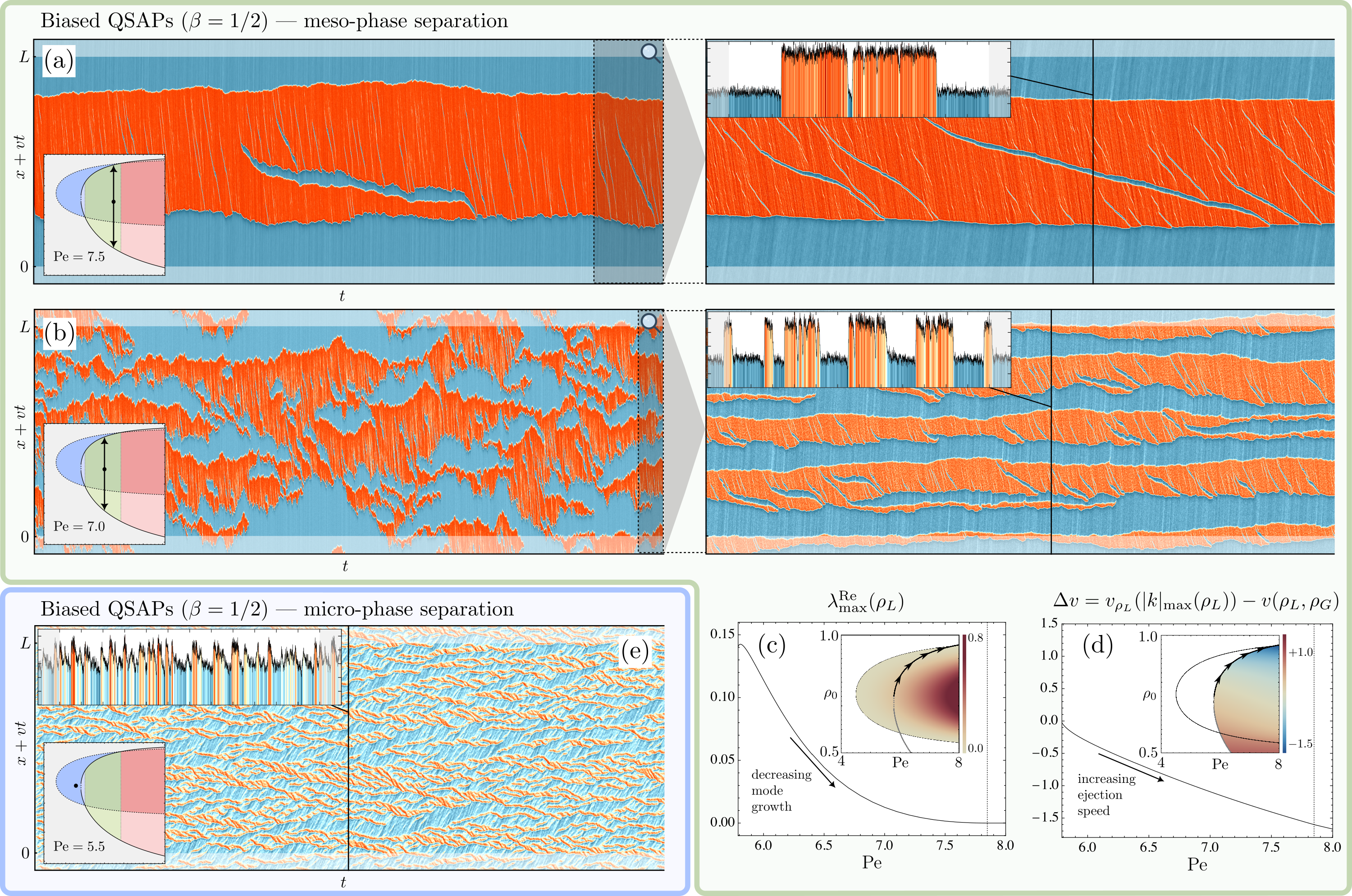} 
   \caption{{\bf Pseudo-Criticality: Meso and Micro-Phase Separation.} 
   Where anything other than two binodals lie outside the spinodals, noise cannot be neglected.
   In conventional MIPS this occurs at the critical point.
   In nonstationary MIPS, it corresponds to an extended region of phase space due to the decoupling of binodals from spinodals.
   This permits access to the interior of the spinodals, where behaviour is not strictly critical but retains several features of criticality.
   Specifically, we identify two types of behaviour: meso- and micro-phase separation.
   Meso-phase separation corresponds to the area with one stable binodal [panels (a)-(d), green bounding box].
   The impact of the fluctuations in this region depends on the position of the unstable binodal.
   As $\text{Pe}$ is increased and the unstable binodal moves towards the spinodal [panels (c) and (d)], the rate of growth of small fluctuations in the liquid phase decreases [panel (c)], whilst the ejection velocity of those fluctuations from the liquid phase envelope increases [panel (d)].
   This allows progressively larger liquid domains to remain coherent, despite the persistent creation and ejection of gaseous defects.
   The length-scale of these liquid domain envelopes can be tuned with $\text{Pe}$ up to arbitrary sizes [\emph{e.g.} panel (a) vs. panel (b)].
   Micro-phase separation corresponds to the area with no stable binodals [panel (e), blue box].
   As a result, the observed behaviour is highly fluctuating.
   However, there does appear to be characteristic emergent length-scales.
   Examples shown use $(\beta=1/2,\text{Pe}\in\{7.5,7,5.5\},\rho_0=0.7)$ using simulation of the underlying particulate SDEs with $L=128\pi$ and $N=320,000$. Time horizons for the density kymographs are $t_h=80,000$, $t_h=400,000$, and $t_h=6,000$ for $\text{Pe}=7.5$, $\text{Pe}=7$, and $\text{Pe}=5.5$, respectively [panels (a), (b), and (e)]. Details of the numerics can be found in Appendix~\ref{sec:simulation}, code available in \cite{spinney_rspinneyamips_2024}.}
   \label{fig:meso_micro}
\end{figure*}

Next, we turn our attention to regions of parameter space where binodals lie within the spinodals or there are no binodal solutions at all [{\it i.e.}, green and blue regions of Fig.~\ref{fig:phase_diagrams}(c)].
In these cases, at least part of the solutions available to the system reside within the region where fluctuations are amplified. Consequently, the system will not satisfy a Ginzburg criterion and a faithful description of the system must incorporate fluctuations. 
In conventional phase separation, as well as stationary MIPS, this only occurs in the vicinity of a single location, where the binodals and spinodals meet--- {\it i.e.}, the critical point.
However, in nonstationary MIPS, it corresponds to an entire region of parameter space. 
Nothwithstanding the exposed spinodal ({\it i.e.}, critical) line, this broader region does not possess certain hallmark features of criticality, such as divergent correlation lengths and critical slowing (see Sec.~\ref{sec:fluctuations}).
It does, nevertheless, experience  uncontrolled growth of fluctuations--- albeit with finite, characteristic length-scales and timescales--- which is a consequence of not being able to satisfy a Ginzburg criterion. 
This causes noise to manifest on large length-scales and we thus term such states `pseudo-critical'. 

\subsubsection{Nonstationary Fluctuations Stabilise the Liquid-Gas Boundary}
If we consider points in phase space where a coexistence solution exists (within the binodals), but the upper binodal lies just within the spinodal line--- {\it i.e.}, rightmost, high Pe areas of the green region, Fig.~\ref{fig:phase_diagrams}(c)--- one might expect the uncontrolled noise in the liquid phase to fundamentally disrupt phase separated solutions.
However, this is not what we observe.
Rather, we still observe two distinct phases, with the system separating into densities that correspond to the coexistence solutions found from the deterministic system.
And broadly, large bulk phases can still persist.
There are several properties of the system that help explain this. Firstly, at higher values of $\text{Pe}$, only the liquid phase density is unstable, with the gaseous phase damping fluctuations as in normal phase separation, helping to stabilise the boundary solutions.
Secondly, and most importantly, the dynamical properties of the fluctuations themselves are implicated in the system's stability.
Specifically, as observed in Sec.~\ref{sec:fluctuations}, fluctuations travel with respect to the characteristic velocity of phase boundaries, $v\propto (1-\rho_G-\rho_L)$.
In particular, the net direction of travelling fluctuations is always away from the (trailing) liquid-to-gas boundary [\emph{cf.} Eq.~(\ref{eq:vdiverge})].
Moreover, in the liquid phase (as opposed to the gas phase) fluctuations in {\it both} species ({\it i.e.}, $\rho^\pm$) travel away from the boundary for the (long) modes which are unstable [\emph{cf.} Fig.~\ref{fig:fluctuations}(h)]. The result is that \emph{all} unstable fluctuations from the liquid phase do not interfere with the trailing phase boundary. Instead, they are transported away until, ultimately, they reach the leading gas-to-liquid boundary, where they are ejected into the stable gas phase, leaving a coherent liquid phase.
\par
This mechanism underpins a hallmark feature of the noisy dynamics in this region--- the continual \emph{creation} and \emph{ejection} of defects in the liquid phase.
Such behaviour does not appear in the deterministic dynamics, but can be clearly observed in Fig.~\ref{fig:meso_micro}(a).
Here, small, growing fluctuations are continuously injected into the liquid bulk since it resides at an unstable density within the spinodals.
But they are immediately, and rapidly, transported away towards the leading boundary as they grow, since $v_{\rm fluct}(\rho_L)<v$.
Some of these fluctuations grow into distinct bubble defects, but because of the coarsening mechanism discussed in Sec.~\ref{sec:coarsening}, they too travel with respect to the bulk phase and are ultimately ejected into the gas phase, which acts as a sink for such unstable fluctuations.
\par
A corollary of this boundary stabilisation effect is that, despite the uncontrolled fluctuating behaviour in the blue and green areas of Fig.~\ref{fig:phase_diagrams}(c), the deterministic solutions to Eq.~(\ref{eq:deterministic}) offer far more insight into the behaviour in these regions of the full fluctuating dynamic than one might expect.
In particular, it causes the coexistence densities to closely resemble those in Fig.~\ref{fig:phase_diagrams}(c), since the stabilised liquid-to-gas boundary is precisely the feature that sets the liquid and gas densities (see Supplemental Material \cite{SM}).
\subsubsection{Unstable Fluctuations Determine the Emergent Length-Scale in Meso-Phase Separation}
The fluctuation ejection/coarsening mechanism, which stabilises both the liquid-to-gas boundary and liquid phase, is only part of the story as Pe decreases (second row, Fig.~\ref{fig:meso_micro}).
Whilst fluctuations and defects are reliably transported through the liquid bulk, sometimes they grow too large before being ejected.
In such cases, a large liquid domain can disintegrate into multiple smaller liquid domains.
These smaller domains can, moreover, be relatively stable, because the distance a fluctuation must travel before being ejected has been effectively reduced.
As such, we understand that the length-scale of liquid features can be related to the distance fluctuations can travel before disrupting the liquid bulk.
In turn, this depends on both their growth rate and their relative speed with respect to the phase boundaries.
It is therefore instructive to formulate a simple heuristic for this approximate length-scale, based on the properties of the fluctuations (see Sec.~\ref{sec:fluctuations}). 
\par
To do so, we consider fluctuations with the fastest growth rate--- those which correspond to $|k|_{\rm max}$ from Fig.~\ref{fig:inside_spinodal} and which, in this regime, grow exponentially with rate $\lambda_{\rm max}^{\rm Re}$. 
The behaviour of $\lambda_{\rm max}^{\rm Re}$ experienced in the liquid phase (\emph{i.e.} for a homogenous density at $\rho=\rho_L$), as a function of $\text{Pe}$, is shown in Fig.~\ref{fig:meso_micro}(c).
Explicitly, it takes the value of the inset heat map of $\lambda_{\rm max}^{\rm Re}$ (recreated from Fig.~\ref{fig:inside_spinodal}) along the contour which corresponds to the liquid binodal.
As $\text{Pe}$ is increased, we observe that the growth rate decreases, culminating with the growth rate vanishing when the liquid binodal lies exactly on the spinodal line, in accord with the critical slowing we expect there.
\par
Similarly, we may characterise the relative velocity of these $|k|_{\rm max}$ fluctuations by considering the quantity $\Delta v = v_{\rho_L}(|k|_{\rm max})-v$, which is the difference between the typical velocity of the $|k|_{\rm max}$ mode at density $\rho_L$ [using Eq.~(\ref{eq:stoch_dispersion})] and the velocity of the phase boundary, $v=\text{Pe}((1-\beta)/(1+\beta))(1-\rho_G-\rho_L)$.
This is shown in Fig.~\ref{fig:meso_micro}(d), where it takes the value of $v_{\rho_0}(|k|_{\rm max}(\rho_0,\text{Pe}))-v$ along the contour of the upper liquid binodal.
As $\text{Pe}$ is increased, we observe that the relative velocity, or ejection speed, of the fluctuations grows.
Whilst we plot the relative velocity of the fastest growing mode in Fig.~\ref{fig:meso_micro}(d), since the typical velocities are reasonably independent of $|k|$ in the liquid phase [\emph{cf.} Fig.~\ref{fig:fluctuations}(h)], we can reasonably approximate $\Delta v\sim v_{\rm fluct}(\rho_L)-v=\text{Pe}((1-\beta)/(1+\beta))(\rho_G-\rho_L)$.
As such, the growth of $\Delta v$ with $\text{Pe}$ is apparent, both  explicitly in the leading pre-factor and implicitly through the difference in coexistence densities (which grows with $\text{Pe}$). 
\par
\begin{figure}[!htp]
   \centering
   \includegraphics[width=0.45\textwidth]{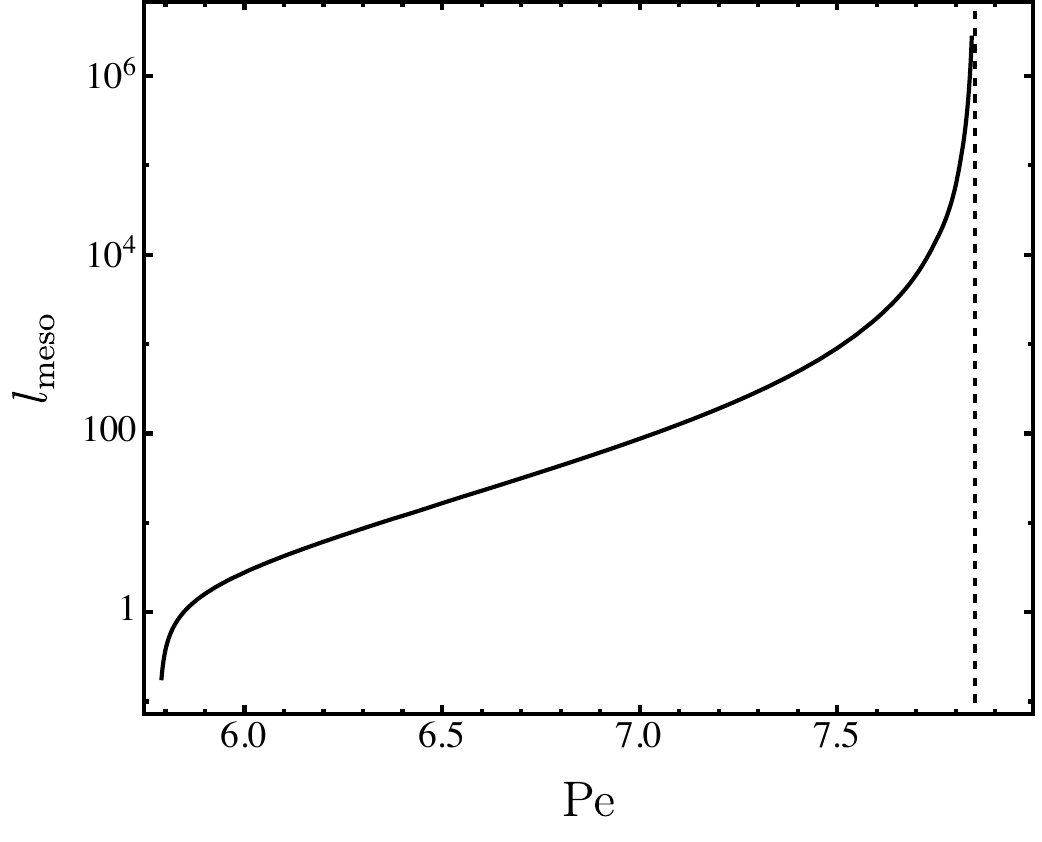} 
   \caption{
        {\bf Heuristic length-scale for meso-phase separation for $\beta=1/2$}. The predicted emergent length-scale with $\text{Pe}$ for $\beta=1/2$ is shown using binodal data from Fig.~\ref{fig:phase_diagrams}(c). The vertical dashed line marks the location where the binodal crosses the spinodal.
    }
   \label{fig:meso_length}
\end{figure}

These two effects then act in concert to define a length-scale $l_{\rm meso}\sim|\Delta v|/\lambda_{\rm max}^{\rm Re}\sim \text{Pe}((1-\beta)/(1+\beta))|\rho_G-\rho_L|/\lambda_{\rm max}^{\rm Re}$ on which we expect to observe emergent features of the system.
A plot of $l_{\rm meso}$ for $\beta=1/2$ is shown in Fig.~\ref{fig:meso_length}.
In particular, we note the response to the two control parameters $\text{Pe}$ and $\beta$.
The heuristic indicates that we should expect larger emergent length-scales with increasing asymmetry. 
In addition, it predicts that the emergent length-scales can effectively be tuned by Pe, with $l_{\rm meso}$ diverging as $\text{Pe}$ takes values approaching that for which the liquid density lies exactly on the spinodal.
At this point, the liquid phase of the system (but not the gas phase) lies on the critical line, again motivating our use of the term `semi-critical lines' in Fig.~\ref{fig:phase_diagrams}(c).

\par
Whilst the accuracy of the heuristic, $l_{\rm meso}$, will invariably be limited due, in part, to its reliance on linear fluctuation properties, its qualitative predictions are borne out in our simulations, where we observe changing length-scales with $\text{Pe}$.
In Fig.~\ref{fig:meso_micro}(a), at $\text{Pe}=7.5$, the emergent length-scale is very large (comparable to the size of the system), such that we see only a single liquid domain with fluctuating defects travelling by means of the ejection mechanism.
By contrast, at a lower value of $\text{Pe}=7$, in panel (b), the length-scale has reduced significantly.
In this case, it is small enough compared to the size of the system that the `pseudo-critical' nature of the dynamics becomes apparent, with the effect of noise manifesting clearly at the scale of the emergent liquid features.
Further density kymographs at lower values of $\text{Pe}$ can be found in the Supplemental Material \cite{SM}, with the emergent length-scale reducing with $\text{Pe}$, in accordance with our heuristic.
\par
We remark that the length-scale which our heuristic, $l_{\rm meso}$, intends to capture is the \emph{largest} scale that emerges from simulation. Below this length-scale we see a complicated fluctuating behaviour where defects of many different sizes form, coalesce, and dissolve. The scaling of the distribution of the size and/or duration of such features remains an open and interesting question.

\subsubsection{Micro-Phase Separation}

At sufficiently low values of $\text{Pe}$, the lower, gas binodal crosses into the spinodal region, and ultimately joins with the upper binodal. To the left of this location, we observe regions of the phase diagram [blue regions in Fig.~\ref{fig:phase_diagrams}(c)] where there are either no stable coexistence densities or no coexistence solutions entirely, but where homogeneous solutions are unstable. These regions lack the stabilising mechanism discussed above, and the linear stability analysis loses its validity; there is no clear homogenous density around which to perturb.
\par
In these regions we must rely entirely on simulation, an example of which is shown in Fig.~\ref{fig:meso_micro}(e). Here the behaviour consists of short peaks which process at an approximately uniform velocity, but which are highly fluctuating, with peaks emerging, disintegrating, and coalescing as the intrinsic noise from the individual particles is expressed on the larger length-scale of the emergent density features. A high resolution density kymograph of this behaviour is included in the Supplemental Material \cite{SM}.

%
\section{Discussion}
%
In this paper, bQSAPs have been used as a prototypical model to better understand the thermodynamics, phase structure, and broader criticallike properties of {\it nonstationary} phenomena, such as that related to driven transport. Our motivation was to show how many of the core ideas from critical phenomena extend (or not) to a paradigmatic and otherwise highly-studied class of non-equilibrium system.

In this context, it is worth remarking on the distinction between bQSAPs and toy models of flocking, which are also nonstationary and have been analysed in terms of dynamic critical phenomena \cite{toner_long-range_1995}. In the latter, nonstationarity emerges due to a {\it spontaneous} symmetry-breaking, whereas bQSAPs have a fundamental asymmetry encoded by a control parameter. Viewed through this lens, flocking is not a model of transport {\it per se}, in that it is not associated with a net current, either in the ensemble average or ($t\to\infty$) time average.
\par
The central contributions of our paper are threefold: 
\begin{enumerate}
    \item By considering bQSAPs in inertial frames where the limiting solutions are stationary, we derive an effective field theory, nsAMB, which allows us to understand nonstationary active phase separation in terms of an \emph{unequal} tangent construction on an effective equilibrium bulk free energy.
    This captures how fluxes between coexisting phases result in the movement of phase boundaries, and explains what it means for phase-coexistence lines to cross spinodals, thermodynamically.
    \item We use a self-consistent numerical scheme to compute the resulting phase-structure, showing how the critical point has expanded to a critical line, and how the region for which a Ginzburg criterion cannot be satisfied is similarly expanded from a single point to an extended region of phase space.
    \item We combine stochastic, particle-based, simulations with a treatment of the system's fluctuating hydrodynamics to analyse the different regimes of nonstationary phase-separation that result.
    \begin{enumerate}
        \item Where both coexistence densities are stable, we observe a type of macro-scale nonstationary phase-separation that displays anomalous coarsening.
        \item Where there is no valid Ginzburg criterion, we observe a type of pseudo-criticality, that has some, but not all, features of criticality. If only one coexistence density is stable, we see meso-scale fluctuating nonstationary phase-separation that has an emergent, characteristic lengthscale tunable by Pe. If there are no stable coexistence densities, we see a chaotic micro-scale form of fluctuating nonstationary phase-separation. 
    \end{enumerate}
\end{enumerate}
Together, these points subvert several deep-rooted expectations from the study of stationary systems, both passive and active, and reflect the potentially wide-ranging consequences of inherent (as opposed to spontaneous) symmetry-breaking in models of active matter.
\par
The use of inertial frames is at the centre of a thermodynamic understanding of  our system. Through their use we have mapped the highly out-of-equilibrium phenomenon of nonstationary phase separation into a flux free and stationary analogue with each bulk phase treated as an equilibriumlike state. That we can treat the bulk phases in this way has relied on an subtle generalisation of the notion of equilibrium--- we have effectively considered a state as at equilibrium if there exists an inertial frame where $k\to 0$ fluctuations appear time reversible ({\it i.e.,} effectively recovering FDT-like behaviour), and the system responds to perturbations according to a free energy, such that it can be associated with a meaningful chemical potential. 
\par
Crucially, however, bQSAPs can possess {\it several} characteristic velocities: a property implicated in a novel coarsening mechanism as well as the stabilisation of phase boundary solutions, particularly where coexistence densities lie inside the traditionally forbidden region within the spinodals.
This means that different facets of equilibrium emerge in distinct frames--- phase boundaries are stationary in one frame [\emph{cf.} Eq.~(\ref{eq:vdual})], whilst bulk states are equilibriumlike in others [\emph{cf.} Eq.~(\ref{eq:vJdegen})]--- raising challenges for constructing a single order-parameter field theory in full quantitative correspondence with the underlying bQSAPs.
On the one hand, our implicit effective
theory [Eq.~(\ref{eq:NSAMB})], requires us to solve self-consistently for the field {\it and} the inertial frame in which bQSAPs are being observed.
On the other hand, candidates for an explicit theory do not take account of inertial frame dependence at all, and therefore do not share solutions with bQSAPs.
\par
Notwithstanding the limitations of either formulation, 
both explict and implict field theories characterise phase-separation in terms of an unequal tangent construction, which permits binodals to cross inside the spinodal lines, and therefore entire regions of phase space that cannot satisfy a Ginzburg criterion.
This not only expands the critical point to a line but identifies a region that exhibits `pseudo-critical' behaviour. 
These features are seemingly important for biology, which is not only the canonical setting of active matter, but also where driven, nonstationary phenomena is commonplace. However, whilst the notion of criticality in biology has been around for some time, it has always proved controversial~\cite{munoz_colloquium_2018}. Proponents remark on the inherent adaptability of wide classes of biological systems, and identify this with  divergent susceptibilities~\cite{mora_are_2011}. They also point to power-law scaling relations, which have been identified at vastly different biological scales, ranging from the brain~\cite{massobrio_criticality_2015,beggs_neuronal_2003} to flocks of birds \cite{cavagna_scale-free_2010}.  Detractors, on the other hand, point to the issue of fine-tuning, which is seemingly required for systems to be close to the critical point, but is not apparent in noisy, biological systems. Whilst in principle this can be circumvented by self-organised criticality (SOC)~\cite{bak_self-organized_1987,tang_critical_1988,tang_mean_1988,bak_how_2013}--- where the critical point is also a fixed point of the dynamics--- mapping biological systems to SOC has not always proved possible~\cite{mora_are_2011}. Scaling relations also provide an area of contention, both from the perspective of identifying them with statistical significance \cite{clauset_power-law_2009} as well as the ability to meaningfully infer underlying mechanisms \cite{stumpf_critical_2012}. Another issue is the notion of critical slowing; in dynamical critical phenomena, the timescale of dynamical changes diverge as the critical point is approached \cite{munoz_colloquium_2018}.
\par
Remarkably, the type of pseudo-criticality that we have reported can circumvent {\it all} of the aforementioned issues: it corresponds to a finite region of phase space, and therefore does not require unrealistic fine tuning; fluctuations manifest at the scale of the macroscopic features, but the system is not formally scale-free, and; the dynamics is characterised by finite timescales. Based on this high-level rationale, the extent to which our findings can have a bearing on biological complexity therefore appears an pertinent and outstanding question. 
\par
Indeed, there are many questions for which we don't currently have answers. For example, what is the nature of higher dimensional versions of bQSAPs, where a continuous symmetry is broken, rather than the discrete one considered in this work? Recent work on active Brownian particles is suggestive that higher dimension can potentially suppress many of the anomalous features reported here \cite{othman_phase_2025}, although further work in the area is needed. Equally, there is considerable interest in non-reciprocal interactions in active matter--- including quorum-sensing particles \cite{duan_phase_2025}--- and it would be interesting to ascertain the extent to which our observations can be recapitulated in those systems, where travelling fronts, pattern-formation, and chaoticlike behaviours have been reported in the context of the non-reciprocal Cahn-Hilliard  \cite{saha_scalar_2020,brauns_phase-space_2020,brauns_nonreciprocal_2024} and related models \cite{mason_dynamical_2025}. Finally, several remaining open questions revolve around the scaling and/or statistics of emergent features. For example, what are the statistics and scaling properties of the liquid and gaseous domains in the meso-phase separated region? What are the dynamical critical exponents, and how do we make sense of them in light of our anomalous phase structure? What are the scaling properties of the coarsening mechanism \cite{bray_coarsening_2003}? And do the distinct regimes (particularly pseudo-criticality) exhibit hyperuniformity and/or giant number fluctuations \cite{zheng_universal_2024,fernandez-nieves_hyperuniform_2024}? As a result, we welcome all further work that aims to shed new light on broad questions at the intersection of criticality and non-equilibrium transport.

\begin{acknowledgments}
RES and RGM acknowledge the EMBL Australia program. RGM acknowledges funding from the Australian Research Council Centre of Excellence for Mathematical Analysis of Cellular Systems (CE230100001). This research includes computations using the computational cluster Katana supported by Research Technology Services at UNSW Sydney.
\end{acknowledgments}
\begin{appendix}
\section{Derivation of non-dimensional SPDEs}
\label{app:SPDEs}
Using the methodology in \cite{spinney_deankawasaki_2025}, based on standard methods due to Dean \cite{dean_langevin_1996}, we may describe the fluctuating one-body probability density functions for particles with orientation $\pm$, taken as shorthand for $\pm 1$, using the following two integro-SPDEs (one for each orientation)
\begin{widetext}
\begin{align}
    \dot{\Phi}^\pm(x,t)&=\mp\nu\partial_x\left[\Phi^\pm(x,t)\left[1-\rho_0 L\int_{0}^L k_\sigma(x-y)\left[\Phi^+(y,t)+\Phi^-(y,t)\right]\,dy\right]\right] \nonumber\\
    &\qquad+ D\partial_x^2\Phi^\pm(x,t)\pm\left(\gamma_+\Phi^-(x,t)-\gamma_-\Phi^+(x,t)\right)\nonumber\\
    &\qquad+\sqrt{\frac{2D}{N}}\partial_x\left(\sqrt{\Phi^\pm(x,t)}\eta_\pm(x,t)\right)\pm\frac{1}{N}\left(\zeta_{-,+}(x,t)-\zeta_{+,-}(x,t)\right).
    \label{initSPDE}
\end{align}
\end{widetext}
Here $\Phi^{\pm}(x,t)$ is the probability density function of finding any particle at position $x$, with orientation $\pm$, at time $t$, such that we have $\int_0^L (\Phi^+(x,t)+\Phi^-(x,t))\,dx=1$. The second line is comprised of spatio-temporal noise terms $\eta_\pm$ and $\zeta_{\pm,\mp}$. The $\eta_+$ and $\eta_-$ are two independent, Gaussian, space time white noises with statistics
\begin{subequations}
\begin{align}
    \mathbb{E}[\eta_\pm(x,t)]&=0,\\
    \mathbb{E}[\eta_\pm(x,t)\eta_\mp(y,s)]&=0,\\
    \mathbb{E}[\eta_\pm(x,t)\eta_\pm(y,s)]&=\delta(x-y)\delta(t-s).
\end{align}
\end{subequations}
The $\zeta_{\pm,\mp}(x,t)$ terms are equal to distribution-sense space-time derivatives of, two-dimensional, compensated \footnote{We understand a compensated Poisson process to be a Poisson process modified to have zero mean by way of a smooth term proportional to its intensity. Explicitly, if a Poisson process has increments $dN\in\{0,1\}$, the compensated process has increments $dN-\mathbb{E}[dN]=dN-\lambda dt$, where $\lambda$ is the intensity of the process.} Poisson processes with intensities
\begin{align}
    \lambda_{\pm,\mp}(x,t)= N\gamma_\mp\Phi^\pm(x,t).
\end{align}
These are also zero mean fields
\begin{align}
    \mathbb{E}[\zeta_{\pm,\mp}(x,t)]&=0,
\end{align}
but are not white noises. Instead, they are defined in the usual manner of a Poisson process, such that the probabilities of unit sized events in the uncompensated underlying 2D Poisson processes, in a small `area' of size $dx\,dt$, are given by $N\gamma_\mp\Phi^\pm(x,t)\,dx\,dt$. 
\par
Using this (exact) SPDE we then make a number of approximations and re-scalings. 
First, however, it is useful to transform these SPDEs into the SPDEs for the total probability density, $\Phi(x,t):=\Phi^+(x,t)+\Phi^-(x,t)$, and the polarisation probability density, $\chi(x,t):=\Phi^+(x,t)-\Phi^-(x,t)$. 
These are governed, up to an equality in distribution, by the SPDEs
\begin{widetext}
\begin{subequations}
\begin{align}
    \dot{\Phi}(x,t)&=-\nu\partial_x\left[\chi(x,t)\left[1-\rho_0 L\int_0^Lk_\sigma(x-y)\Phi(y,t)\,dy\right] \right] + D\partial_x^2\Phi(x,t)+\sqrt{\frac{2D}{N}}\partial_x\left(\sqrt{\Phi(x,t)}\eta_\Phi(x,t)\right),\\
    \dot{\chi}(x,t)&=-\nu\partial_x\left[\Phi(x,t)\left[1-\rho_0 L\int_0^Lk_\sigma(x-y)\Phi(y,t)\,dy\right]\right]  + D\partial_x^2\chi(x,t)-(\gamma_++\gamma_-)\chi(x,t)+(\gamma_+-\gamma_-)\Phi(x,t)\nonumber\\
    &\qquad+\sqrt{\frac{2D}{N}}\partial_x\left(\sqrt{\Phi(x,t)}\eta_\chi(x,t)\right)+\frac{2}{N}\left(\zeta_{-,+}(x,t)-\zeta_{+,-}(x,t)\right).
\end{align}
\end{subequations}
\end{widetext}
Here $\eta_\Phi(x,t)$ and $\eta_\chi(x,t)$ are two, new, zero mean, Gaussian, space time white noises, but are correlated, such that they have statistics
\begin{subequations}
    \begin{align}
        \mathbb{E}[\eta_\Phi(x,t)]&= \mathbb{E}[\eta_\chi(x,t)]=0,\\
        \mathbb{E}[\eta_\Phi(x,t)\eta_\chi(y,s)]&=\frac{\chi(x,t)}{\Phi(x,t)}\delta(x-y)\delta(t-s),\\
        \mathbb{E}[\eta_\Phi(x,t)\eta_\Phi(y,s)]&=\mathbb{E}[\eta_\chi(x,t)\eta_\chi(y,s)]\nonumber\\
        &=\delta(x-y)\delta(t-s).
    \end{align}
    \label{cov}
\end{subequations}
Next, we make our first approximation. We stipulate that the width of the interaction kernel, $\sigma$, is small compared to both $L$ and emergent length-scales of the fields, such that $k_\sigma$ behaves as a Dirac delta function so that the integrals in Eq.~(\ref{initSPDE}) asymptotically approach the sifting identity, assuming that the field is regular enough on these length-scales for such an operation. 
Under such an approximation these equations become
\begin{widetext}
\begin{subequations}
    \label{dimensionfuleqs}
    \begin{align}
    \dot{\Phi}(x,t)&=-\nu\partial_x\left[\chi(x,t)\left(1-\rho_0 L\Phi(x,t)\right)\right]  + D\partial_x^2\Phi(x,t)+\sqrt{\frac{2D}{N}}\partial_x\left(\sqrt{\Phi(x,t)}\eta_\Phi(x,t)\right),\\
    \dot{\chi}(x,t)&=-\nu\partial_x\left[\Phi(x,t)\left(1-\rho_0 L\Phi(x,t)\right)\right]  + D\partial_x^2\chi(x,t)-(\gamma_++\gamma_-)\chi(x,t)+(\gamma_+-\gamma_-)\Phi(x,t)\nonumber\\
    &\qquad+\sqrt{\frac{2D}{N}}\partial_x\left(\sqrt{\Phi(x,t)}\eta_\chi(x,t)\right)+\frac{2}{N}\left(\zeta_{-,+}(x,t)-\zeta_{+,-}(x,t)\right).
\end{align}
\end{subequations}
\end{widetext}
Next we seek to non-dimensionalise our equations, and in doing so introduce our phase space parameters.
\par
First, we define $\bar{\gamma}\coloneq 2\gamma_+\gamma_-/(\gamma_++\gamma_-)\geq 0$ and $\beta\coloneq \gamma_-/\gamma_+\in[0,1]$ which characterise the inverse mean persistence time and mean population symmetry, respectively, under the assumption, without loss of generality, that $\gamma^+\geq \gamma^-$. 
This allows us to describe the system in terms of the conventional P\'eclet number, $\text{Pe}=\nu/\sqrt{\bar{\gamma}D}$, and the new parameter, $\beta$, which has no analogue in terms of the system's dynamical timescales. 
\par
We then combine this reparametrisation with the non-dimensionalisation, $x\to x'=x/l$, $t\to t'=t/\tau$, and $\Phi\to\rho=\Phi/\bar{\Phi}$, $\chi\to\psi=\chi/\bar{\Phi}$, where $l=\sqrt{D/\bar{\gamma}}$, $\tau=1/\bar{\gamma}$, and $\bar{\Phi}=(\rho_0 L)^{-1}$ are a characteristic length-scale, time scale, and probability density, respectively. 
Noting the Gaussian space-time white noise rescaling property ${\eta}_\cdot(x',t')=(l\tau)^{1/2}{\eta}_\cdot(x'l,t'\tau)=(l\tau)^{1/2}{\eta}_\cdot(x,t)$ and the requisite scaling of the Poisson noise increments (which natively possess units probability per unit space per unit time) $\zeta_{\cdot,\cdot}\to\hat{\zeta}_{\cdot,\cdot}=(l\tau)\zeta_{\cdot,\cdot}$, we arrive at dynamics in terms of the unit-less particle and polarisation  densities (no longer probability densities), $\rho$ and $\psi$, which vary on the scale of $\rho_0$,
\begin{widetext}
\begin{subequations}
\begin{align}
\partial_{t'} \rho(x',t')&=-\text{Pe}\partial_{x'} \left[\psi(x',t')\left(1-\rho(x',t')\right)\right] +\partial_{x'}^2\rho(x',t')+\partial_{x'}\left(\sqrt{\frac{2N_L\rho_0\rho(x',t')}{N}}\eta_{\rho}(x',t')\right),\\
\partial_{t'} \psi(x',t')&=-\text{Pe}\partial_{x'} \left[\rho(x',t')\left(1-\rho(x',t')\right)\right] +\partial_{x'}^2\psi(x',t')-\frac{(1+\beta)^2}{2\beta}\psi(x',t')+\frac{1-\beta^2}{2\beta}\rho(x',t')\nonumber\\
&\qquad+\partial_{x'}\left(\sqrt{\frac{2N_L\rho_0\rho(x',t')}{N}}\eta_{\psi}(x',t')\right)+\frac{2\rho_0 N_L}{N}\left(\hat{\zeta}_{-,+}(x',t')-\hat{\zeta}_{+,-}(x',t')\right).
\end{align}
\label{prefinalSPDE}
\end{subequations}
\end{widetext}
Here $N_L=L/l$ is the rescaled size of the domain, \emph{i.e.} a dimensionless length defined as a multiple of the characteristic length, $l$. 
In addition, the Poisson fields' intensities must also be rescaled such that
$\lambda_{\pm,\mp}(x,t)\to \hat{\lambda}_{\pm,\mp}(x',t')=l\tau \lambda_{\pm,\mp}(x'l,t'\tau)=l\tau \lambda_{\pm,\mp}(x,t)$. 
Explicitly, the intensities of the Poisson fields appearing in Eq.~(\ref{prefinalSPDE}) are given by
\begin{align}
    \hat{\lambda}_{\pm\mp}&=l\tau \lambda_{\pm\mp}=l\tau N\gamma_\mp \Phi^\pm=l\tau N\gamma_\mp \frac{\Phi\pm\chi}{2}\nonumber\\
    &=\frac{Nl\gamma_\mp}{\rho_0L\bar{\gamma}}\frac{\rho\pm\psi}{2}=\frac{N\gamma_\mp}{2\rho_0N_L\bar{\gamma}}(\rho\pm\psi),
\end{align}
such that they, also, are dimensionless.
\par
Finally, we consider these equations in the $N\gg 1$ limit such that we can consider the Poisson fields well represented by suitable Gaussian white noises with strength given by the square root of their intensity \cite{spinney_deankawasaki_2025}.
This allows us to make the approximation (relabelling $x'\to x$, $t'\to t$ for convenience)
\begin{widetext}
\begin{subequations}
\begin{align}
\partial_t \rho(x,t)&=-\text{Pe}\partial_x \left[\psi(x,t)\left(1-\rho(x,t)\right)\right] +\partial_x^2\rho(x,t)+\partial_x\left(\sqrt{\frac{2N_L\rho_0\rho(x,t)}{N}}\eta(x,t)\right),\\
\partial_t \psi(x,t)&=-\text{Pe}\partial_x \left[\rho(x,t)\left(1-\rho(x,t)\right)\right] +\partial_x^2\psi(x,t)-\frac{(1+\beta)^2}{2\beta}\psi(x,t)+\frac{1-\beta^2}{2\beta}\rho(x,t)+\partial_x\left(\sqrt{\frac{2N_L\rho_0\rho(x,t)}{N}}\eta_{\psi}(x,t)\right)\nonumber\\
&\qquad+\sqrt{\frac{2\rho_0 N_L\gamma_-(\rho(x,t)+\psi(x,t))}{N\bar{\gamma}}}\eta_{+,-}(x,t)-\sqrt{\frac{2\rho_0 N_L\gamma_+(\rho(x,t)-\psi(x,t))}{N\bar{\gamma}}}\eta_{-,+}(x,t),
\end{align}
\end{subequations}
\end{widetext}
where $\eta_{\pm,\mp}(x,t)$ are uncorrelated, unit, Gaussian space-time white noises. 
We can express this, up to an equivalence in distribution, in terms of the single unit Gaussian space-time white noise, $\eta_\leftrightarrow(x,t)$, whilst eliminating the $\gamma_\pm$ in favour of $\bar{\gamma}$ and $\beta$, finally arriving at Eq.~(\ref{eq:SPDE}) in the main text. 
Following non-dimensionalisation, the crucial $\rho_0$ parameter is then set through the (conserved) spatial average value of $\rho(x,t)$, $\rho_0={N_L}^{-1}\int_0^{N_L}dx\;\rho(x,t)$, as might the average physical density in a strongly interacting system. 
The noise covariances retain the same forms as in Eq.~(\ref{cov}), with the simple relabelling ${\Phi}\to\rho$ and ${\chi}\to\psi$. 
Finally, we remark that the equivalent single species fields are similarly constructed as $\rho^\pm=\Phi^\pm/\bar{\Phi}=(\rho\pm\psi)/2$, leading to the alternative non-dimensional formulation
\begin{widetext}
\begin{align}
\label{eq:SPDEs2}
\partial_t \rho^\pm(x,t)&=-\text{Pe}\partial_x\left[ \rho^\pm(x,t)\left(1-\rho^+(x,t)-\rho^-(x,t)\right)\right]+\partial_x^2\rho^\pm(x,t)\pm\frac{(1+\beta)}{2\beta}(\rho^-(x,t)-\beta\rho^+(x,t))\nonumber\\
&\qquad+\partial_x\left(\sqrt{\frac{2N_L\rho_0\rho^\pm(x,t)}{N}}\eta_{\rho^\pm}(x,t)\right)\pm\sqrt{\frac{N_L\rho_0(1+\beta)(\rho^-(x,t)+\beta\rho^+(x,t))}{2N\beta}}\eta_{\leftrightarrow}(x,t),
\end{align} 
\end{widetext}
with all of $\eta_{\rho^+}$, $\eta_{\rho^-}$,  and $\eta_{\leftrightarrow}$, unit, zero-mean, Gaussian space time white noises, and mutually independent.


\subsection{Symmetric (\texorpdfstring{$\beta = 1$}{beta = 1}) and Totally Asymmetric (\texorpdfstring{$\beta=0$}{beta = 0}) Limits} 

\label{b10limits}
It is instructive to consider the dynamics in the opposing limits of symmetry and total asymmetry. 
In the first case we have $\beta=1$ and thus the dynamics reduce to  
\begin{subequations}
\begin{align}
\partial_t \rho(x,t)=&-\text{Pe}\partial_x\left[ \psi(x,t)\left(1-\rho(x,t)\right)\right]+\partial_x^2\rho(x,t)\nonumber\\
&+\partial_x\left(\sqrt{\frac{2N_L\rho_0\rho(x,t)}{N}}\eta_{\rho}(x,t)\right),\\
\partial_t \psi(x,t)=&-\text{Pe}\partial_x \left[\rho(x,t)\left(1-\rho(x,t)\right)\right]+\partial_x^2\psi(x,t)\nonumber\\
&-2\psi(x,t)+\sqrt{\frac{4N_L\rho_0\rho(x,t)}{N}}\eta_{\leftrightarrow}(x,t)\nonumber\\
&+\partial_x\left(\sqrt{\frac{2N_L\rho_0\rho(x,t)}{N}}\eta_{\psi}(x,t)\right).
\end{align} 
\end{subequations}
The mean behaviour and large $N$ limit of this model matches the mean behaviour and hydrodynamic limit of a very closely related (and strongly interacting) on-lattice exclusion process \cite{agranov_exact_2021,kourbane-houssene_exact_2018,erignoux_hydrodynamic_2021}, 
however it does not possess the same noise structure. 
Notably, here, since the particles are off lattice, there is no bound from above on the local density and thus, for example, the noise strength is monotonic increasing in $\rho$, rather than being proportional to $\sqrt{\rho(x,t)(1-\rho(x,t))}$ as in \cite{agranov_exact_2021}, which emerges from the upper bound on density due to the lattice.
\par
In contrast, for $\beta=0$ all density is in the $+$ state with probability $1$, such that $\psi=\rho$, and we lose a degree of freedom (at each $x\in[0,L)$). 
Making the substitution, then taking the limit $\beta\to 0$ causes both equations of motion to coincide with
\begin{align}
\partial_t \rho(x,t)&=-\text{Pe}\partial_x \left[ \rho(x,t)\left(1-\rho(x,t)\right)\right]+\partial_x^2\rho(x,t)\nonumber\\
&+\partial_x\left(\sqrt{\frac{2N_L\rho_0\rho(x,t)}{N}}\eta_{\rho}(x,t)\right).
\label{burgers}
\end{align} 
This is identifiable as a stochastic viscous Burgers' equation under a frame shift at velocity $\text{Pe}$, or a stochastic refinement of the Lighthill-Whitham-Richards (LWR) model \cite{lighthill_kinematic_1997}, as studied, for example, up to a choice of constants, in application to traffic modelling in \cite{worsfold_density_2023}. 
See, also \cite{sznitman_propagation_1986}.
%
%
\section{Linear Stability of Eq.~\ref{eq:NSAMB}}
\label{app:NSAMB_stability}

As discussed in the main text, the triple limiting solution of Eq.~(\ref{eq:NSAMB}) is underdetermined in the homogeneous case of $\phi=\rho_0$. We can deduce the values of $v$ and $J^{\rm st}_v$ for which stability properties between Eq.~(\ref{eq:deterministic}) and Eq.~(\ref{eq:NSAMB}) are equivalent, simultaneously assuring an absence of procession in the response to small perturbations, by performing a linear stability analysis.
\par
To do so we enforce the homogeneous condition $F=0$ [Eq.~(\ref{eq:F=0})], thus setting 
\begin{align}
    \label{eq:J_homogeneous}
    J^{\rm st}_v&=\text{Pe}\frac{1-\beta}{1+\beta}\rho_0(1-\rho_0)-v\rho_0=J_{\rm bulk}(\rho_0)-v\rho_0,
\end{align}
in Eq.~(\ref{eq:NSAMB}). 
Then, as in Sec.~\ref{sec:linear_stability}, we expand to first order in fluctuations around the homogeneous state, $\delta\phi\coloneq \phi-\rho_0$, Fourier-transform in space, then expand in orders of $k$ to obtain
\begin{widetext}
\begin{align}
    \frac{d}{dt}\delta\tilde{\phi}(k,t)&=\frac{i  M(\rho _0)(\beta +1)  \left((\beta -1) \text{Pe} \left(2 \rho _0-1\right)-(\beta +1) v\right)}{2 \beta  \text{Pe} \left(\rho _0-1\right)}k\nonumber\\
    &+\frac{ M(\rho _0) \left(\beta  \left(\beta  (\beta +3)+2 (\beta +1) \text{Pe}^2 \left(\rho _0-1\right) \left(2 \rho _0-1\right)+2 (\beta -1) \text{Pe} \rho _0 v-2 (\beta +1) v^2+3\right)+1\right)}{2 \beta  (\beta +1)
    \text{Pe} \left(\rho _0-1\right)}k^2+\mathcal{O}(k^3).
\end{align}
\end{widetext}
Choosing 
\begin{align}
    v&=\text{Pe}\frac{1-\beta}{1+\beta}(1-2\rho_0),
    \label{vchoice}
\end{align}
equal to Eq.~(\ref{eq:vfluct}), causes the (imaginary) term linear in $k$ to vanish, allowing us to conclude that large scale fluctuations in this one-parameter system are purely dispersive.  
And, crucially, the remaining expression, at $\mathcal{O}(k^2)$, becomes
\begin{align}
    \frac{d}{dt}\delta\tilde{\phi}(k,t)=&\frac{k^2 M(\rho _0) \left(\frac{(\beta +1)^4}{\rho _0-1}+16 \beta ^2 \text{Pe}^2 \rho _0-8 \beta ^2 \text{Pe}^2\right)}{2 \beta  (\beta +1)^2 \text{Pe}}\nonumber\\
    &+\mathcal{O}(k^3).
\end{align}
Solving for the $\mathcal{O}(k^2)$ term vanishing then yields precisely the same stability line as Eq.~(\ref{eq:spinodal}), \emph{i.e.} the stability line for Eq.~(\ref{eq:deterministic}). 
Thus, homogeneous solutions to Eq.~(\ref{eq:NSAMB}), under this choice of $v$, have the same macroscopic stability properties as Eq.~(\ref{eq:deterministic}) and do so with a purely dispersive damping/growth of fluctuations as in effective equilibrium theories (those without steady current) such as, for example, AMB ($\beta=1$). The matched value of $J^{\rm st}_v$ in Eq.~(\ref{eq:vJdegen}) then immediately follows from Eq.~(\ref{eq:J_homogeneous}).
\par
Finally, we remark that these $\{v,J^{\rm st}_v\}$ are unique for a given $\rho_0$. This follows from noting that $J_{\rm bulk}$ is concave, whilst $v=J_{\rm bulk}'(\rho_0)$, with $J^{\rm st}_v$ given by the Legendre transform of $J_{\rm bulk}$, $\sup_{\rho_0}\rho_0v-J_{\rm bulk}(\rho_0)$, owing to the linear construction Eq.~(\ref{eq:linearJ}).
%

\section{Duality Between $\{v,J^{\rm st}_v\}$ and $\{\rho_L,\rho_G\}$}
\label{sec:duality_figure}

\begin{figure}[!htp]
    \centering
    \includegraphics[width=0.48\textwidth]{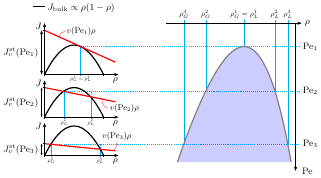} 
 
    \caption{{\bf Graphical relationship between $\{v,J^{\rm st}_v\}$ and $\{\rho_L,\rho_G\}$.} The velocity of the co-moving frame, $v$, and homogeneous current in the co-moving frame, $J^{\rm st}_v$, uniquely determine the co-existence densities, should they exist, and \emph{vice versa}. As the control parameter ($\text{Pe}$) is varied (here taking values $\text{Pe}_1$, $\text{Pe}_2$, and $\text{Pe}_3$), $v$ and $J^\rho_v$ change, intersecting distinct points in $J_{\rm bulk}$, defining the coexistence densities $\rho_G$, $\rho_L$, mapping out a (cartoon of a) phase diagram. \label{fig:dualvfig}}
\end{figure}
The duality between $v$, $J^{\rm st}_v$ and $\rho_G$, $\rho_L$ is illustrated in Fig.~\ref{fig:dualvfig}.
\section{Numerics and Visualisation}
\label{sec:simulation}

We use numerical simulation to confirm much of the predicted behaviour, and to discern behaviour that is beyond the reach of analytics. 
Code for the simulation of both the underlying SDEs, and limiting deterministic PDEs, can be found in \cite{spinney_rspinneyamips_2024}. 
Large simulations were performed on the Katana computing cluster \cite{noauthor_katana_nodate}. 
Pertinent details are described below.


\subsection{Numerical Integration of the \texorpdfstring{$N\to \infty$}{N to infinity} Limiting PDEs}

%
%
These equations are solved numerically with the FenicsX finite element package \cite{baratta_dolfinx_2023}. 
We utilise a Crank-Nicholson, mid-point, time discretisation scheme and a Newton non-linear solver to converge the fields at each time step. 
This is used in conjunction with the DolfinX multi-point constraints (MPC) package to implement periodic boundaries \cite{dokken_dolfinx-mpc_2023}. 


\subsubsection{Parameters Used}

%
In all simulations, unless otherwise specified, we use an equal-spaced mesh containing $3\times 10^5$ elements on a (non-dimensional) domain length $N_L = 256\pi$, and a time discretisation of $\Delta t = 5\times10^{-4}$.
%

\subsection{Numerical Integration of the Particulate Jump-SDEs}

%
To empirically investigate the full stochastic behaviour of the system described by the SPDEs given in Eqs.~(\ref{eq:SPDE}), we simulate the underlying particle dynamics as written in Eqs.~(\ref{eq:SDEs}). 
Obtaining approximate solutions of Eq.~(\ref{eq:SDEsA}) entails a forward Euler-Maryama integration scheme with a fixed time-step, $\Delta t$, setting the temporal resolution of the simulation. 
Meanwhile, integration of the $\mathcal{Y}_i$ in Eq.~(\ref{eq:SDEsB}) is performed through use of the Gillespie algorithm, since all transition rates are constant. 
This is performed on an interval of length $L$, with periodic boundary conditions, in order to simulate the ring $S^1(L/(2\pi))$.
\par
To complete the model we must choose an interaction kernel, which we specify to be a simple top-hat function
\begin{align}
    \label{ksigma}
    k_\sigma(X_i-X_j)\equiv k^{\rm PBC}_\sigma(\Delta x) &=
    \begin{cases}
        (2\sigma)^{-1}&|\Delta x| < \sigma,\\
        0&|\Delta x| \geq \sigma,    
    \end{cases}
\end{align}
where $\sigma$ is the maximum interaction distance, and $\Delta x$ is the minimum absolute distance between the particles on $S^1(L/(2\pi))$ [\emph{i.e.} accounting for periodic boundary conditions, \emph{cf.} the superscript in Eq.~(\ref{ksigma})].
%

\subsubsection{Empirical Coarse-Graining}

%
The simulations naturally deal with point particles, but generically, we wish to deal with (continuous) density fields. 
As such we coarse-grain the atomic raw data into such a field by defining it in terms of Gaussian basis functions, $\rho^{\varsigma}_{X_i(t)}$,  with width $\varsigma$, \emph{viz.}
\begin{subequations}
\begin{align}
    \rho^{\pm}_{\rm sde}(x,t)&:=\frac{\rho_0 L}{N}\sum_{i=1}^{N}\delta_{\pm,\mathcal{Y}_i(t)}\rho^{\varsigma}_{X_i(t)}(x),\\
    \rho^{\varsigma}_{X_i(t)}(x)&:=\frac{1}{2\sqrt{\pi\varsigma^2}}\exp\left[-(2\varsigma)^{-1}(X_i(t)-x)^2\right],
\end{align}
\end{subequations}
from which we can construct $\rho_{\rm sde}=\rho^{+}_{\rm sde}+\rho^{-}_{\rm sde}$ and $\psi_{\rm sde}=\rho^{+}_{\rm sde}-\rho^{-}_{\rm sde}$. 
Finally, we note the pre-factor in the above is used so as to satisfy $L^{-1}\int_0^Ldx\;\rho_{\rm sde}(x,t)=\rho_0$, allowing $\rho_{\rm sde}$ to be contrasted with the results from Eq.~(\ref{eq:deterministic}).
%

\subsubsection{Parameters Used}

%
Simulations, unless otherwise specified, use $N=320,000$ particles on a domain of length $L=128\pi$ using a temporal resolution of $\Delta t=10^{-3}$. In all other cases the same average density is used (\emph{e.g.} $N=80,000$ particles on a domain of length $32\pi$).
The interaction distance is taken as $\sigma = 0.005$, such that $\sigma/L=5.0\times 10^{-5}\ll 1$, satisfying our condition for approximating the interaction kernel as a delta function. 
We then finally specify $\nu=6.0$ and $D=1.0$, with $\gamma^{\pm}$ varied as required to achieve the desired control parameter values for $\text{Pe}$ and $\beta$. 
A fixed $\nu$ is used in order to avoid integration accuracy depending on control parameters. 
Conversion formulae between dimensionful and dimensionless quantities are provided where relevant.
\par 
The continuous field is then constructed using a coarse-graining length-scale of $\varsigma/2=0.0375=7.5\sigma$, chosen to balance overall fidelity  without resolving features on the length-scale of the interaction kernel, and effective communication of macroscopic phenomena. 


\subsection{Visualisation}


%
We present the behaviour of the system using density kymographs, such that colours represent density, with the spatial extent of the system taking up the vertical aspect of the plot as it evolves horizontally in time to the right. 
In these kymographs bluer colours are used to represent lower densities, whilst redder colours represent higher densities. 
We must, however, make clear, this is only rigorously the case within individual plots. 
Colours are not mapped consistently to values of density between plots, and are chosen to best illustrate the phenomena at hand, which might require diminished/augmented colour variation/contrast between or within a given plot, dependent on the situation.
\section{Self-Consistent Shooting Method}
\label{app:shooting}
In this appendix we fully describe the methodology behind the self-consistent shooting algorithm used to calculate the binodals in Fig.~\ref{fig:phase_diagrams}(c). We reiterate that we aim to solve the implicit boundary value problem prescribed by the third order ODE
\begin{align}
    {\mathcal{L}}_{\rho_L,\rho_G}[\phi]=0,
\end{align}
where we seek the the boundary values of $\phi$ at $x=-\infty$ and $x=+\infty$, constrained to be equal to the values of $\rho_L$ and $\rho_G$, used in the operator ${\mathcal{L}}_{\rho_L,\rho_G}$ itself, respectively.
\par
As described in the main text we convert this into an initial value problem by means of a shooting method. As such, given that ${\mathcal{L}}_{\rho_L,\rho_G}[\phi]=0$ defines a third order ODE, we thus, as with a conventional shooting method, integrate from a near boundary using initial values of $\phi$, $\phi'$, and $\phi''$, and aim to hit a target value of $\phi$ at the far boundary. There are, however, two notable distinctions between the present method and a conventional shooting method. Firstly, seeking perfectly flat bulk-state boundaries located at $x=\pm\infty$ is not possible numerically. Instead, we integrate from an \emph{approximately} flat solution at $x=0$, and integrate until a plausible solution envelope is left at some $x=x_{\rm max}$. We then search along the generated function and seek the most plausible point to consider as the far boundary at a value $x=l\in[0,x_{\rm max}]$ --- \emph{i.e.} the location of the far boundary is unknown and is a result of the computation. And, secondly, we do not know the values the function should take on the near and far boundary, as these are the unknown $\rho_L$ and $\rho_G$ that we seek in ${\mathcal{L}}_{\rho_L,\rho_G}$. Consequently, these are treated as unknowns in the shooting method, on equal footing with the unknown initial $\phi(0)$, $\phi'(0)$ and $\phi''(0)$.
\par
We are thus presented, at first, with five unknowns $\{\phi(0),\phi'(0),\phi''(0),\rho_L,\rho_G\}$ that are required to integrate ${\mathcal{L}}_{\rho_L,\rho_G}[\phi]=0$ and produce trial solutions. However, in order to be self-consistent we require $\phi(0)\simeq \rho_L$. In practice, we stipulate that we integrate from $\phi(0)=\rho_L-\delta$, where $0<\delta\ll 1$ is an (unoptimised) hyper-parameter accounting for the fact that we cannot reach a true bulk phase with finite computation, in turn allowing us to integrate from non-zero (but small) values of $\phi'(0)$ and $\phi''(0)$. Consequently, we thus integrate from an initial condition $\phi(0)=\rho_L-\delta$, $\phi'(0)$, $\phi''(0)$ based on the specification of four unknowns $\{\rho_L,\rho_G,\phi'(0),\phi''(0)\}$. We then seek the values of these four unknowns such that the resulting solution is self-consistent (\emph{i.e.} $\phi(0)\simeq\rho_L$ and $\phi(l)\simeq \rho_G$ with vanishing derivatives at $x=0$ and $x=l$ such that the solution represents the liquid and gaseous bulk phases at $x=0$ and $x=l$, respectively).
\par 
To find these values, the four quantities thus become parameters in an optimisaton problem and our task becomes one of specification of an appropriate objective function so as to converge on the correct solutions. Once a trial solution has been generated by integrating from $x=0$ to $x=x_{\rm max}$, we specify such an objective function as
\begin{align}
    \label{eq:objFn}
   \varepsilon&= \min_{x\in[0,x_{\rm max}]}[(\rho_G-\phi(x))^2+\alpha_1(\phi'(0))^2+\alpha_2(\phi''(0))^2\nonumber\\
   &\qquad\qquad+\alpha_3(\phi'(x))^2+\alpha_4(\phi''(x))^2+\alpha_5 (\phi'''(x))^2],
\end{align}
in terms of the unoptimised hyper-parameters $\alpha_1$ through $\alpha_5$. The returned value of $\varepsilon$ will then be small when the function reaches a flat profile at the self-consistent gaseous density, $\rho_G$, starting from an approximately flat bulk configuration at $\rho_L$. 
This in turn also determines the effective location of the far boundary as per
\begin{align}
    l&=\argmin_{x\in[0,x_{\rm max}]}[(\rho_G-\phi(x))^2+\alpha_1(\phi'(0))^2+\alpha_2(\phi''(0))^2\nonumber\\
   &\qquad\qquad+\alpha_3(\phi'(x))^2+\alpha_4(\phi''(x))^2+\alpha_5 (\phi'''(x))^2],
\end{align}
as the $x=l$ which determines $\varepsilon$ coincides with the location in the generated trial solution which best resembles the bulk gaseous phase. Following this minimisation we then only consider the function to be valid on $[0,l]$. 
Any result of the integration for $x>l$, in general, eventually leads to uncontrolled divergences owing to a lack of infinite precision and the stiff nature of the underlying ODE.
\par
By then varying $\{\rho_L,\rho_G,\phi'(0),\phi''(0)\}$ such that we minimise the above objective function we converge on a boundary solution that satisfies ${\mathcal{L}}_{\rho_L,\rho_G}[\phi]=0$. 
This can be done iteratively with a variety of optimisation methods, \emph{e.g.} gradient descent, conjugate gradient, \emph{etc.} 
When such a minimum is reached, we thus assert that $\rho_L$ and $\rho_G$ are the coexistence densities for the values of $\beta$ and $\text{Pe}$ being utilised. 
\par
In practice, we utilised a small value of $\delta=10^{-10}$, utilised equal values of $\alpha_1=\alpha_2=\alpha_3=\alpha_4=\alpha_5=1$, integrated in the shooting step using the \emph{NDSolve} function in the \emph{Mathematica}\textsuperscript{\tiny\textregistered} software package \cite{wolfram_research_inc_mathematica_nodate},  and demanded the lowest value of the objective function that can be found for a given set of parameters and initial conditions using a conjugate gradient optimisation method. The entire method for determining $\{\rho_G,\rho_L\}$ is encapsulated into Alg.~1, which appears in the Supplemental Material \cite{SM}. A \emph{Mathematica}\textsuperscript{\tiny\textregistered} notebook containing the results shown in Fig.~\ref{fig:phase_diagrams}(b) and (c), alongside illustration of the method, is released alongside code in \cite{spinney_rspinneyamips_2024}.
\par
We note that selecting initial guesses for the optimised parameters is an important part of how the coexistence solutions were found, as the initial values need to be close enough to the true values in order to allow for suitable convergence (and, for example, to avoid convergence to the trivial homogeneous solution $\rho_L=\rho_G=\rho_0$). In practice this was achieved by starting with known coexistence solutions (found from integrating Eq.~(\ref{eq:deterministic}) at suitably high $\text{Pe}$), confirming convergence and self-consistency of these solutions using the present method (Alg.~1 in the Supplemental Material \cite{SM}), then using the resulting optimised parameters as the initial parameter values to use in the algorithm at some small deviation $\text{Pe}-d\text{Pe}$ along the proposed binodals. This procedure was then repeated to find the full binodal lines. Such a procedure is illustrated in Alg.~2 in the Supplemental Material \cite{SM}.
\par
In practice, this procedure is fairly computationally intensive and so calculation of the full binodal line involves solution of Eq.~(\ref{eq:deterministic}) at high values of $\text{Pe}$, with the present method only used for lower values of $\text{Pe}$. At the value of $\text{Pe}$ where we exchange solution methods, both methods agree on $\rho_L$ and $\rho_G$ to four decimal places. Finally, we note that at the lowest values of $\text{Pe}$, near the bifurcation point, convergence of the algorithm would either fail or fall into the basin centred on $\rho_L=\rho_G=\rho_0$ as part of the exploration process, explaining the absence of data in that region.
\par
\begin{figure*}[!htp]
   \centering
   \includegraphics[width=\textwidth]{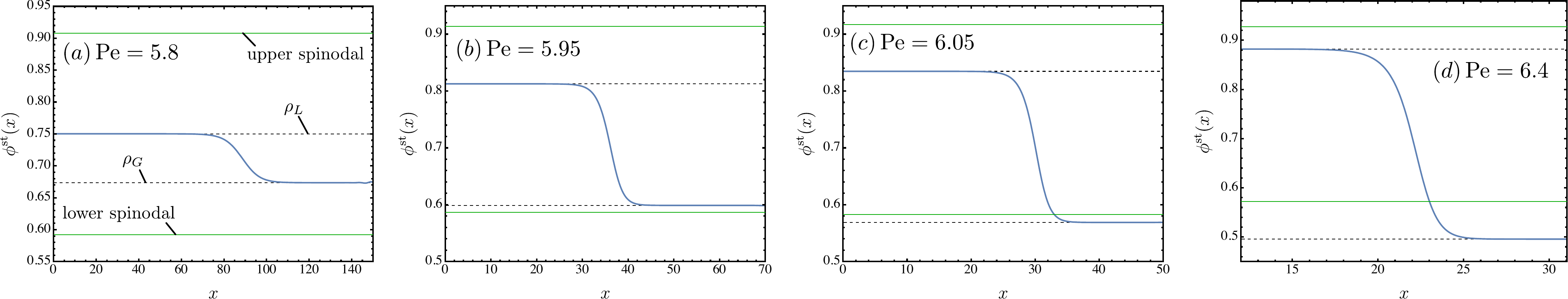} 
   \caption{{\bf Phase boundary solutions obtained using the self-consistent shooting method for $\beta=1/2$.} By utilising the self-consistent shooting method we may obtain sigmoidal liquid-to-gas phase boundary solutions (blue lines) even when the bulk coexistence densities lie within the spinodals (green lines). All plots are shown for $\beta=1/2$.}
   \label{fig:shooting} 
\end{figure*}

Example solutions that correspond to points on the binodals in Fig.~\ref{fig:phase_diagrams}(b) and (c) are shown in Fig.~\ref{fig:shooting}.

\section{Full Fluctuation Formulae}
\label{app:fluctuations}
Calculating $C_{\rho,\rho}(k,\omega)$ from Eq.~(\ref{eq:psd}) yields the full power spectral density
\begin{align}
\label{eq:app_psd}
&\mathbb{E}[\overline{\delta\hat{\rho}(k,\omega)}\delta\hat{\rho}(q,\upsilon)]=\varepsilon\frac{C_1}{ C_2}\delta_{k,q}\delta(\omega-\upsilon),
\end{align}
where
\begin{widetext}
\begin{subequations}
\begin{align}
C_1&=2 k^2 \xi _+ \left(\xi _+ \left(k^2+\xi _+\right) \left(k^2+\xi _++\text{Pe}^2 \left(1-\rho _0\right)^2\right)+2 k \xi _- \text{Pe} \left(1-\rho _0\right) \omega +\xi _+ \omega ^2-\xi _-^2 \text{Pe}^2 \left(1-\rho _0\right)^2\right),\\
C_2&=\xi _+^4 \left(k^4+\omega ^2\right)+2 k \xi _+^3 \left(k^3 \text{Pe}^2 \left(1-2 \rho _0\right) \left(1-\rho _0\right)+k^5+k \omega ^2+\xi _- \text{Pe} \left(2 \rho _0-1\right) \omega \right)\nonumber\\
&\quad+\xi _+^2 \Big[k^2 \text{Pe}^2 \left(2 \rho _0-1\right) \left(\xi _-^2 \left(2 \rho _0-1\right)+2 \left(1-\rho
   _0\right) \omega ^2\right)+k^4 \left(\text{Pe}^4 \left(1-\rho _0\right)^2 \left(2 \rho _0-1\right)^2+2 \omega ^2\right)\nonumber\\
   &\qquad\qquad+2 k^6 \text{Pe}^2 \left(1-2 \rho _0\right) \left(1-\rho _0\right)+4 k^3 \xi _- \text{Pe} \left(2 \rho _0-1\right) \omega +k^8+\omega ^4\Big]\nonumber\\
   &\qquad+2 k \xi _- \xi _+ \text{Pe} \rho _0
   \left(k^2 \omega  \left(k^2+\text{Pe}^2 \left(1-\rho _0\right) \left(2 \rho _0-1\right)\right)+k^3 \xi _- \text{Pe} \left(2 \rho _0-1\right)+\omega ^3\right)+k^2 \xi _-^2 \text{Pe}^2 \rho _0^2 \left(k^4+\omega ^2\right),
\end{align}
\end{subequations}
\end{widetext}
in terms of collected parameters $\xi_+\coloneq (1+\beta)^2/(2\beta)$ and $\xi_-\coloneq (1-\beta^2)/(2\beta)$, and recalling $\varepsilon=\rho_0^2N_L/N$.
\par
To obtain correlation and cross correlation functions between the individual positive and negatively oriented species fields, we Fourier transform and linearise Eq.~(\ref{eq:SPDEs2}) yielding
\begin{align}
\begin{bmatrix}
\delta\hat{\rho}^{+}(k,\omega)\\
\delta\hat{\rho}^{-}(k,\omega)
\end{bmatrix}=-\sqrt{\varepsilon}\left(i\omega\mathbb{I}+\mathbf{B}(k)\right)^{-1}\hat{\boldsymbol{\eta}}(k,\omega).
\end{align}
Here the linearised deterministic operator is given by
\begin{widetext}
\begin{align}
\mathbf{B}(k)&=
\begin{bmatrix}
-\frac{1+\beta}{2} - k^2+\frac{ i k \text{Pe} \left((\beta +2) \rho _0-\beta -1\right)}{\beta +1} & \frac{1+\beta}{2\beta }+\frac{ i k \text{Pe} \rho _0}{\beta +1} \\
 \frac{1+\beta}{2} -\frac{ i \beta  k \text{Pe} \rho _0}{\beta +1} & -\frac{1+\beta }{2 \beta }-k^2-\frac{i k \text{Pe} \left(2 \beta  \rho _0-\beta +\rho _0-1\right)}{\beta +1} \\
\end{bmatrix},
\end{align}
\end{widetext}
with noise vector
\begin{align}
\hat{\boldsymbol{\eta}}(k,\omega)&=
\begin{bmatrix}
\sqrt{2}ik\hat{\eta}_{\rho^+}(k,\omega)+\hat{\eta}_\leftrightarrow(k,\omega)\\
\sqrt{2}ik\hat{\eta}_{\rho^-}(k,\omega)-\hat{\eta}_\leftrightarrow(k,\omega)
\end{bmatrix},
\end{align}
where all constituent noise terms are delta correlated and mutually independent.
\par
This allows us to calculate the following correlation, and cross-correlation, power spectra
\begin{subequations}
\begin{align}
&\mathbb{E}[\overline{\delta{\hat{\rho}^{-}}(k,\omega)}\delta\hat{\rho}^{-}(q,\upsilon)]=\varepsilon\frac{C_L}{ C_D}\delta_{k,q}\delta(\omega-\upsilon),\\
&\mathbb{E}[\overline{\delta{\hat{\rho}^{+}}(k,\omega)}\delta\hat{\rho}^{+}(q,\upsilon)]=\varepsilon\frac{C_R}{ C_D}\delta_{k,q}\delta(\omega-\upsilon),\\
&\mathbb{E}[\overline{\delta{\hat{\rho}^{+}}(k,\omega)}\delta\hat{\rho}^{-}(q,\upsilon)]=\varepsilon\frac{C_{LR}}{ C_D}\delta_{k,q}\delta(\omega-\upsilon),
\label{fullLR}
\end{align}
\end{subequations}
where
\begin{widetext}
\begin{align}
   C_L=&2 \beta ^2 \Big[(\beta +1) \left(2 \omega ^2 \left(\beta +2 \beta  k^2+1\right)+4 \beta  k^6+k^4 \left(2 \beta  \left(2 \beta +2 \text{Pe}^2+3\right)+2\right)+(\beta +1) k^2 \left((\beta +1)^2+2 \text{Pe}^2\right)\right)\nonumber\\
   &\qquad-4 k
   \text{Pe} \omega(\beta+1)  \left(\beta +2 \beta  k^2+1\right)+4 k \text{Pe} \rho _0 \left(k \text{Pe} \rho _0 \left(\beta  (\beta +4) k^2+2\right)-2 \left(\beta +\beta  (\beta +2) k^2+1\right) (k \text{Pe}-\omega )\right)\Big],
   \end{align}
   \begin{align}
   C_R=&2 \Big[(\beta +1) \left(2 \beta ^2 \omega ^2 \left(\beta +2 k^2+1\right)+k^2 \left(\beta  \left(4 \beta  k^4+2 k^2 \left(\beta  \left(\beta +2 \text{Pe}^2+3\right)+2\right)+\beta  \left(\beta +2 (\beta +1)
   \text{Pe}^2+3\right)+3\right)+1\right)\right)\nonumber\\
   &\,+4 (\beta +1)\beta ^2 k \text{Pe} \omega  \left(\beta +2 k^2+1\right)+4 \beta ^2 k \text{Pe} \rho _0 \left(k \text{Pe} \rho _0 \left(2 \beta ^2+(4 \beta +1) k^2\right)-2 \left(\beta ^2+\beta
   +2 \beta  k^2+k^2\right) (k \text{Pe}+\omega )\right)\Big],
   \end{align}
   \begin{align}
   C_{LR}&=-2 \beta   \Big[2 \beta  (\beta +1)^2 \omega ^2-2 \beta  k^4 \left(\beta  \left(\beta +2 \text{Pe}^2 \left(2-3 \rho _0\right) \rho _0+2\right)+1\right)\nonumber\\
   &\qquad\qquad\qquad-k^2 \left(\beta  (\beta +2) (\beta  (\beta +2)+2)+2 \beta 
   \text{Pe}^2 \left(-\beta +2 \rho _0-1\right) \left(\beta  \left(2 \rho _0-1\right)-1\right)+1\right)\nonumber\\
   &\qquad\qquad\qquad+8 i \beta ^2 k^5 \text{Pe} \rho _0+4 i \beta  (\beta +1)^2 k^3 \text{Pe} \rho _0-4 \beta  \left(\beta ^2-1\right) k \text{Pe}
   \rho _0 \omega \Big],
   \end{align}
   \begin{align}
   C_D=& 4 \beta ^2 (\beta +1)^2 \omega ^4+(\beta +1)^6 \omega ^2+4 \beta ^2 (\beta +1)^2 k^8+(\beta +1)^2 k^4\nonumber\\
   &+\beta (\beta +1)^2 k^4 \left(\beta  \left(\beta  (\beta +4)+8 \omega ^2+6\right)+4 \text{Pe}^2 \left(2 \rho _0-1\right)
   \left(2 \left(\beta ^2+1\right) \rho _0-(\beta +1)^2\right)+4 \beta  \text{Pe}^4 \left(2 \rho _0^2-3 \rho _0+1\right)^2+4\right)\nonumber\\
   &+k^2 \left(4 \beta  (\beta +1)^4 \omega ^2+\text{Pe}^2 \left((\beta -1)^2 (\beta +1)^4
   \left(1-2 \rho _0\right)^2-4 \beta ^2 \omega ^2 \left(-6 (\beta +1)^2 \rho _0+(\beta +3) (3 \beta +1) \rho _0^2+2 (\beta +1)^2\right)\right)\right)\nonumber\\
   &+8 \beta  \left(\beta ^2-1\right) k^3 \text{Pe} \left(2 \rho _0-1\right)
   \omega  \left(\beta  \text{Pe}^2 \left(\rho _0-1\right) \rho _0-(\beta +1)^2\right)\nonumber\\
   &+4 \beta  k^6 \left(\text{Pe}^2 \rho _0 \left(\beta  (\beta  (5 \beta +6)+5) \rho _0-6 \beta  (\beta +1)^2\right)+(\beta +1)^2 \left(\beta 
   \left(\beta +2 \text{Pe}^2+2\right)+1\right)\right)\nonumber\\
   &-8 \beta ^2 \left(\beta ^2-1\right) k^5 \text{Pe} \rho _0 \omega +2 (\beta -1) (\beta +1) k \text{Pe} \omega  \left((\beta +1)^4-2 \rho _0 \left(2 \beta ^2 \omega ^2+(\beta
   +1)^4\right)\right).
\end{align}
\end{widetext}
The power spectral density of the equal time two point correlation function in Sec.~\ref{sec:fluct_corrlength} is calculated to be
\begin{widetext}
\begin{subequations}
\begin{align}
    &\mathbb{E}[\overline{\delta\tilde{\rho}(k,t)}{\delta\tilde{\rho}(q,t)}]=\delta_{k,q}\varepsilon\frac{C_N}{C_D},\\
    C_N&=-(2 c_2^2 c_5 + c_5^3) \nonumber\\
    &\quad + (1+\beta)^{-1}\Big[ -((c_1^2 c_5 + 5 c_5^2) (1 + \beta )) - c_2^2 (4 + 2 c_4 + c_5 + 4 \beta  - 2 c_4 \beta  + c_5 \beta ) \nonumber\\
    &\qquad\qquad\qquad\qquad- c_2 (2 c_1 c_5 (-1 + \beta ) - c_1 c_4 (1 + \beta ) + c_3 c_5 (1 + \beta ))\Big]k^2 \nonumber\\
    &\quad+ (1 + \beta )^{-1}\left[ -2 c_2^2 (1 + \beta ) - (c_1^2 + 8 c_5) (1 + \beta ) - 2 c_2 (-c_1 + c_3 + c_1 \beta  + c_3 \beta )\right]k^4 - 4 k^6,\\
   C_D&=(1+\beta)\left[(c_2^2 c_4^2 - c_5^3 - 
   c_2 (c_1 c_4 c_5 + c_3 c_5^2)) - (c_1^2 c_5 + 4 c_2 c_3 c_5 + 
    5 c_5^2) k^2 - (c_1^2 + 4 c_2 c_3 + 8 c_5) k^4 - 4 k^6\right],
    \label{eq:corr_denom}
\end{align}
\end{subequations}
\end{widetext}
with $c_1=\text{Pe}\rho_0(1-\beta)/(1+\beta)$, $c_2=\text{Pe}(1-\rho_0)$, $c_3=\text{Pe}(1-2\rho_0)$, $c_4=(1-\beta^2)/2\beta$, and $c_5=(1+\beta)^2/2\beta$.
\end{appendix}


%

\end{document}


\title{\texorpdfstring{Nonstationary Critical Phenomena: Expanding The Critical Point\\ Supplementary Information}{Nonstationary Critical Phenomena: Expanding The Critical Point. Supplementary Information}}

\author{Richard E. Spinney}
\affiliation{School  of  Physics, University  of  New  South  Wales  -  Sydney  2052,  Australia}
\affiliation{EMBL-Australia  node  in  Single  Molecule  Science, School of Medical Sciences, University  of  New  South  Wales  -  Sydney  2052,  Australia}

\author{Richard G. Morris}
\affiliation{School  of  Physics, University  of  New  South  Wales  -  Sydney  2052,  Australia}
\affiliation{EMBL-Australia  node  in  Single  Molecule  Science, School of Medical Sciences, University  of  New  South  Wales  -  Sydney  2052,  Australia}
\affiliation{ARC Centre of Excellence for the Mathematical Analysis of Cellular Systems, UNSW Node, Sydney, NSW 2052, Australia}
 
\date{\today}

\maketitle

\vspace{-2em}

\begingroup
\setlength{\parskip}{0pt plus .2pt}
\tableofcontents
\endgroup

\clearpage
\section{Additional fluctuation results/simulation}

\subsection{Dimensions}
\label{sec:dimensions}

We note, ahead of comparison with simulation, that the analytical spectra presented here and in the main text come from the non-dimensionalised SPDEs, whereas a simulated or ``real" system, based on the underlying (dimensional) SDEs, can depend on all model parameters not present in the analytical spectra. 
As such we note that all of the power spectra presented can be converted to those of a dimensional system via the rescaling formula
\begin{align}
   \mathbb{E}[\overline{\delta\hat{\rho}_{\rm dim}(k,\omega)}\delta\hat{\rho}_{\rm dim}(q,\upsilon)]=l\tau\mathbb{E}[\overline{\delta\hat{\rho}(lk,\tau\omega)}\delta\hat{\rho}(lq,\tau\upsilon)],
\end{align}
where $l=\sqrt{D/\bar{\gamma}}$ and $\tau=1/\bar{\gamma}$, with the expression on the L.H.S. now using dimensional $x$ and $t$, but still in terms of the unit-less model density $\rho_{\rm dim}\equiv \rho_0 L\Phi$.
Where (graphical) comparison with simulation is considered, the re-dimensionalised forms are used throughout this document and the main text.
\par
A corollary of this is that velocities in the dimensionful simulations of the underlying SDEs, $v_{\rm sde}$, can be compared with velocities in the non-dimensional analysis, $v$, through the transform $v_{\rm sim}=v\cdot l/\tau$. Consequently, non-dimensional velocities $v\sim\text{Pe}$ correspond to dimensionful velocities $v_{\rm sde}\sim \nu$, where we recall $\nu$ is the motile strength (since $\text{Pe}=\nu/\sqrt{\bar{\gamma}D}$). Consequently, the phase boundary procession velocity from the main text $v=\text{Pe}((1-\beta)/(1+\beta))(1-\rho_G-\rho_L)$ becomes $v_{\rm sde}=\nu((1-\beta)/(1+\beta))(1-\rho_G-\rho_L)$ in the dimensionful simulation. Similarly, the typical ($k\to 0$) fluctuation velocity $v_{\rm fluct}=\text{Pe}((1-\beta)/(1+\beta))(1-2\rho)$ becomes $v_{\rm fluct}^{\rm sde}=\nu ({(1-\beta)}{(1+\beta)})(1-2\rho)$.
\par
In all simulations we control $\text{Pe}$ through variation in $\gamma^\pm$, utilising a fixed $\nu=6$, rendering velocities such as the frame boost $v$ and fluctuation velocity $v_{\rm fluct}$ dependent only on non-dimensional densities in the system (\emph{e.g.} $\rho_0$, $\rho_G$, $\rho_L$).

\subsection{Totally asymmetric, \texorpdfstring{$\beta = 0$}{beta = 0}, case --- density dependent fluctuation procession}
\begin{figure}[!htp]
    \centering
    \includegraphics[width=\textwidth]{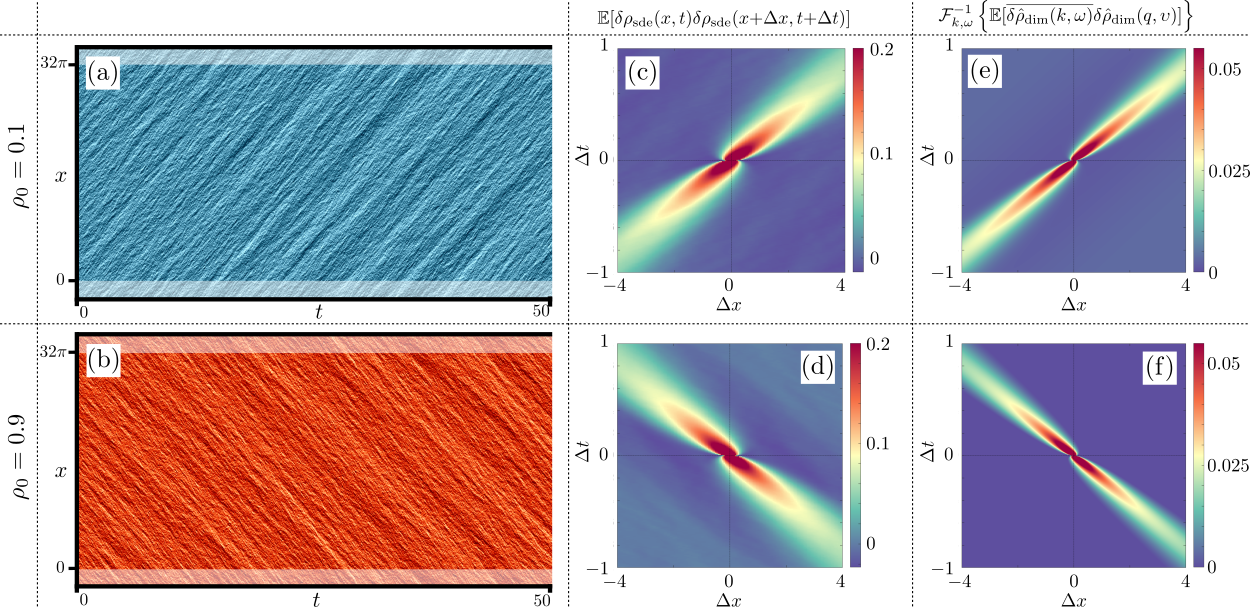} 
    \caption{\label{beta0fig} {\bf Fluctuation behaviour around homogeneous fixed point states in the $\beta\to 0$ limit.} Here we show density/$\rho_0$ dependent behaviour in the fluctuations, illustrated for $\rho_0=0.1$ [panels (a), (c), and (e), top row] and $\rho_0=0.9$ [panels (b), (d), and (f), bottom row]. Panels (a) and (b) are density kymographs illustrating the travelling wave nature of the fluctuations with direction/velocity determined by density/$\rho_0$. Panels (c) and (d) are empirical correlation functions of the coarse grained density from simulation of the SDEs, denoted $\rho_{\rm sde}$. These can be compared with panels (e) and (f) which illustrate numerical inversion of the analytical power spectra (re-dimensionalised, as denoted by $\hat{\rho}_{\rm dim}$) for the linearised system. Full simulation details are described in a main text appendix. Analytics predict a fluctuation velocity of $\nu(1-2\rho_0)$, see Sec.~\ref{sec:dimensions} (where here, $\nu=6$), giving velocities $\pm 4.8$ in excellent agreement with the above plots. Both simulations utilise $L=32\pi$ and $N=80,000$.}
 \end{figure}

Here we illustrate the behaviour of fluctuations in the $\beta\to 0$ case. We re-iterate that in this case we only have one field variable ($\rho=\rho^+$) and the power spectra density is given by
\begin{align}
    \mathbb{E}[\overline{\delta\hat{\rho}(k,\omega)}\delta\hat{\rho}(q,\upsilon)]&=\delta_{k,q}\delta(\omega-\upsilon)\frac{2\varepsilon k^2}{k^2 \text{Pe}^2 \left(1-2 \rho _0\right)^2+k^4+2 k \text{Pe} \left(2 \rho _0-1\right) \omega +\omega ^2}.
\end{align}
Behaviour of the system, at different densities, is shown in Fig.~\ref{beta0fig}. We observe fluctuations that travel in a particular direction. This breaks time reversal symmetry as can be observed in the characteristically asymmetric correlation functions. In other words, if the kymographs are flipped in time, we can easily distinguish it from the original.

\subsection{Totally symmetric, \texorpdfstring{$\beta = 1$}{beta = 1}, case --- density dependent species coupling}

\label{b1spectra}
In the opposite, totally symmetric, limit corresponding to conventional run-and-tumble behaviour ($\beta=1$) the general expressions presented in the appendix to the main text simplify to
\begin{subequations}
\begin{align}
\mathbb{E}[\overline{\delta\hat{\rho}(k,\omega)}\delta\hat{\rho}(q,\upsilon)]&=\delta_{k,q}\delta(\omega-\upsilon)\frac{2\varepsilon k^2 \left(\left(k^2+2\right) \left(k^2+\text{Pe}^2 \left(\rho _0-1\right)^2+2\right)+\omega
   ^2\right)}{\left(k^2 \left(\text{Pe}^2 \left(\rho _0-1\right) \left(2 \rho _0-1\right)+2\right)+k^4-\omega
   ^2\right)^2+4 \left(k^2+1\right)^2 \omega ^2},\\
   \mathbb{E}[\overline{\delta{\hat{\rho}^{\pm}}(k,\omega)}\delta\hat{\rho}^{\pm}(q,\upsilon)]&=\delta_{k,q}\delta(\omega-\upsilon)\nonumber\\
   &\quad\times\frac{\varepsilon \left(2 \left(k^2{+}1\right) \left(k^2 \left(\text{Pe}^2+2\right){+}k^4{\pm}2 k \text{Pe} \omega {+}\omega ^2\right){+}k \text{Pe} \rho _0 \left(k \left(5 k^2{+}2\right) \text{Pe} \rho _0{-}2 \left(3 k^2{+}2\right) (k
   \text{Pe}{\pm}\omega )\right)\right)}{\left(k^2 \left(\text{Pe}^2 \left(\rho _0-1\right) \left(2 \rho _0-1\right)+2\right)+k^4-\omega ^2\right){}^2+4 \left(k^2+1\right)^2 \omega ^2},\\
   \mathbb{E}[\overline{\delta{\hat{\rho}^{+}}(k,\omega)}\delta\hat{\rho}^{-}(q,\upsilon)]&=\delta_{k,q}\delta(\omega-\upsilon)\nonumber\\
   &\quad\times\frac{\varepsilon  \left(2 k^2 \left(k^2+\text{Pe}^2+2\right)+k^2 \text{Pe} \rho _0 \left(\left(2-3 k^2\right) \text{Pe} \rho _0+2 \left(k^2-2\right) \text{Pe}-2 i k \left(k^2+2\right)\right)-2 \omega ^2\right)}{\left(k^2 \left(\text{Pe}^2 \left(\rho _0-1\right) \left(2 \rho _0-1\right)+2\right)+k^4-\omega ^2\right){}^2+4 \left(k^2+1\right)^2 \omega ^2}.
\end{align}
\end{subequations}

\begin{figure}[!htp]
   \centering
   \includegraphics[width=\textwidth]{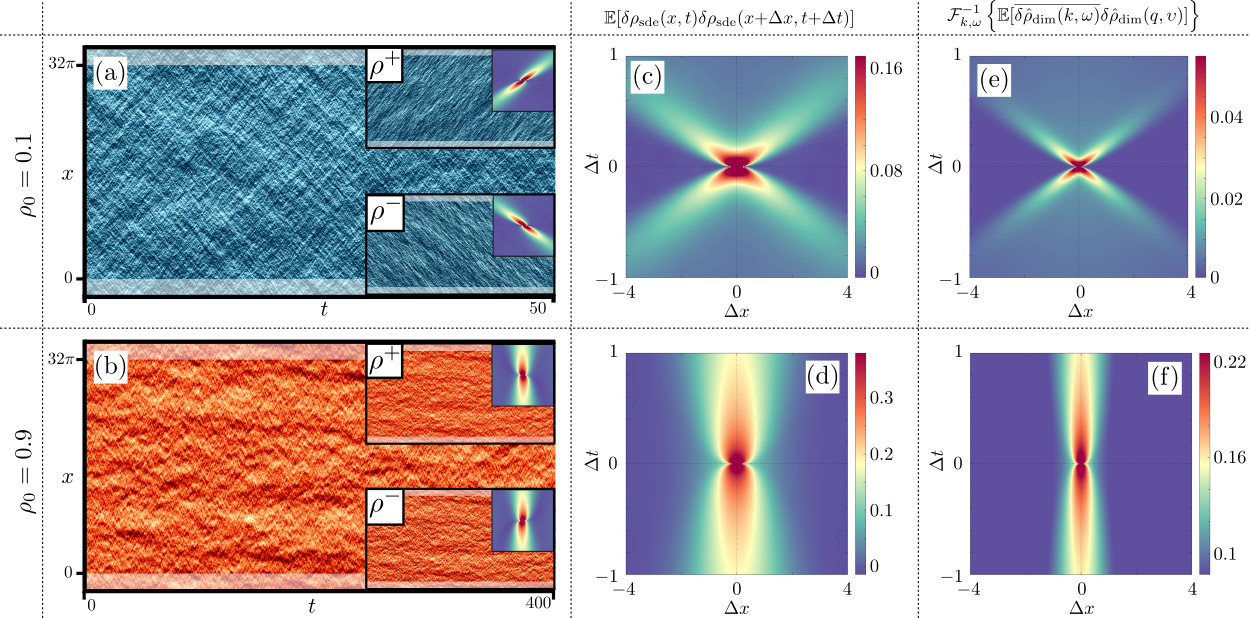}
   \caption{\label{beta1fig} {\bf Fluctuation behaviour around homogeneous fixed point states in the symmetric $\beta=1$ case.} Here we show density/$\rho_0$ dependent behaviour in the fluctuations, illustrated for $\rho_0=0.1$ [panels (a), (c), and (e), top row] and $\rho_0=0.9$ [panels (b), (d), and (f), bottom row]. Panels (a) and (b) are total density kymographs, with inset single species kymographs and empirical correlation functions. Panels (c) and (d) are empirical correlation functions of the coarse grained density from simulation of the SDEs, denoted $\rho_{\rm sde}$. These can be compared with panels (e) and (f) which illustrate numerical inversion of the analytical power spectra (re-dimensionalised, as denoted by $\hat{\rho}_{\rm dim}$) for the linearised system. The $\rho_0=0.1$ case exhibits characteristic species decoupling, whilst $\rho_0=0.9$ exhibits, coherent (and stationary) fluctuation structure. Control parameters used are $\text{Pe}=4.0$ for $\rho_0=0.9$ and $\text{Pe}=7.0$ for $\rho_0=0.1$. Both simulations utilise $L=32\pi$ and $N=80,000$.}
\end{figure}

\begin{figure}[!htp]
   \centering
   \includegraphics[width=\textwidth]{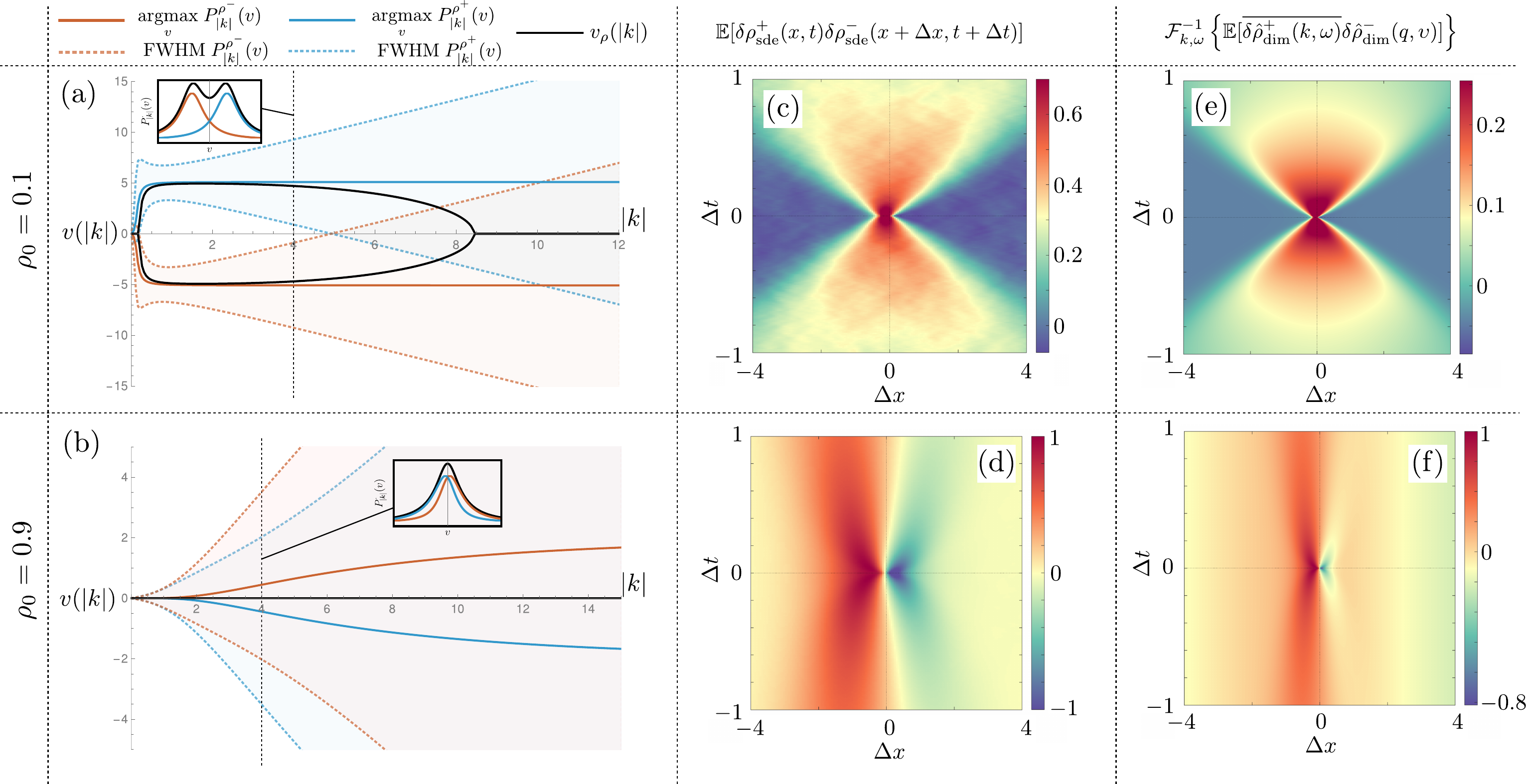}
   \caption{\label{beta1fig2} {\bf Density dependent coupling and decoupling around homogeneous fixed point states in the symmetric $\beta=1$ case.} Here we show density/$\rho_0$ dependent correlation between the species, illustrated for $\rho_0=0.1$ [panels (a), (c), and (e), top row] and $\rho_0=0.9$ [panels (b), (d), and (f), bottom row]. Panels (a) and (b) illustrate the behaviour of the power spectra over phase velocities at given $|k|$, with solid and dashed blue lines illustrating the phase velocity with peak power, and the FWHM of the spectra, respectively, for the $+$ve species, and red lines the same for the $-$ve species. Black lines then illustrate the typical phase velocities associated with local maxima in the total density power spectra. Inset are the full power spectra at an illustrative value of $|k|=4$. The statistical independence/coupling is reflected in the species cross-correlation functions on the right-hand side. Panels (c) and (d) are empirical correlation functions from simulation of the SDEs ($\rho_{\rm sde}$), whilst panels (e) and (f) show numerical inversion of the re-dimensionalised analytical spectra. The $\rho_0=0.1$ case exhibits broad species independence, whilst $\rho_0=0.9$ exhibits, coherent (and stationary) fluctuation structure between the species. Parameters used are $\text{Pe}=4.0$ for $\rho_0=0.9$ and $\text{Pe}=7.0$ for $\rho_0=0.1$.}
\end{figure}

Behaviour of the system, at different densities, is shown in Figs.~\ref{beta1fig} and \ref{beta1fig2} where we plot density kymographs and corroborate the analytical power spectra. Note, in contrast to the $\beta=0$ case above, and general $\beta<1$ case in the main text, fluctuations are symmetric in both space and time \emph{i.e.} they appear time reversal symmetric as can be explicitly seen in the correlation functions. In other words, if the kymographs are flipped in time, we could not distinguish it from the original.

\par
In this symmetric, $\beta=1$, case it is possible to calculate the value of $v_\rho(|k|)$ exactly. Explicitly, we find
\begin{align}
v_{\rho}(|k|)&=\pm\Big[|k|^{-1}\text{Pe} \sqrt{ \left(\rho _0-1\right) \left(\text{Pe}^2 \left(\rho _0-1\right) \left(|k|^2 \left(3 \rho _0-2\right)+2 \left(\rho _0-1\right)\right){}^2+4 \left(|k|^2+1\right) \left(|k|^2+2\right) \left(|k|^2 \left(2 \rho _0-1\right)+\rho
   _0-1\right)\right)}\nonumber\\
   &\qquad-|k|^{-1}\left(|k|^2+2\right) \left(|k|^2+\text{Pe}^2 \left(\rho _0-1\right){}^2+2\right)\Big]^{\frac{1}{2}}, 
\end{align}
with $v_\rho(|k|)=0$ when no such real roots exist, revealing a well-defined envelope of $|k|$ where these symmetric travelling fluctuations can be observed.
\par
We can then use this result to determine the phase space dependence of this species decoupling on $\rho_0$ and $\text{Pe}$. Solving for the existence of the above root in $\text{Pe}$ reveals a ($|k|$-dependent) minimum P\'eclet number to observe species separation, $\text{Pe}_d$, given by
\begin{align}
&(\text{Pe}_d)^2=\nonumber\\
&-\frac{ |k|^4 (3 \rho_0{-}1){+}2 \left(1{+}\left(|k|^2{-}1\right)\rho_0\right){+}\sqrt{\left(17 |k|^8+52 |k|^6+56 |k|^4+24 |k|^2+4\right) \rho _0^2+4 \left(|k|^2+1\right)^4-8 \left(2 |k|^2+1\right) \left(|k|^2+1\right)^3 \rho _0}}{|k|^2 \left(|k|^2+2\right)^{-1}\left(\rho _0-1\right) \left(2 \rho _0-1\right) \left(|k|^2 \left(4 \rho _0-3\right)+4 \left(\rho _0-1\right)\right)}.
\end{align}
In short, the required $\text{Pe}$ to observe the two species' fluctuations diverges as $|k|\to 0$ such that is not observed on thermodynamic length scales, and, in linearly stable regions where this analysis is valid, only for values of $\rho_0<1/2$.

\clearpage
%

\section{Solutions and behaviour of the \texorpdfstring{$N\to\infty$}{N to infinity} deterministic limit of bQSAPs}

%
\label{binodalsol}
Here we seek to understand the limiting behaviour of the full deterministic system in $\{\rho,\psi\}$ [Eq.~(3) of the main text], as well as the approach to such solutions in some specific cases. 
The observed behaviour is far richer than in conventional phase separation, requiring separate discussion of numerous regions of the phase space. 
We will broadly classify these regions based on the location of the coexistence densities should they exist (which correspond to the unequal tangent construction) relative to the position of the spinodal, stability line.

\subsection{Two stable coexistence densities (Large \texorpdfstring{$\text{Pe}$}{Pe}) --- macro-phase separation}

%
\label{macro_phase_det}
We first consider the nature of these domain solutions in a region of phase space where we expect stable behaviour akin to that in conventional, stationary, phase separation, where the binodal lines lie outside the spinodals, by utilising a relatively large value of the P\'eclet number. 
%

\subsubsection{Limiting state}

\begin{figure}[!htp]
    \centering
    \includegraphics[width=0.99\textwidth]{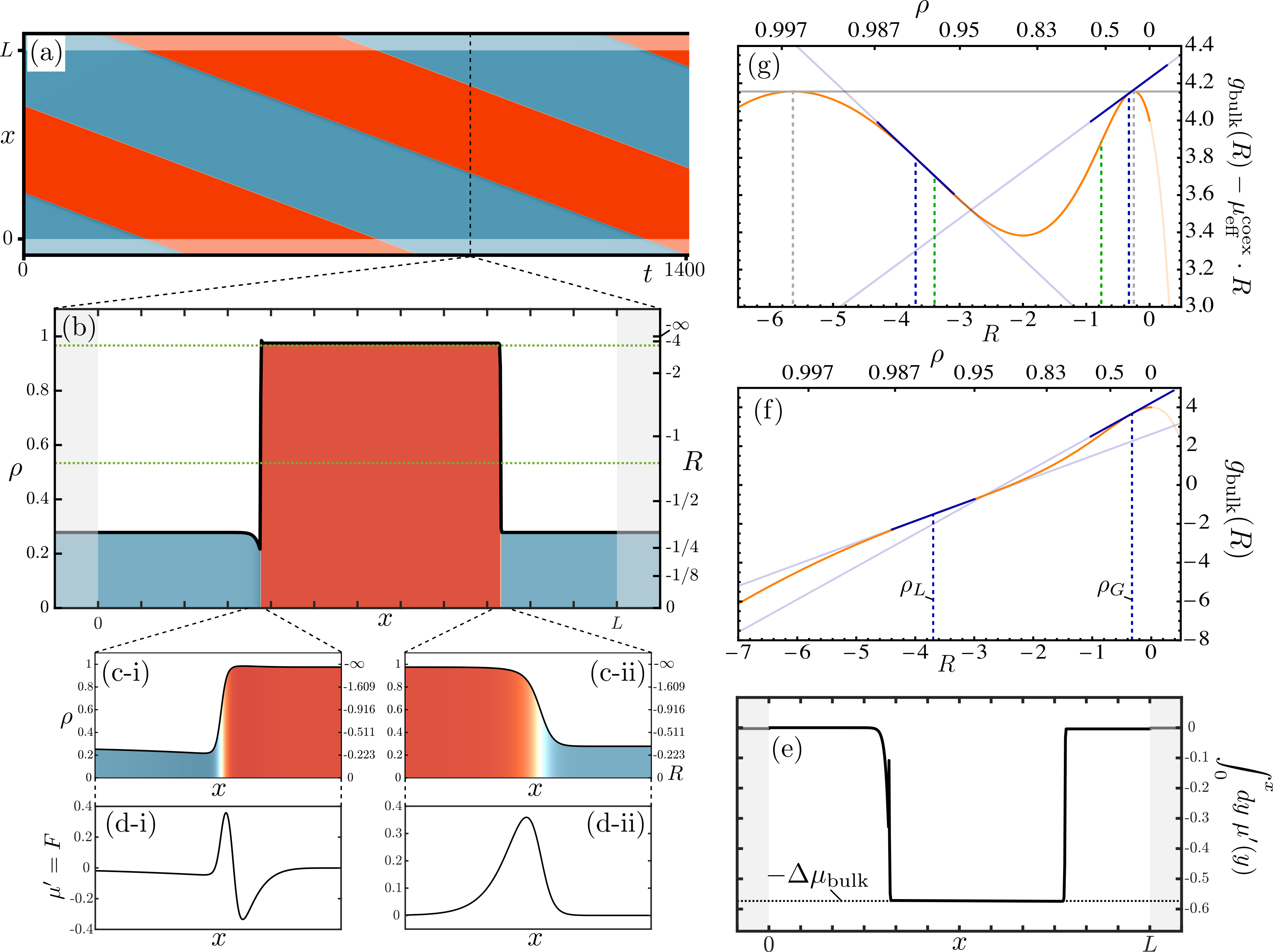} 
    \caption{\label{macro_det} {\bf Deterministic solutions in the $N\to\infty$ limit --- high $\text{Pe}$:} Sub-panel (a) is a density kymograph of the system in a phase separated state. The phases demonstrably translate in the negative direction, whilst maintaining phase boundary profiles. A slice of this kymograph is shown below it in panel (b) revealing the asymmetry between the leading and following domain boundaries, and with spinodal line densities marked as dashed green lines. Below the global density plot are detailed figures of the density profiles of the two domain boundaries in panels (c-i) and (c-ii) alongside the empirically determined $\partial_x\mu=F$ in panels (d-i) and (d-ii). Panel (e) shows the integral of this quantity over the whole domain, demonstrating the equal and opposite net chemical potential changes over the boundaries. Panel (f) illustrates the implied unequal tangent construction on $g_{\rm bulk}$ using the two empirically determined phase densities, $\rho_G\simeq 0.2781$ and $\rho_L\simeq 0.9752$. Panel (g) shows the unequal tangent construction with an additional linear factor with coefficient equal to the coexistence chemical potential under a common tangent construction, for clarity. Coexistence densities corresponding to the unequal tangents are shown in blue, lying outside of the spinodal lines (green), but within the hypothetical coexistence densities corresponding to a common tangent construction (gray). Example shown uses $\beta=1/2$, $\text{Pe}=9$, $\rho_0=0.6$, $L=256\pi$.}
 \end{figure}

%
\label{limit_macro_det}
We first investigate the behaviour for a system with $\text{Pe}=9$, $\beta=1/2$ and $\rho_0=0.6$, using the full underlying system  in $\{\rho,\psi\}$, in the $N\to\infty$ limit, through numerical simulation. 
The resulting behaviour, alongside pertinent theoretical aspects, are shown in Fig.~\ref{macro_det}. 
Large domains are stable, and process in the direction counter to particle motion. 
Further, as predicted, leading and trailing domain boundaries are distinct --- the liquid to gas boundary is monotonic, whereas the gas to liquid boundary is non-monotonic and overshoots beyond $[\rho_G,\rho_L]$. 
\par
Here we illustrate the unequal tangent construction on $g_{\rm bulk}(R)$ defined such that $g'_{\rm bulk}(R)=f'_{\rm bulk}(\phi)=\mu_{\rm bulk}(\phi)$, with $R=\ln(1-\phi)$, \emph{viz.}
\begin{align}
g_{\rm bulk}(R)&=-\frac{(\beta +1)^2 R^2}{4 \beta  \text{Pe}}-\frac{2 \beta  \text{Pe}
\left(e^R-1\right)^2}{(\beta +1)^2},
\end{align}
such that it may be easily contrasted with the common tangent construction that would normally be associated with active model B. 
\par
As such, we see that the coexistence densities reflect an unequal tangent construction on $g_{\rm bulk}$, with the coexistence densities lying \emph{within} the densities which correspond to an equal tangent construction on $g_{\rm bulk}$. For this value of $\text{Pe}$ the coexistence densities lie \emph{outside} of the spinodals, as in normal phase separation, allowing for stable domain sizes on the scale of the system, even in the thermodynamic limit $L\to\infty$. 
 \par
\begin{figure}[!htp]
    \centering
    \includegraphics[width=0.99\textwidth]{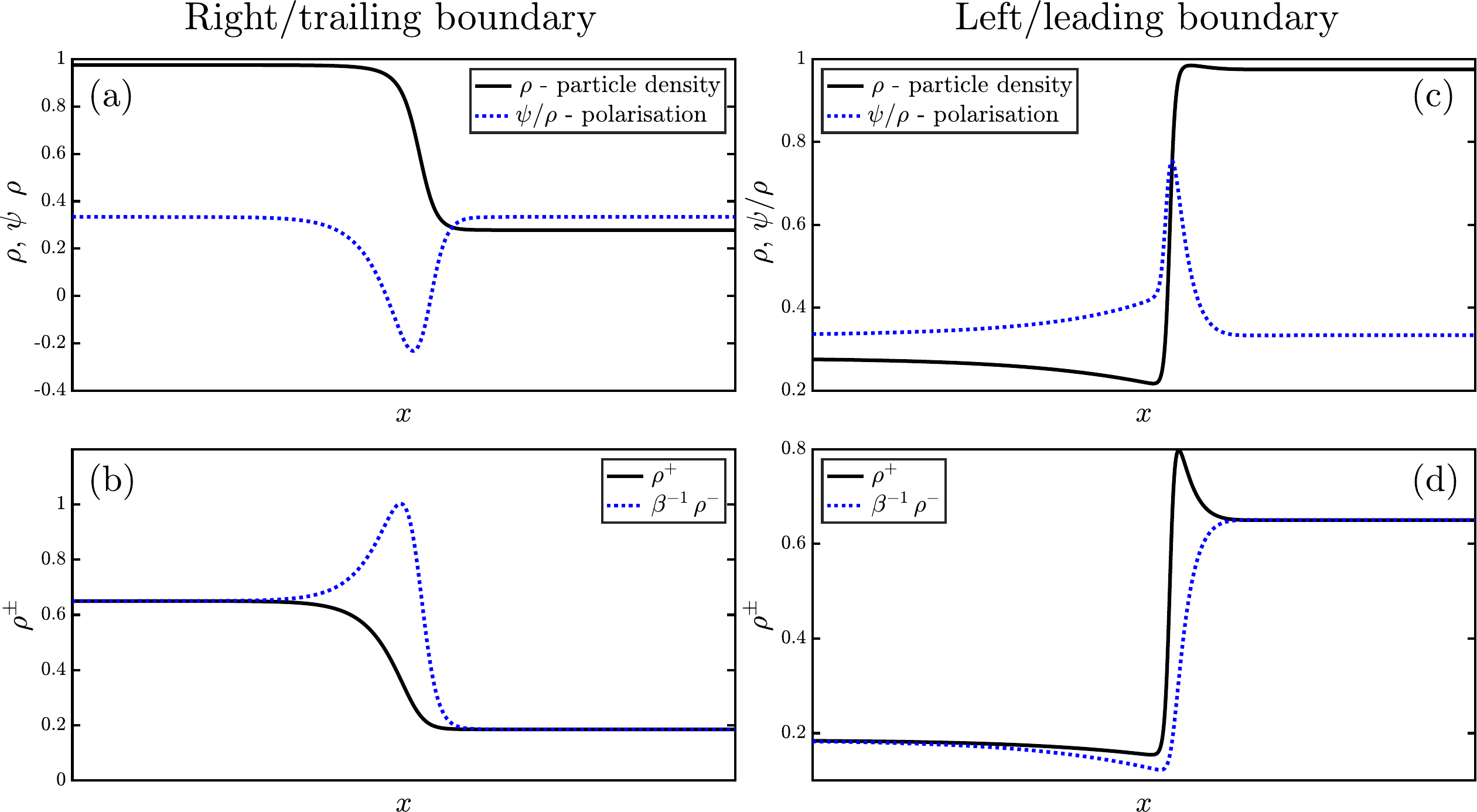} 
    \caption{\label{macro_det_bound} {\bf Constituent fields for deterministic left and right boundaries in the $N\to\infty$ limit --- high $\text{Pe}$:} 
    Sub-panel (a) contains density and polarisation (polarisation density on density) profiles for the trailing, right-most, boundary from a liquid to gaseous phase. This boundary is monotonic in density and is characterised by a decrease in polarity  at the boundary. Sub-panel (b) shows density profiles for the same boundary, but for individual left and right moving species. Sub-panel (c) contains density and polarisation profiles for the leading, left-most, boundary from a gaseous to liquid phase. This boundary is non-monotonic in density and is characterised by a very slow increase in polarity from within the gaseous bulk until it sharply peaks at the boundary. Sub-panel (d) show density profiles for the same boundary, but for individual left and right moving species. Example shown uses $\beta=1/2$, $\text{Pe}=9$, $\rho_0=0.6$, $L=256\pi$.}
 \end{figure}

 In Fig.~\ref{macro_det_bound} we illustrate the nature of the asymmetric phase boundaries in terms of both $\{\rho,\psi\}$ and the individual species densities $\{\rho^+,\rho^-\}$. We observe that the monotonic liquid to gaseous boundary is associated with a decrease in polarisation due to a relative build up of left moving particles. In contrast, the non-monotonic gaseous to liquid boundary is associated with an increase in polarisation due to a relative build up of right moving particles. This effect, however, is augmented on the leading boundary since the right moving particles are more numerous and flowing towards an approaching boundary, whilst it is diminished on the trailing boundary because left moving particles are less numerous and flowing towards a receding boundary.
\par


\subsubsection{Transient behaviour - the trailing boundary sets coexistence}

%
\label{macrotransient}
\begin{figure}[!htp]
    \centering
    \includegraphics[width=0.99\textwidth]{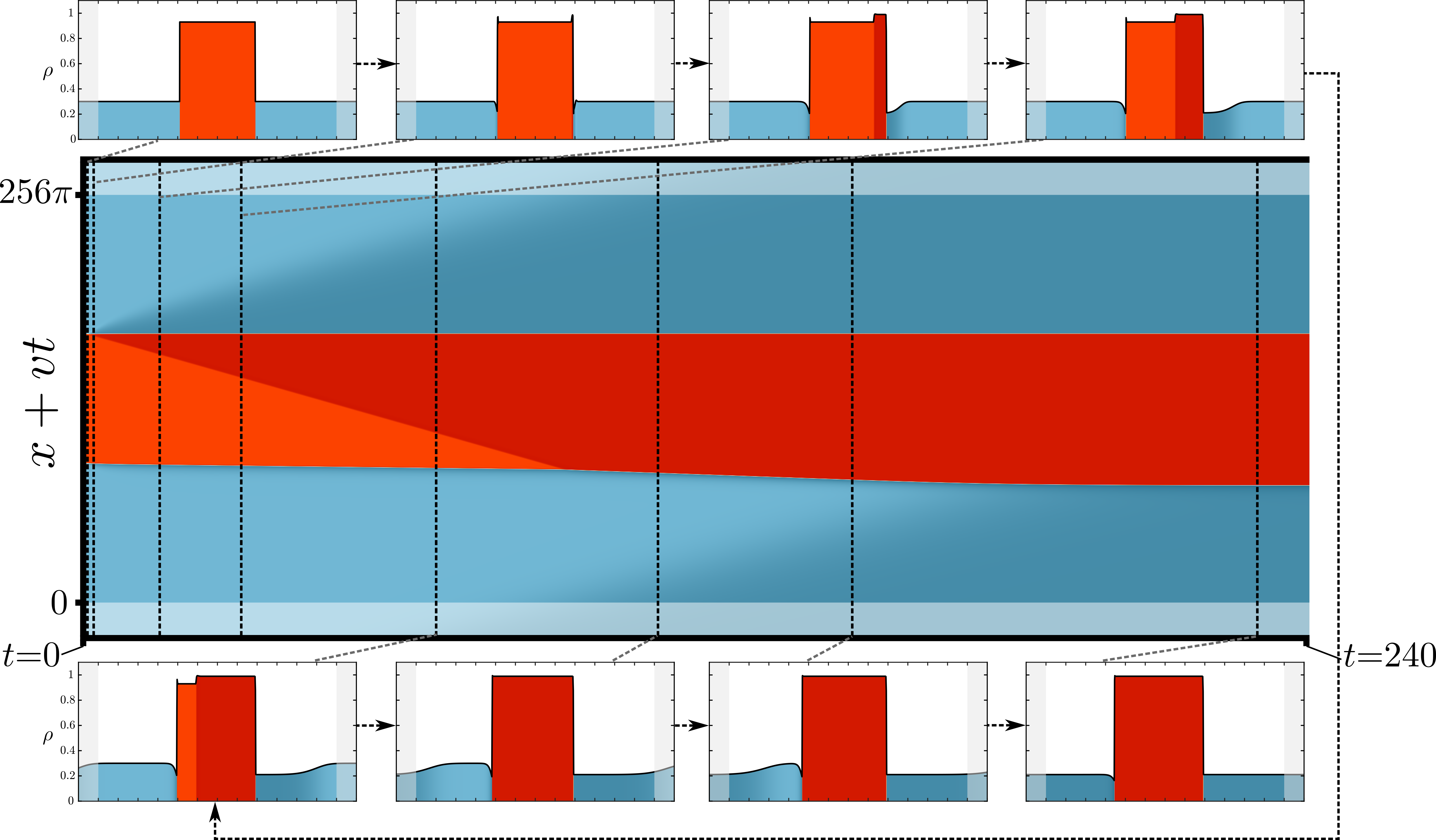} 
    \caption{\label{macro_det_quench}{\bf Density kymographs showing asymmetric boundary stability and phase evolution in the $N\to\infty$ limit --- high $\text{Pe}$:} Centre is a density kymograph of a high $\text{Pe}$ behaviour, Galilean shifted so that the liquid phase appears centred. The coexistence densities for the parameters used are $\rho_G\simeq0.21$, $\rho_L\simeq0.99$, and the system is initially prepared in a two-domain piece-wise constant state at densities $\rho\in\{0.3,0.93\}$. This causes the system to evolve towards a phase separated state at the coexistence densities. The density profile at key points in the evolution are explicitly plot in the smaller peripheral plots, evolving forward in time from top left to top right, and then bottom left to bottom right. 
    Example shown uses $\beta=1/2$, $\text{Pe}=11$, $\rho_0=0.5$, $L=256\pi$.}
\end{figure}
Here we present characteristic transient behaviour of the model, beyond the coarsening mechanism illustrated in the main text, in Fig.~\ref{macro_det_quench}. 
Here the system is initially prepared in a state formed of two bulks at densities $\rho\in\{0.3,0.93\}$, whilst the coexistence densities for the given parameters are approximately equal to $\{\rho_G,\rho_L\}\simeq\{0.21,0.99\}$. At the start of the simulation the boundaries immediately begin to reconfigure in an approximately symmetric manner. 
However, the left, leading, boundary rapidly stops evolving, forming a characteristic non-monotonic profile on the initial (not coexistence) bulk densities--- \emph{i.e.}, the leading boundary can stabilise around densities which are not the coexistence densities. 
The right, trailing, boundary on the other hand continues to reconfigure around the coexistence densities $\rho_G$ and $\rho_L$, with these values of the bulk propagating out to the rest of the system until the whole profile appears as in Fig.~\ref{macro_det_bound} (with features in the liquid phase moving in the negative direction and features in the gaseous phase moving in the positive direction, as per the expected behaviour of small fluctuations \emph{cf.} the main text). 
\par
The overall picture is one where the left, leading, boundary is a highly resilient structure, seemingly able to adapt its characteristic `overshoot' peak and void in order to stabilise the boundary and match a range of chemical potential differences. 
In contrast, the right, trailing, boundary cannot adapt in this way and sets the coexistence densities which then propagate out to the whole system.


 \subsubsection{Phase behaviour and stability}
 
\begin{figure}[!htp]
    \centering
    \includegraphics[width=0.99\textwidth]{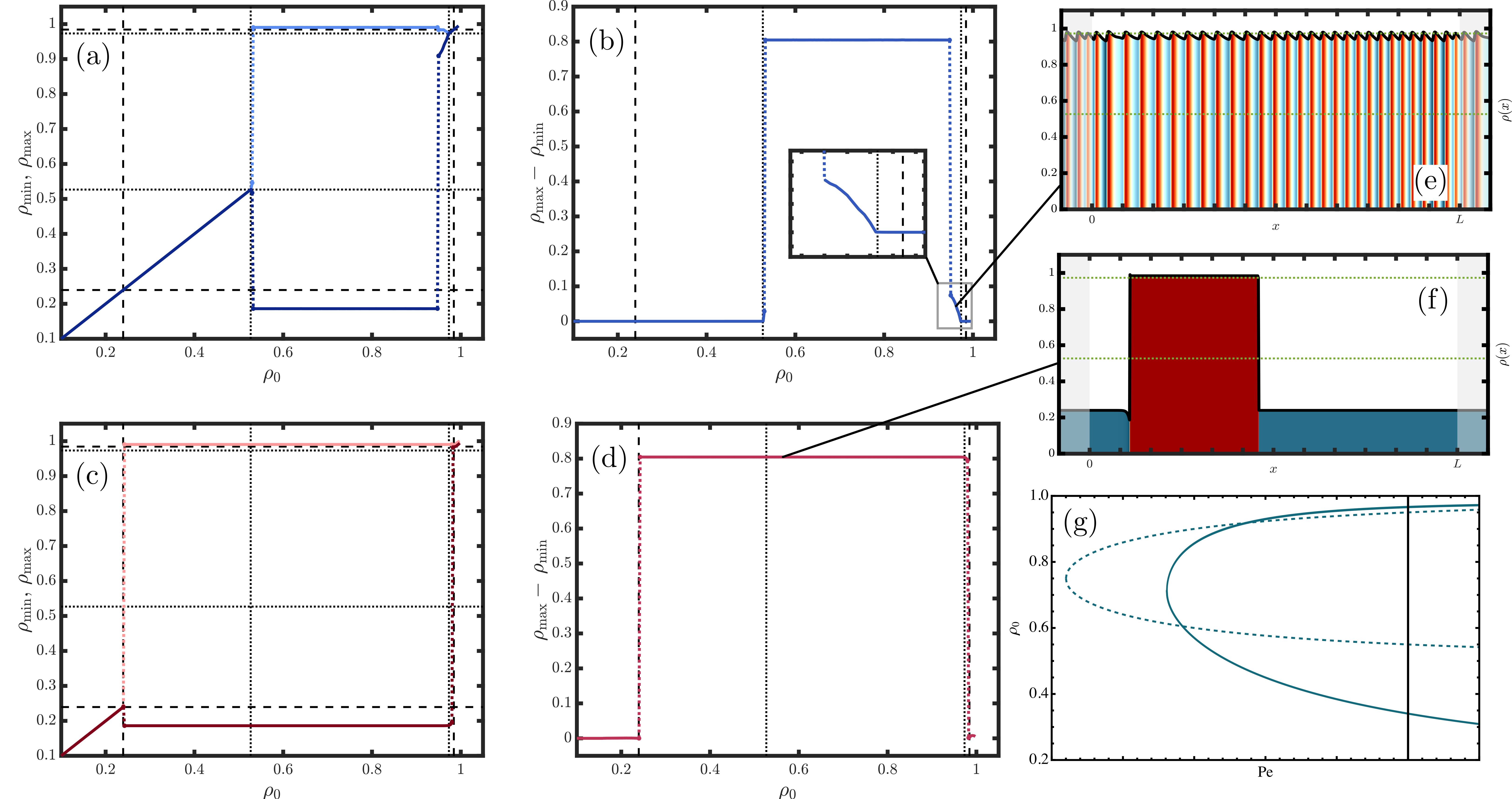}  
    \caption{\label{structure_highPe} {\bf Phase structure for high $\text{Pe}$. Limiting states resulting from distinct initial conditions over a range of $\rho_0$ with all other control parameters fixed ($\text{Pe}=10$, $\beta=1/2$):} In sub-panels (a) and (b) we characterise the limiting solutions resulting from an initial condition consisting of a homogeneous state with a small Gaussian perturbation. In sub-panels (c) and (d) we characterise limiting solutions resulting from an initial condition consisting of an artificially phase-separated piece-wise constant state ($\rho(0)\in\{0.1,0.9999\}$). All such plots characterise solutions over a range of $\rho_0$ at a value of $\text{Pe}$ where binodal/coexistence solutions lie outside of the spinodals/unstable region as indicated in sub-panel (g). Dotted black lines in sub-panels (a)-(d) indicate spinodal densities whilst dashed black lines indicate coexistence densities. Dotted green lines in sub-panels (e) and (f) indicate the spinodals. 
    Sub-panels (a) and (c) show the maximum and minimum values of the $\rho(x)$ in the limiting state, whilst sub-panels (b) and (d) show the difference between the maximum and minimum values of $\rho(x)$ in the limiting state. Starting from a perturbative state, outside of the spinodals the system converges on a homogeneous solution, whilst inside it tends to other limiting states, consisting of a macroscopically phase separated state (\emph{e.g.} sub-panel (f)) and an oscillatory state (sub-panel (e)) immediately inside the upper spinodal. Starting from a piece-wise constant state the system reaches a phase separated state (sub-panel (f)) within the binodal/coexistence lines, and tends towards a homogeneous state outside of them.  Parameters and domain used are $\beta=0.5$, $\text{Pe}=10$, $L=256\pi$.}
 \end{figure}

 %
 \label{macro_phase}
In this high $\text{Pe}$ regime, we have coexistence densities, or binodals, lying outside the region of linear instability, bounded by the spinodals. 
 As such, much of the expected phase behaviour in this regime thus reflects that of a conventional phase structure. 
 We illustrate this behaviour in Fig.~\ref{structure_highPe} wherein we plot results from numerically solving the full model in $\{\rho,\psi\}$ over a range of the parameter $\rho_0$, at high $\text{Pe}$, for two different initial conditions. 
 The two initial conditions we consider are those of a small $\delta\rho$ Gaussian perturbation to an otherwise homogeneous state, and of an artificially piece-wise constant phase separated state such that $\rho\in\{0.1,0.9999\}$. 
 The former is designed to probe the nature of spinodal decomposition, whilst the latter is designed to explore the more general bounds of coexistence and the nature of phase coexistence, without any concern for metastability.
 \par
 The illustration of these results in Fig.~\ref{structure_highPe} is performed through the plotting of the maximal and minimal values of $\lim_{t\to\infty}\rho(x,t)$, alongside their difference, in order to indicate convergence on $\rho_0$ or those solutions associated with the phase separated coexistence densities.  
 We note that because the leading boundary is neither monotonic nor confined to $[\rho_G,\rho_L]$, the maximal and minimal values do \emph{not} correspond to the values of $\rho_G$ and $\rho_L$ as in conventional phase separation, but instead lie a characteristic amount below and above these densities, respectively.
 \par
 The solutions arising from the artificial initial phase separated state follow exactly as expected from the behaviour of conventional phase separation and are shown in panels (c) and (d) (red lines). 
 For values $\rho_0\notin[\rho_G,\rho_L]$, such that we consider regions of the phase diagram outside of the coexistence densities, the system converges towards a one phase, constant $\rho=\rho_0$, mixture state since phase separation is not compatible with mass conservation. 
 In contrast, for values $\rho_0\in [\rho_G,\rho_L]$, such that we consider regions of the phase space within the coexistence densities, the system converges on a two phase, phase separated, solution. 

 \par
 The solutions arising from the perturbative initial condition (panels (a) and (b), blue lines) also align very closely with that expected from conventional phase separation. 
 Writing the low and high density spinodal densities $\rho_{s_1}$ and $\rho_{s_2}$, we observe that for values of $\rho_0\notin [\rho_{s_1},\rho_{s_2}]$, such that we consider regions outside of the spinodal densities, the system converges towards a one phase, constant $\rho=\rho_0$, mixture state. 
 This occurs even when the value of $\rho_0$ lies within the coexistence densities ($\rho_0\in[\rho_G,\rho_L]$)  characteristic of the one phase state being meta-stable as in conventional phase separation. 
 For densities within the spinodal densities, $\rho_0\in[\rho_{s_1},\rho_{s_2}]$, the perturbation is not stable, by definition, and so the system converges on a distinct solution. 
 For the vast majority of this region the system ultimately converges on the phase separated solution, however, near the spinodal densities themselves, the solution that they converge on is not the conventional phase separated state formed of bulk densities at $\rho_G$ and $\rho_L$, but rather a third kind of state, consisting of oscillatory density profiles. 
 The existence of this third kind of state is very important to the wider phase structure at lower values of $\text{Pe}$, and its existence can broadly be inferred from the imaginary components of the eigenvalues of the linearised dynamical matrix that appears in the main text.  
 Since this state is converged upon near the spinodal densities, where the growth of unstable modes whilst uncontrolled is relatively weak and restricted to longer length-scales, the behaviour hints at a three-tiered metastability structure: Between the spinodal and binodal densities homogeneous states and phase separated states are both locally stable to perturbations, but with the latter globally preferable. 
 Similarly, on the interior of the spinodals two such states are locally stable, an oscillatory state and the phase separated state, with the former meta-stable with respect to the latter which, again, appears to be globally preferable.
%


\subsection{Zero stable coexistence densities (Low \texorpdfstring{$\text{Pe}$}{Pe}) --- micro-phase separation}

%
\label{micro_para}
Next, we consider behaviour at the other extreme ---  that where $\text{Pe}$ is low enough that there exist no stable coexistence densities (but with $\rho_0$ inside the spinodals). 
We note that this characterisation encompasses cases where there are no coexistence densities at all and cases where two coexistence densities exist, but both are unstable. 
We consider this equivalent for our purposes here as we observe no distinction in their limiting behaviour.
\par
In these regimes both conventional solutions associated with phase separation, namely a homogeneous mixture state and a liquid/gas coexistence state, are unavailable to the system due to them being either unstable or non-existent. 
Consequently, we must expect the system to arrive at alternative states. We have seen evidence of such states above, whereby we observed that on the interior of the spinodal region, near the spinodal line, an initial state consisting of a small perturbation around a homogeneous state converged on oscillatory states distinct from macroscopic phase separation. 
These solutions can be contextualised by identifying that in crossing the spinodal line from an initially stable homogeneous state, (emulated by our peturbative Gaussian initial condition), the system undergoes a Hopf bifurcation owing to the presence of a non-zero imaginary component in the maximum eigenvalue of its linearised dynamical matrix.
\par
Consequently, with the spinodal line exposed in this manner, without an associated binodal solution, it acts as a critical line separating two phases by means of a continuous phase transition between homogeneous and oscillatory states in the manner of a Hopf bifurcation. 
%

\subsubsection{Limiting and transient behaviour}

\begin{figure}[!htp]
    \centering
    \includegraphics[width=0.99\textwidth]{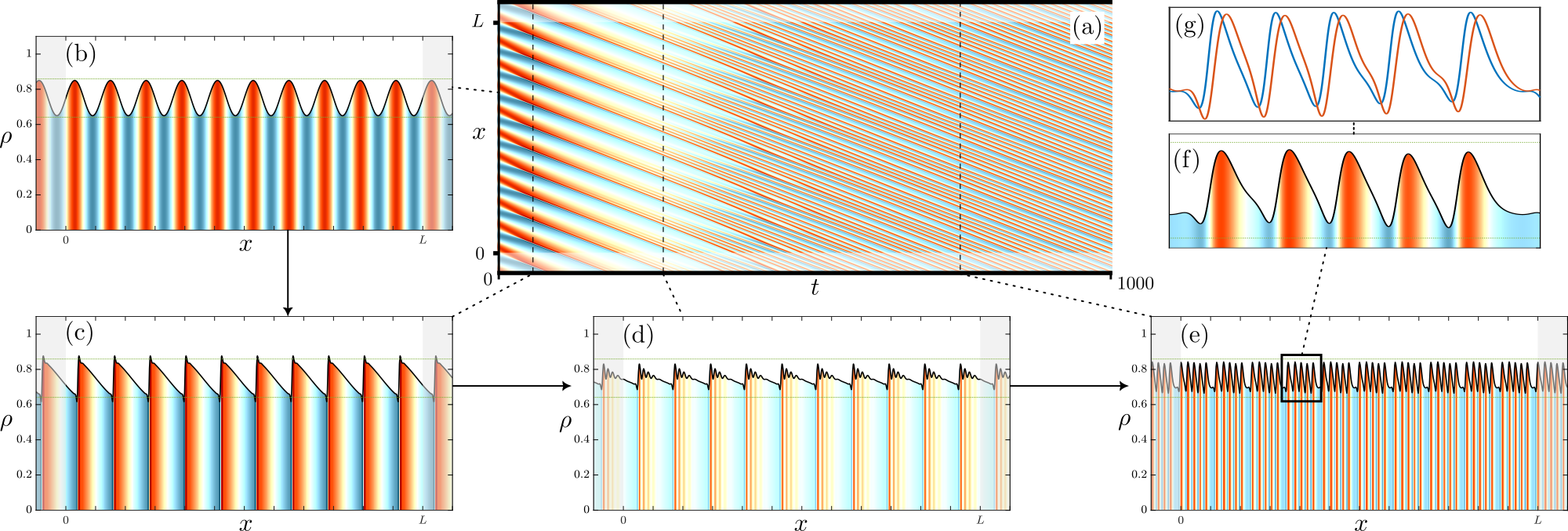} 
    \caption{\label{micro_fig} {\bf System behaviour in the unstable region without binodal solutions:} Behaviour of the system starting from a sinusoidal initial condition evolving towards a processive micro-phase separated solution is shown by means of density profiles and kymographs. Sub-panel (a) shows a density kymograph of the system evolution. Sub-panel (b) illustrates the system's initial condition. Initially the system starts to converge on a shock like solution (sub-panel (c)), but this soon destabilises (sub-panel (d)), before settling into a broadly oscillatory processive state formed of short sharp peaks (sub-panel (e)). A close up of these density features is shown in sub-panel (f). Finally, sub-panel (g) shows the behaviour of the two component species --- the blue line shows the density profile of the positively oriented species, whilst the red line shows (twice) the density profile of the negatively oriented species. 
    In all cases dashed green lines indicate the location of the spinodal densities. Example shown uses $\beta=1/2$, $\text{Pe}=5$, $\rho_0=0.75$, $L=256\pi$.}
 \end{figure}

%
An example of behaviour characteristic of this region is shown in Fig.~\ref{micro_fig} wherein the system, initially prepared in a sinusoidal state, evolves towards a state of travelling micro-phase separation, characterised by small sharp peaks which process at a single, well-defined, velocity. 
Interestingly, the system starts to evolve towards a shock solution, hinting at the similarities of the system with the Burgers' equation, but these structures soon destabilise into the micro-phase separation which characterises this region. Also shown is the behaviour of the distinct positive and negatively oriented species in these micro-phase solutions. 
Here we see that the features in the density of the positively oriented species distinctly lead the features found in the negatively oriented density profiles. 
Both such profiles process in the negative direction meaning that the positively oriented profiles are moving counter to their particles' direction, whilst the negatively oriented profiles are moving with their direction. 
%


\subsubsection{Phase behaviour}

\begin{figure}[!htp]
    \centering
    \includegraphics[width=0.99\textwidth]{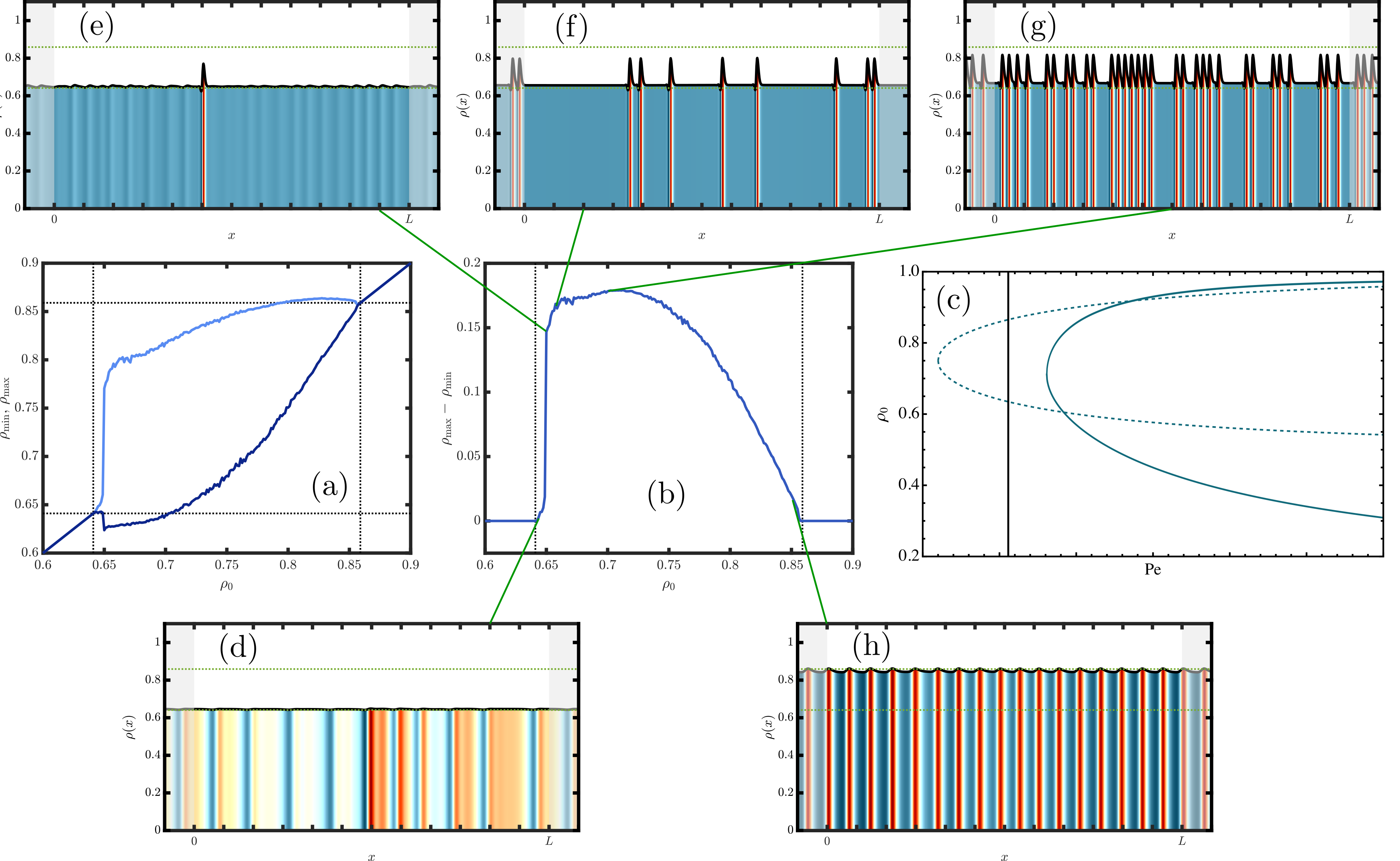}  
    \caption{\label{structure_lowPe} {\bf Phase structure for low $\text{Pe}$. Limiting states resulting from a perturbative initial condition over a range of $\rho_0$ with all other control parameters fixed ($\text{Pe}=5$, $\beta=1/2$):} Starting from an initial homogeneous state with a small Gaussian perturbation, the system is allowed to reach a limiting state. The maximal and minimal densities are then plot in sub-panel (a), whilst their difference is plot in sub-panel (b). Dotted lines indicate the position of the spinodal densities. Sub-panel (c) shows a cartoon indicating the position of the parameter sweep (vertical black line), relative to the spinodals (dashed line) and coexistence solutions (solid line). Sub-panels (d)-(h) show solutions for $\rho_0\in\{0.645,0.65,0.66375,0.7,0.85\}$, respectively, where dotted green lines indicate the spinodal densities. Parameters and domain used are $\beta=0.5$, $\text{Pe}=5$, $L=256\pi$.}
 \end{figure}

%
The low $\text{Pe}$ regime where there are no stable coexistence densities represents a first aspect of novelty in the phase structure offered by this model. 
Here we expect there to be no binodals or for them to reside within the spinodals and thus have no role in the system's limiting behaviour. 
As such we are unconcerned with questions of meta-stability, but simply the nature of the transition across the spinodal line which can now be crossed at points in a whole region of phase space, not just at the critical point as with conventional (or $\beta=1$) phase separation. 
Owing to the complex eigenvalues of the system's linearised dynamical matrix, and with their stability changing (\emph{i.e.} their real component changing sign) along the spinodal, we expect a Hopf bifurcation in the absence of a more stable phase separated solution. 
We thus expect general oscillatory solutions within the spinodal region that we characterise, broadly, as a flavour of micro-phase separation, owing to the absence of bulk, or bulk-like, features. The behaviour here is shown in Fig.~\ref{structure_lowPe} whereby maximal and minimal values of the density for the limiting solutions are plot for a slice of varying $\rho_0$ and fixed $\text{Pe}=5$ across the phase diagram for $\beta=1/2$. This is above the critical, or spinodal bifurcation, point in $\text{Pe}$, but low enough that there are no stable coexistence densities. 
Unlike in the case of conventional phase separation, we do not see sharp jumps to constant coexistence densities as seen in Fig.~\ref{structure_highPe}. 
Instead, we see the emergence of oscillatory solutions on the interior of the densities, whose extrema vary continuously across the spinodal region. On the interior solutions are characterised by short sharp peaks whose frequency depends on $\rho_0$. 
Outside of the spinodals the system converges on homogeneous solutions, as expected.
%

 \subsection{One stable coexistence density (Intermediate \texorpdfstring{$\text{Pe}$}{Pe}) --- meso-phase separation}

 %
 \label{mesopara}
 With the extremes of micro and macro phase separation characterised, we turn to regions of the phase diagram at intermediate values of $\text{Pe}$, where coexistence solutions exist, but where one density lies within the spinodal region, and one lies outside it, such that only one is expected to be stable.  In particular, we seek to characterise behaviour at such values of $\text{Pe}$ where $\rho_0\in[\rho_G,\rho_L]$, such that we expect the system to converge on a state that is not either totally stable nor unstable, in that one coexistence density is linearly stable, whilst the other is not.
 %

 \subsubsection{Limiting behaviour at higher intermediate \texorpdfstring{$\text{Pe}$}{Pe} --- semi-stable domains}
 
\begin{figure}[!htp]
    \centering
    \includegraphics[width=0.99\textwidth]{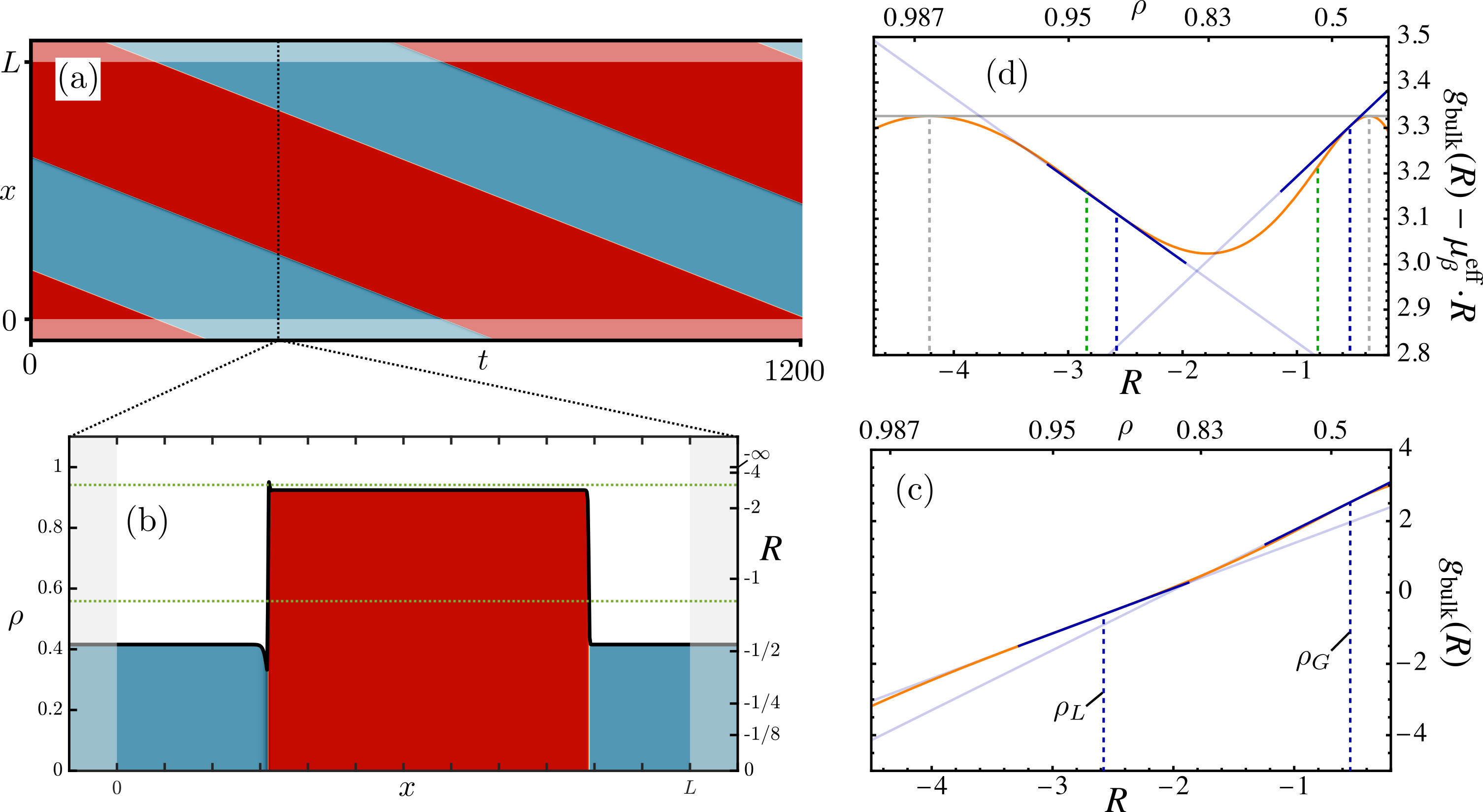} 
    \caption{\label{meso_det} {\bf Deterministic solutions in the $N\to\infty$ limit --- intermediate $\text{Pe}$:} Sub-panel $(a)$ is a density kymograph of the system in a meso-phase separated state with large domains, found from simulating the deterministic Eqs. for $\{\rho,\psi\}$. The phases demonstrably translate in the negative direction, whilst maintaining phase boundary profiles. A slice of this kymograph is shown below it in panel $(b)$ revealing the asymmetry between the leading and following domain boundaries, and with spinodal line densities marked as dashed green lines. Note, here the liquid density lies \emph{within} the spinodal densities. Panel $(c)$ illustrates the implied unequal tangent construction on $g_{\rm bulk}$ using the two empirically determined phase densities, $\rho_G\simeq 0.4156$ and $\rho_L\simeq 0.9244$. Panel $(d)$ shows the unequal tangent construction with an additional linear factor such that a common tangent solution (gray lines) appear horizontal, for clarity. Coexistence densities corresponding to the unequal tangents are shown in blue. The gaseous coexistence density lies outside of the spinodal lines (green), whilst the liquid density lies within them. All such densities lie within the coexistence densities associated with a common tangent construction (gray). Example shown uses $\beta=1/2$, $\text{Pe}=7$, $\rho_0=0.7$, $L=256\pi$.}
 \end{figure}

 %
 \label{mesohighPe}
 We first consider a scenario where the value of $\text{Pe}$ is relatively large such that the unstable coexistence density is only a small distance into the spinodal region with a key example shown in Fig.~\ref{meso_det}. Here, despite the liquid phase lying within the spinodal, a large, system sized, phase separated state appears to both exist and remain stable, such that the system's solution is indistinguishable from the idealised coexistence solution. 
 As such we can characterise such a state with an unequal tangent construction illustrating the relative positions of the spinodal and coexistence densities characteristic of this region of phase space.
 \par
 \begin{figure}[!htp]
    \centering 
    \includegraphics[width=0.99\textwidth]{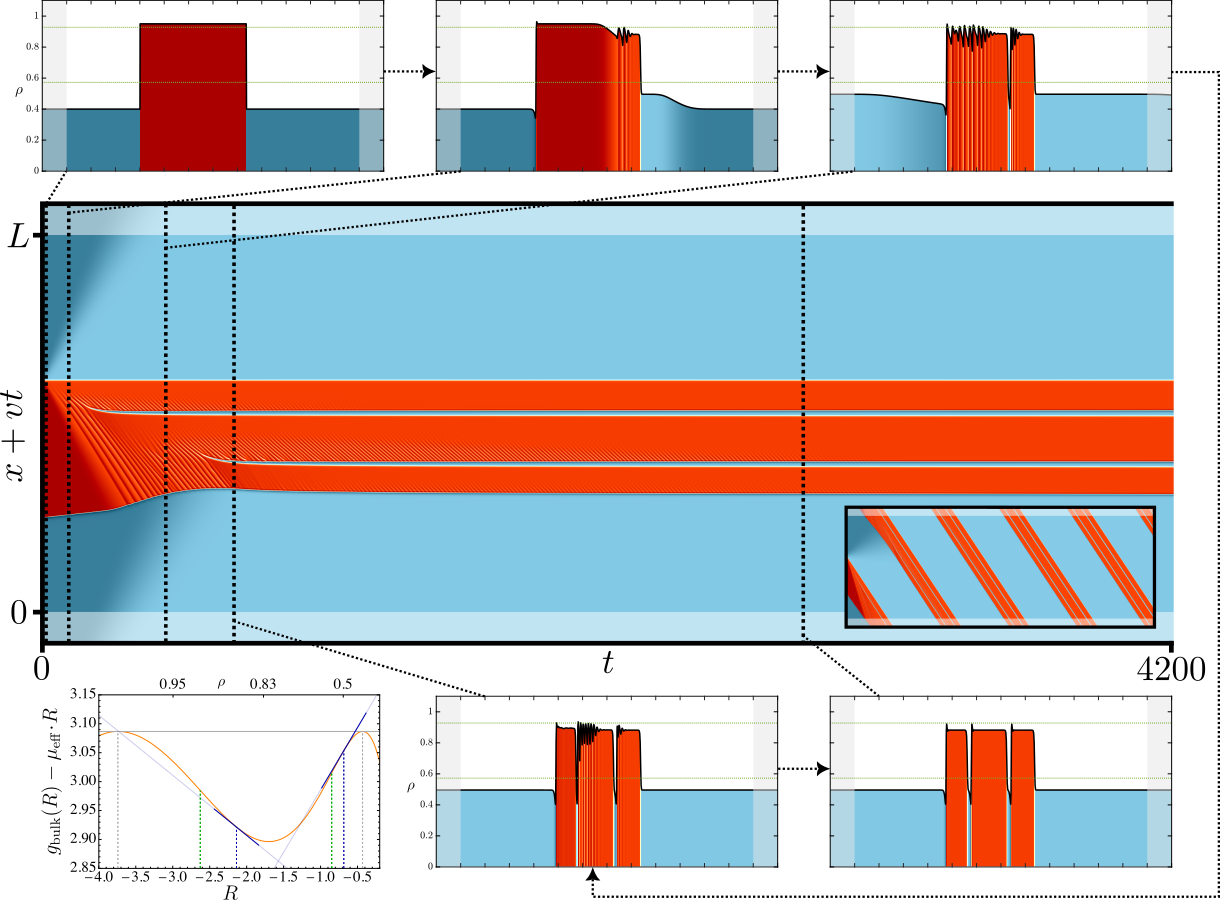} 
    \caption{\label{meso_transient}{\bf Density kymographs showing semi-stable liquid domains evolve and separate at lower-intermediate $\text{Pe}$:} Centre is a density kymograph of an evolving intermediate $\text{Pe}$ meso-phase separated state, Galilean shifted so that the liquid phase appears centred. The coexistence densities for the parameters used are $\rho_G\simeq0.496$, $\rho_L\simeq0.882$, and the system is initially prepared in a two-domain piece-wise constant state at densities $\rho\in\{0.4,0.95\}$. This causes the system to evolve towards a phase separated state at the coexistence densities. The density profile at key points in the evolution are explicitly plot in the smaller peripheral plots, evolving forward in time from top left to top right, and then bottom centre to bottom right. Bottom left shows the corresponding unequal tangent construction on $g_{\rm bulk}$. As the system configures around the liquid coexistence density, its unstable nature causes large oscillations to form which are transported to the left. These oscillations can then develop into large voids causing the liquid phase to be broken up into a `train' of subdomains of an approximately characteristic length.
    Example shown uses $\beta=1/2$, $\text{Pe}=6.4$, $\rho_0=0.6$, $L=256\pi$.}
 \end{figure}


 \subsubsection{Limiting and transient behaviour at lower \texorpdfstring{$\text{Pe}$}{Pe} --- reduced stability implies smaller domains}

%
Next, we consider lower values of $\text{Pe}$ whilst remaining in the region with one stable coexistence density. 
In general, we observe that as we progressively reduce the value of $\text{Pe}$, such that the unstable coexistence density moves further into the spinodal region, perturbations that diverge from the liquid to gas boundary grow and cause the liquid phase to split into distinct domains of a characteristic length.
\par
This phenomenon is shown for a lower value of $\text{Pe}$ in Fig.~\ref{meso_transient}. 
In this figure we initialise the system in an artificially phase separated, piece-wise constant, state consisting of a single liquid-gas domain pair such that $\rho\in\{0.4,0.95\}$. 
As the system configures around the coexistence densities (with one inside, and one outside, the spinodal region), strong oscillations start to grow and diverge away from the trailing boundary, such that the liquid bulk starts to destabilise despite the liquid to gas boundary remaining intact.  
Eventually these oscillations grow into voids, separating the system into several smaller liquid domains which do not coarsen over time.
\par
\begin{figure}[!htp]
    \centering
    \includegraphics[width=0.99\textwidth]{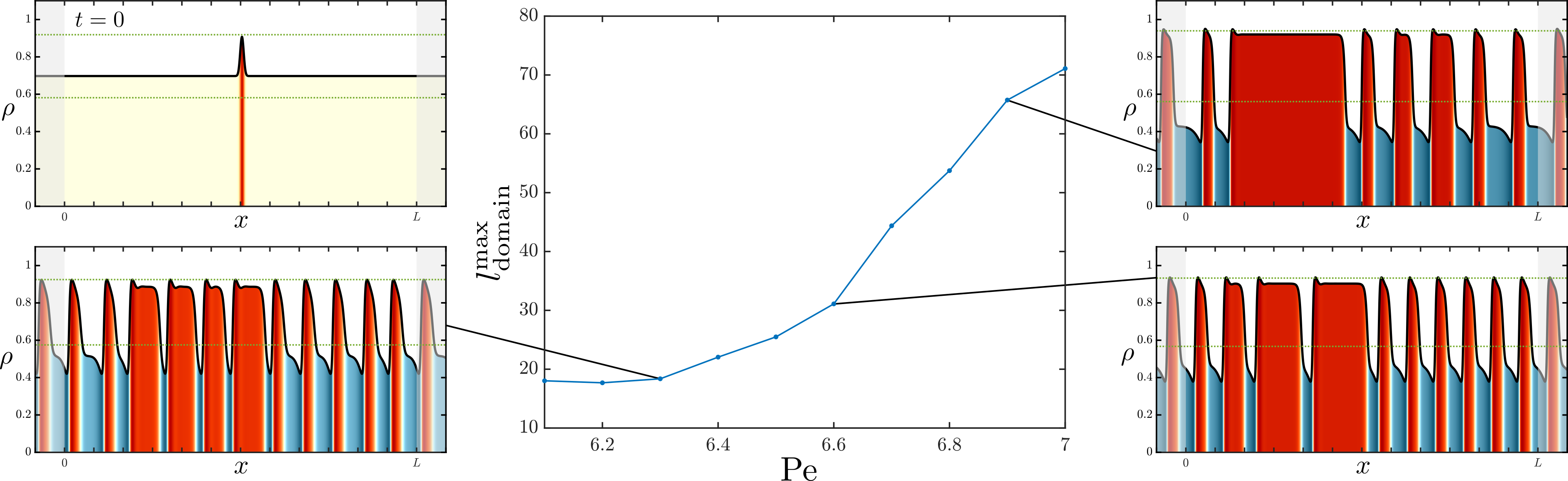} 
    \caption{\label{meso_domain_lengths} {\bf Domain length behaviour with $\text{Pe}$ for meso phase separation:} Starting from a common initial condition (top left sub-panel) the system was evolved until $t=5000$ for a range of $\text{Pe}$. After a burn in of $t=1000$ the largest single domain, defined as contiguous regions of density above $(\max \rho+\min\rho)/2$, was recorded. The length of this domain against $\text{Pe}$ is plot in the central sub-panel. The bottom left sub-panel shows the state of the system coinciding with the occurrence of the maximum domain length for $\text{Pe}=6.3$, the bottom right for $\text{Pe}=6.6$, and the top right for $\text{Pe}=6.9$. Simulations utilised $\beta=1/2$, $\rho_0=0.7$, $L=64\pi$.}
 \end{figure}

We observe that as $\text{Pe}$ is reduced the characteristic length-scale tends to become smaller. 
To illustrate this we consider the system at distinct values of $\text{Pe}$, simulating them from a common initial state of a Gaussian deviation around a homogeneous state, and recording the maximum domain that forms in the subsequent evolution. 
In practice, we simulate the system for $5000$ seconds and consider domains after a burn in period of $1000$ seconds. 
The resulting maximum domain sizes are plot in Fig.~\ref{meso_domain_lengths} where we see those at lower values of $\text{Pe}$ possessing distinctly smaller maximum domain sizes. As $\text{Pe}$ is reduced to around $6.3$ the domains approach the intrinsic length-scale of the system (\emph{i.e.} the length of a characteristic domain boundary/micro-phase structure) and cannot reduce further. 
%

\subsubsection{Phase behaviour and stability}

%
\label{meso_phase}
As with our previous treatment of macro and micro phase separating behaviour regimes we can systematically investigate the phase and stability behaviour of the system in this intermediate $\text{Pe}$, meso phase separation regime.
 \par
\begin{figure}[!htp]
    \centering
    \includegraphics[width=0.99\textwidth]{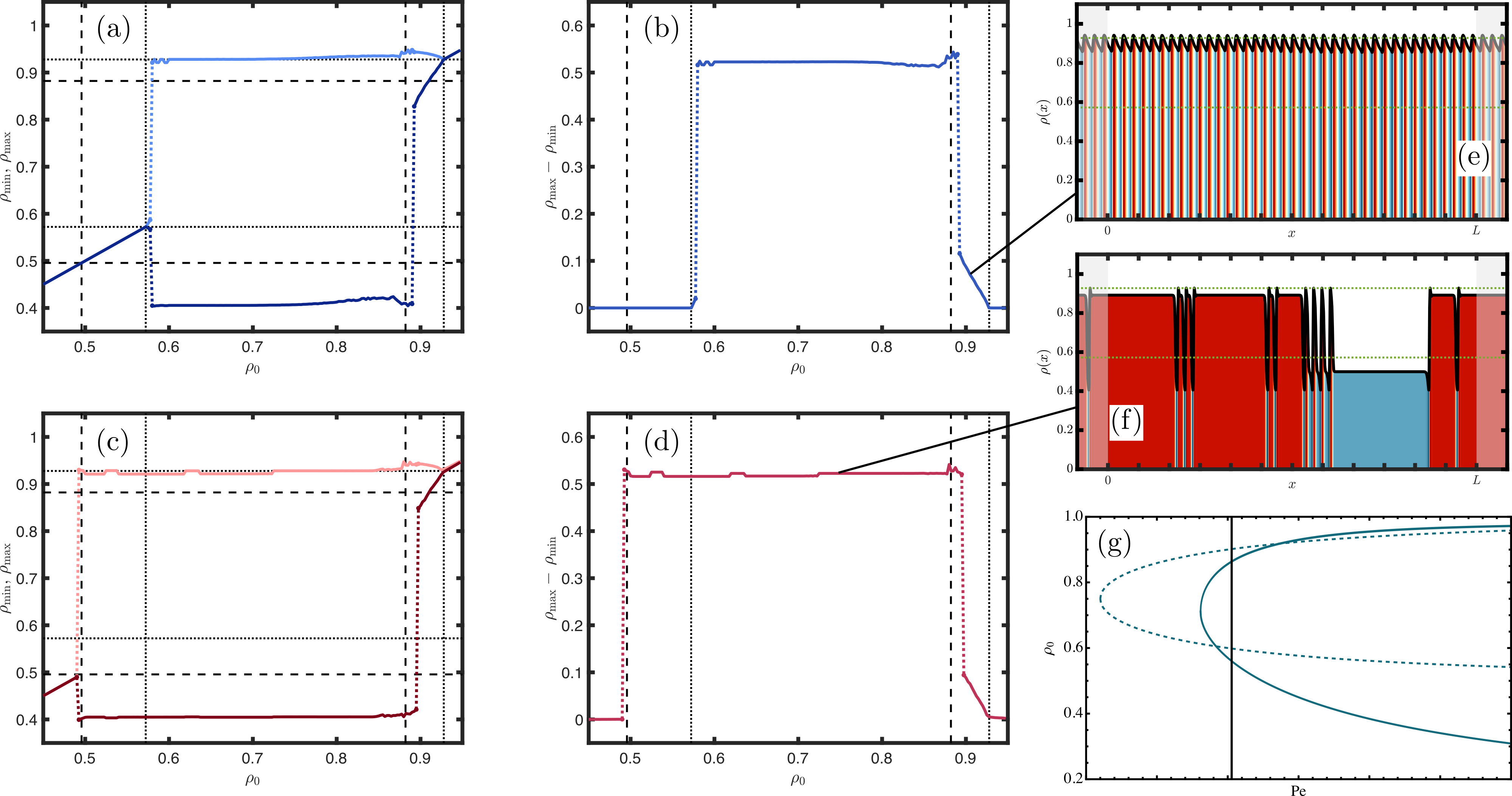}  
    \caption{\label{structure_midPe} {\bf Phase structure for intermediate $\text{Pe}$. Limiting states resulting from distinct initial conditions over a range of $\rho_0$ with all other control parameters fixed ($\text{Pe}=6.4$, $\beta=1/2$):} In sub-panels (a) and (b) we characterise the limiting solutions resulting from an initial condition consisting of a homogeneous state with a small Gaussian perturbation. In sub-panels (c) and (d) we characterise limiting solutions resulting from an initial condition consisting of an artificially phase-separated piece-wise constant state ($\rho(0)\in\{0.1,0.9999\}$). Dotted black lines in sub-panels (a)-(d) indicate spinodal densities whilst dashed black lines indicate coexistence densities. All such plots characterise solutions over a range of $\rho_0$ at a value of $\text{Pe}$ where binodal/coexistence solutions lie outside of the spinodals/unstable region as indicated in sub-panel (g). Sub-panels (a) and (c) show the maximum and minimum values of the $\rho(x)$ in the limiting state, whilst sub-panels (b) and (d) show the difference between the maximum and minimum values of $\rho(x)$ in the limiting state. Starting from a perturbative state, outside of the spinodals the system converges on a homogeneous solution, whilst inside it tends to other limiting states, consisting of a macroscopically phase separated state (\emph{e.g.} sub-panel (f)) and an oscillatory state (sub-panel (e)) immediately inside the upper spinodal. Starting from a piece-wise constant state the system reaches a phase separated state (sub-panel (f)) within the binodal/coexistence lines, and tends towards a homogeneous state outside of them. Dotted green lines in sub-panels (e) and (f) indicate the spinodals. Parameters and domain used are $\beta=0.5$, $\text{Pe}=6.4$, $L=256\pi$.}
\end{figure}

As demonstrated above, in this intermediate $\text{Pe}$ regime, we have coexistence densities, or binodals, placed such that one lies within the spinodal line, and one lies outside it. 
We investigate the range of this behaviour in Fig.~\ref{structure_midPe} wherein we plot results from numerically solving the full model in $\{\rho,\psi\}$ over a range of the parameter $\rho_0$, at intermediate $\text{Pe}$, for varying initial conditions. The two initial conditions we consider are those from before: namely a small $\delta\rho$ perturbation to an otherwise homogeneous state, and of an artificially piece-wise constant phase separated state such that $\rho\in\{0.1,0.9999\}$. 
Again, the former is designed to probe the nature of spinodal decomposition, whilst the latter is designed to explore the more general bounds of coexistence and the nature of phase coexistence, without any concern for metastability.
\par
The solutions we find are comparable to conventional phase separation where the structure is of the same form. Following the black line in panel (g) of Fig.~\ref{structure_midPe} we have four distinct regions
\begin{enumerate}
    \item Where points are outside both the spinodals and binodals, the system converges on the homogeneous mixed phase state for both initial conditions.
    \item Where points lie within both the spinodals and binodals, the system converges on a phase separated state for both initial conditions.
    \item Where points lie within the binodals, but outside the spinodals, the homogenous state is meta-stable. Consequently, the initially piece-wise state converges on a phase separated state (red lines, panels (c) and (d)), but the initial perturbation state converges on the homogeneous solution.
    \item Where points are outside the binodals, but within the spinodals, there are no phase separated states and the homogeneous state is unstable. As such the system converges on the micro-phase oscillatory solution for both initial conditions.
\end{enumerate}

\newpage
\section{Dynamics of implicit nsAMB}

%
Here we consider the behaviour of our implicit nsAMB (non-stationary active model B), when used directly in simulation, rather than for the inference of thermodynamic properties of bQSAPs as set out in the main text. 
We note that the $v$ and $J^{\rm st}_v$ that cause nsAMB to be implicit are still just constants, so that if known, solutions to nsAMB can be found from ordinary integration as a PDE. To do so here we obtain $v$ and $J^{\rm st}_v$ from the equivalent system of bQSAPs (the dynamic in $\rho$ and $\psi$). 
When $v$ and $J^{\rm st}_v$ are properly specified, by construction, nsAMB shares solutions with the full model in $\rho$ and $\psi$. 
As such the relevant points of distinction in its behaviour concern its transient evolution towards the shared limiting state and its stability properties.  
\subsection{Effective spinodals with phase separation}

\begin{figure}[!htp]
    \centering
    \includegraphics[width=0.6\textwidth]{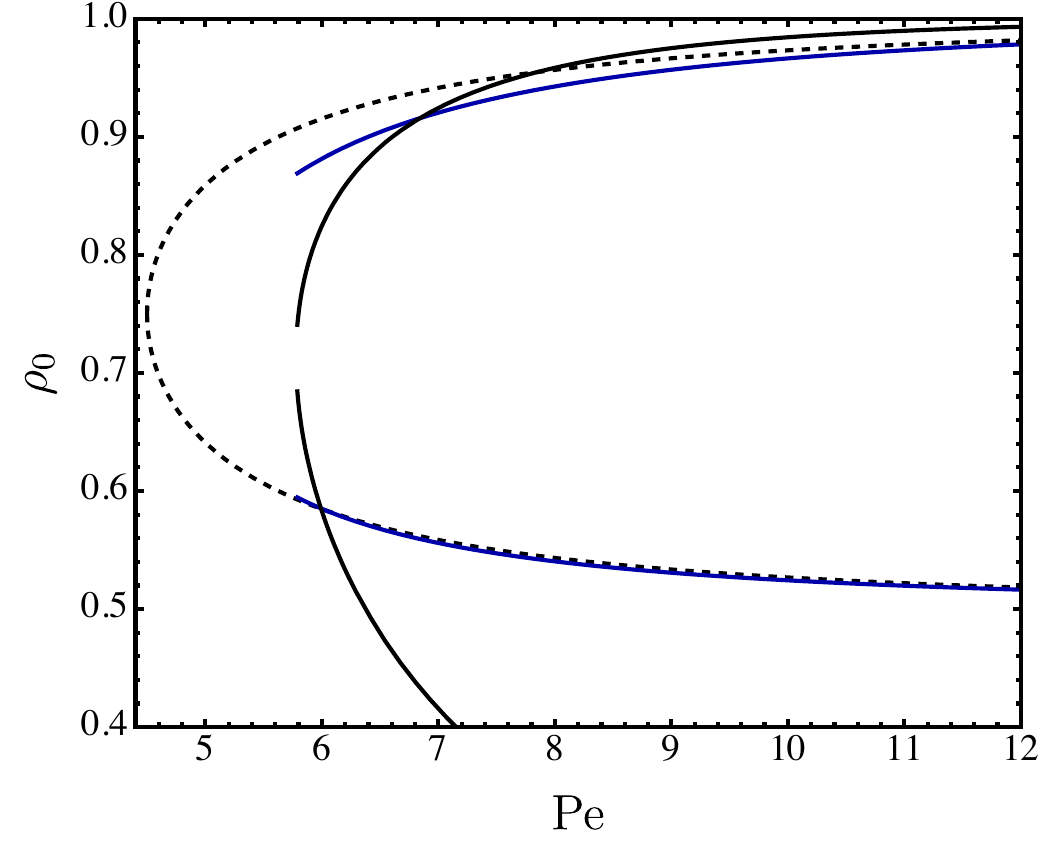}
    \caption{\label{effective_spinodals} {\bf Effective spinodals of nsAMB:} The dotted line indicates the true binodal of the system of bQSAPs whilst the solid black line indicates the binodal densities as found from the self-consistent shooting method. Solid blue lines indicate the bounds of the effective spinodal region when simulating nsAMB with $v$ and $J^{\rm st}_v$ associated with the binodal solutions.}
\end{figure}

One of the peculiarities of our implicit nsAMB is that different properties of the underlying system of bQSAPs are captured in different frames. For example, for a steady state in the context of phase separation we must have $v=\text{Pe}((1-\beta)/(1+\beta))(1-\rho_G-\rho_L)$ and $J^{\rm st}_v=\text{Pe}((1-\beta)/(1+\beta))\rho_G\rho_L$, whereas in each bulk phase we must have $v=\text{Pe}((1-\beta)/(1+\beta))(1-2\rho)$ and $J^{\rm st}_v=\text{Pe}((1-\beta)/(1+\beta))\rho^2$ for the same stability properties as the system of bQSAPs. Where nsAMB is used to understand the effective thermodynamics of the system of bQSAPs this is something of a technicality. However, if one wishes to simulate nsAMB directly, the consequence is that, in a phase separated state, the stability of the two bulks will not be determined by the true spinodal line, but rather by the linear stability properties of nsAMB itself with the phase separated $v$ and $J^{\rm st}_v$. These lines are shown in Fig.~\ref{effective_spinodals}. The net result is that the effective spinodal region is marginally reduced with the upper spinodal effective lying at smaller values of $\rho$. The result does not qualitatively change the overall phase structure, with the same demarcation into micro, meso, and macro-phase separation as before, but their quantitative locations on the phase diagrams are marginally shifted. 

\subsection{Macro-phase separation}
\begin{figure}[!htp]
    \centering
    \includegraphics[width=0.99\textwidth]{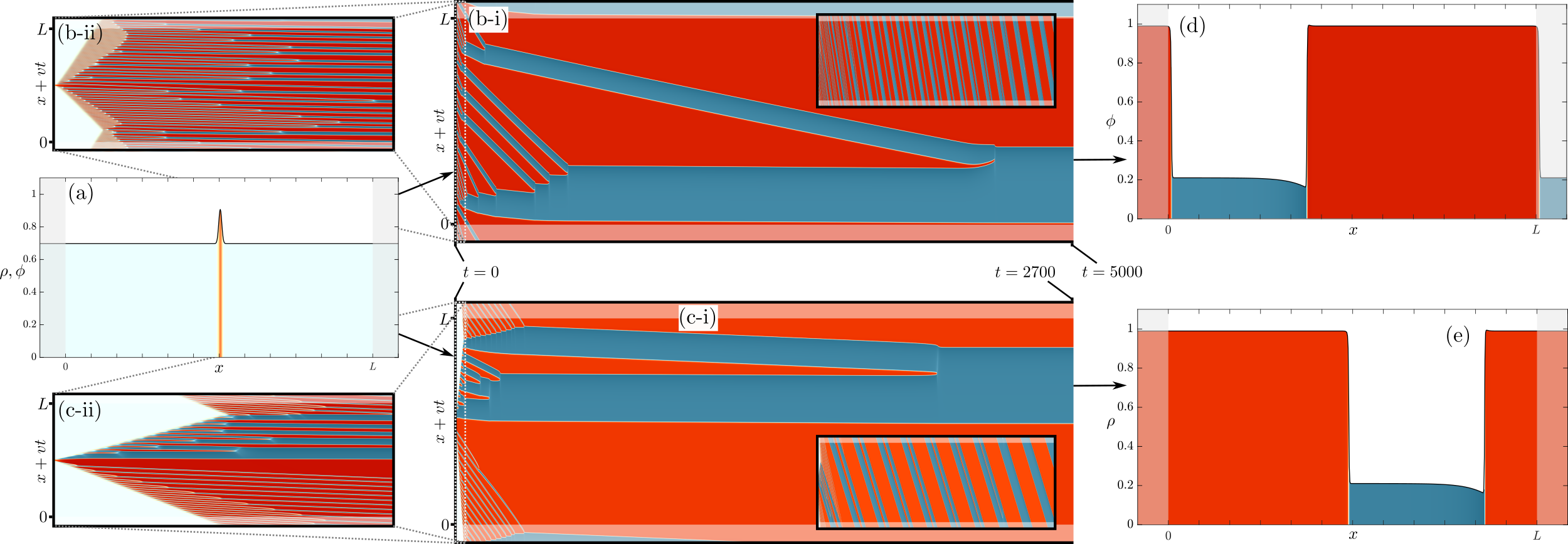}
    \caption{\label{macro_det_eject_NSAMB} {\bf Density kymographs comparing defect ejection and coarsening between the full model in $\{\rho,\psi\}$ and of the single order parameter ($\phi$) nsAMB, for $\beta=1/2$ in the $N\to\infty$ limit:} Two systems, one simulated with the full model (bottom) and one using NSAMB (top) are prepared in an identical state shown in sub-panel (a). Their evolution is shown in the central density kymographs --- sub-panel (b-i) shows a Galilean shifted NSAMB (non-frame-shifted kymograph inset), with (b-ii) showing early time evolution, and similarly sub-panel (c-i) shows a Galilean shifted density kymograph for the full model (non-frame-shifted kymograph inset), with (c-ii) early time evolution. After suitable evolution in time, the nsAMB system is in a state illustrated in sub-panel (d), whereas the full model system is in a state illustrated by sub-panel (e).
    Both systems use $\text{Pe}=11$, $\rho_0=0.7$, $L=64\pi$.}
\end{figure}

We compare the result of integrating nsAMB with the full system in $\{\rho,\psi\}$ in Fig.~\ref{macro_det_eject_NSAMB}, where we see agreement in the limiting states (as they must)  in addition to broad qualitative agreement in their transient evolution. 
The constants $v$ and $J_v^{\rm st}$ were derived from the limiting behaviour of the full model, taking values $v\simeq-0.73409$ and $J_v^{\rm st}\simeq0.76365$. 
 %


\subsection{Micro-phase separation}

\begin{figure}[!htp]
    \centering
    \includegraphics[width=0.99\textwidth]{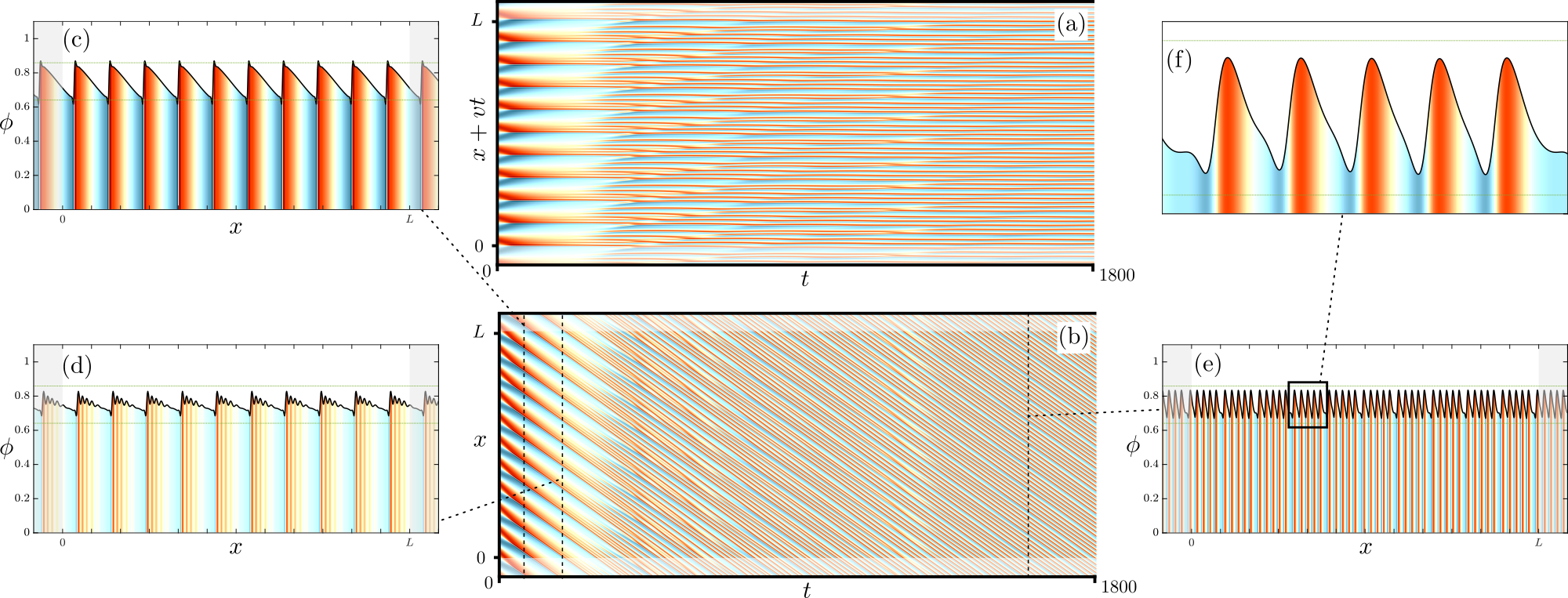} 
    \caption{\label{micro_fig_NSAMB} {\bf Behaviour of nsAMB in the unstable region without binodal solutions:} In this figure we replicate the contents of Fig.~\ref{micro_fig} but using non-stationary active model B using deduced values $v=-0.83$ and $J_v^\rho=0.9339$. 
    Sub-panel (a) shows a Galilean shifted density kymograph of the systems's evolution at frame speed $v=-0.83$ indicating that the solution is self-consistent. Sub-panel (b) shows the unmodified density kymograph of the evolution. The transient solution has the same qualitative features as the full model, including approach to a shock solution (sub-panel (c)), before destabilisation (sub-panel (d)) and eventual stable micro-phase separated state (sub-panel (e)), with detail shown in sub-panel (f). 
    In all cases dashed green lines indicate the location of the spinodal densities. Example shown uses $\beta=1/2$, $\text{Pe}=5$, $\rho_0=0.75$, $L=256\pi$.}
 \end{figure}

%
As with the large $\text{Pe}$, macro-phase separating case in Sec.~\ref{macro_phase_det} we can investigate the behaviour of the proposed implicit nsAMB in this case. 
As with the previous case we deduce the values of $v$ and $J^{\rm st}_v$ from the full simulation, here recognising that $v=\partial_x J/\partial_x\rho$, $J_v^{\rm st} = J - v\rho$ and $J = \text{Pe}\psi(1-\rho)-\partial_x\rho$ once the system has settled into a limiting, processing, state. 
We replicate the behaviour shown in Fig.~\ref{micro_fig} in Fig.~\ref{micro_fig_NSAMB} using deduced values $v=-0.83$ and $J_v^{\rm st}=0.9339$. 
Here NSAMB replicates the qualitative behaviour of the full model extremely well with minor differences in the evolution. 
 %

\subsection{Meso-phase separation}

\begin{figure}[!htp]
    \centering
    \includegraphics[width=0.99\textwidth]{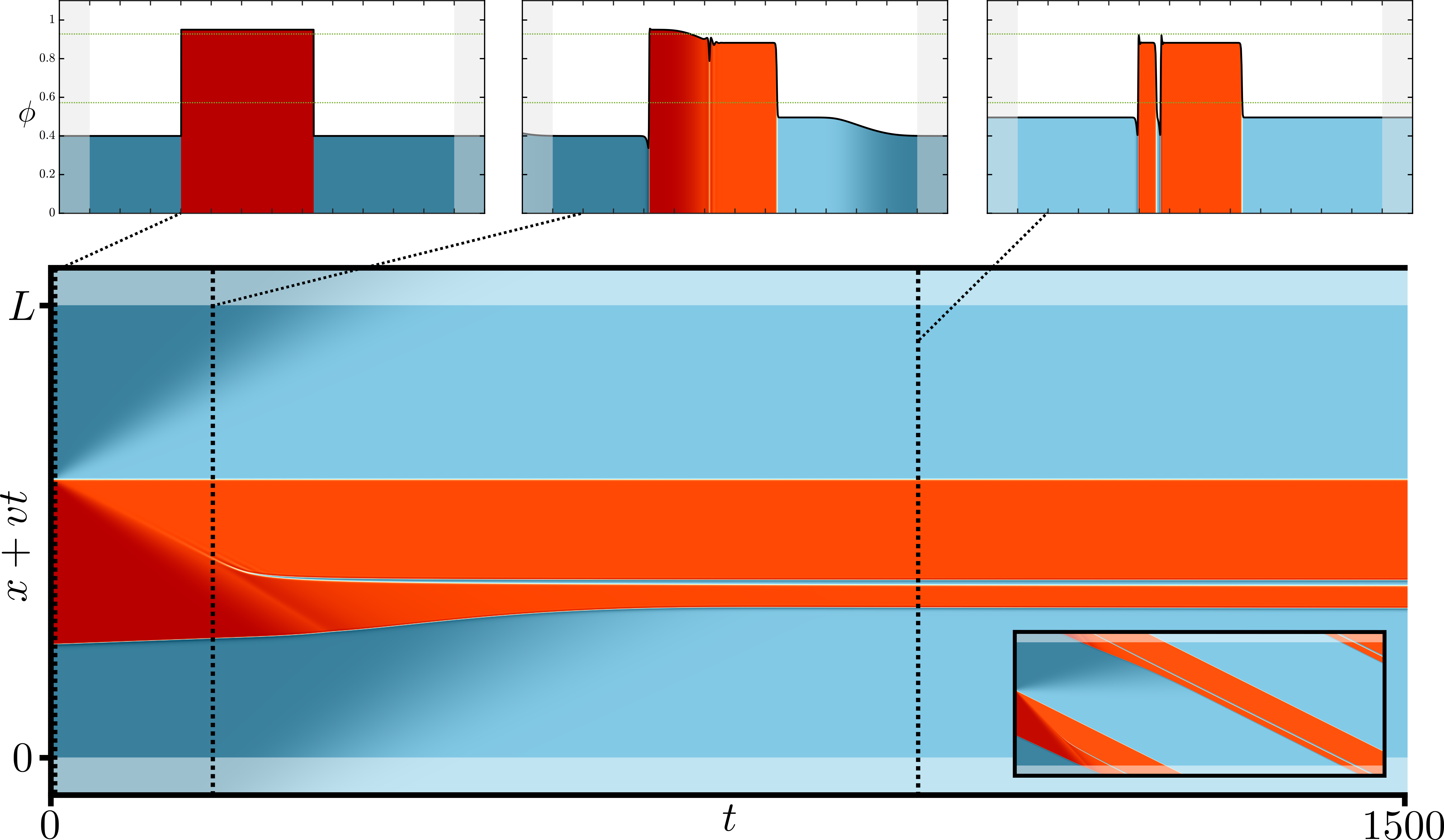} 
    \caption{\label{meso_fig_NSAMB} {\bf Behaviour of NSAMB with one stable coexistence density:} In this figure we replicate the contents of Fig.~\ref{meso_transient} but using non-stationary active model B rather than the full model using a value of $v=-0.8057$ and $J_v^{\rm st}=0.9326$ obtained from that solution. Here we see very close agreement between the full model (Fig.~\ref{meso_transient}) and NSAMB, but with decreased instability and thus larger domain length-scales. The stationary nature of the solution in a co-moving frame at velocity $v$ indicates that the system is self-consistent. Example shown uses $\beta=1/2$, $\text{Pe}=6.4$, $\rho_0=0.75$, $L=256\pi$.}
 \end{figure}

%
As with the macro and micro-phase separating cases we can investigate the behaviour of the proposed implicit nsAMB in this case. As with the previous case we deduce the values of $v$ and $J^{\rm st}_v$ from the full simulation once the system has settled into a limiting, processing, state. We replicate the behaviour shown in Fig.~\ref{meso_transient} in Fig.~\ref{meso_fig_NSAMB} using deduced values $v=-0.8057$ and $J_v^{\rm st}=0.9326$. 
Here NSAMB replicates the qualitative behaviour of the full model extremely well, however the emergent length-scale of the formed domains are larger, due to the distinct location of the effective spinodals in nsAMB.
\clearpage
\section{Pseudocode for the self-consistent shooting method}
Here we present pseudo-code for the two algorithms discussed in the main text appendix.

\begin{widetext}
    
    \RestyleAlgo{ruled}
    \begin{minipage}{.9\linewidth}
        \vspace{2em}
    \begin{algorithm}[H]
        \caption{SelfConsistentShootingSolve}
        \label{alg:one}
        \tcp{Hyper parameters:}
        \KwData{$\alpha_1>0,\,\alpha_2>0,\,\alpha_3>0,\,\alpha_4>0,\,\alpha_5>0,\,\delta>0$}
        \tcp{Phase space control parameters:}
        \KwData{$\text{Pe} > 0,\,\beta\in(0,1)$}
        \tcp{Parameter initial values:}
        \KwData{$\rho_L^{\rm init}\in[0,1],\,\rho_G^{\rm init}\in[0,\rho_L^{\rm init}),\,\phi'_{\rm init}(0)<0,\,\phi''_{\rm init}(0)<0$}
        \tcp{Output:}
        \KwResult{$\rho_L,\,\rho_G,\,\phi'(0),\,\phi''(0)$}
        \tcp{}
        \textbf{Function} \textsc{SelfConsistentShootingSolve}($\text{Pe}$, $\beta$, $\rho_L^{\rm init}$, $\rho_G^{\rm init}$, $\phi'_{\rm init}(0)$, $\phi''_{\rm init}(0)$)\\
        \tcp{Initialise parameters:}
        $\rho_L,\rho_G,\phi'(0),\phi''(0)\gets \rho_L^{\rm init},\rho_G^{\rm init},\phi'_{\rm init}(0),\phi''_{\rm init}(0)$\;
        $\phi(0)\gets \rho_L - \delta$\;
        $parameterHistory\gets\{\}$\;
        $objectiveHistory\gets\{\}$\;
        $converged\gets False$\;
        \tcp{Loop whilst minimisation algorithm has not converged}
        \While{converged = False}{
            \tcc{Shooting step:}
            \tcc{Integrate $\partial_x\mu=F$, with $v=\text{Pe}(1-\rho_G-\rho_L)((1-\beta)/(1+\beta))$ and $J_v^{\rm st}=\text{Pe}\rho_G\rho_L((1-\beta)/(1+\beta))$, from an initial condition $\phi(0)$, $\phi'(0)$, $\phi''(0)$ until a halting condition is met at some $x=x_{\rm max}$.\\E.g. the condition might be $\phi(x)$ leaves $[\rho_G,\rho_L]$, and/or the profile becomes non-monotonic (\emph{e.g.}$\phi'(x)\gtrapprox 0$), and/or $x$ reaches a maximal value}
            $\phi_I(x)\gets integrate(\text{Pe},\beta,\rho_L,\rho_G,\phi(0),\phi'(0),\phi''(0))$\;
            \tcp{Calculate objective function quantifying distance from expected bulk gaseous phase}
            $\varepsilon \gets \text{min}_{x\in[0,x_{\rm max}]}[(\rho_G-\phi_I(x))^2+\alpha_1(\phi'(0))^2+\alpha_2(\phi''(0))^2+\alpha_3(\phi_I'(x))^2+\alpha_4(\phi_I''(x))^2+\alpha_5(\phi_I'''(x))^2]$\;
            \tcp{Store parameter and objective function values}
            $parameterHistory.\text{append}(\{\rho_L,\rho_G,\phi'(0),\phi''(0)\})$\;
            $objectiveHistory.\text{append}(\varepsilon)$\;
            \tcc{Optimisation:}
            \tcp{Update parameter values with chosen optimisation algorithm}
            $\rho_L,\rho_G,\phi'(0),\phi''(0)\gets \text{optimisationAlgorithm}(parameterHistory,objectiveHistory)$\;
            \tcp{Update initial condition}
            $\phi(0)\gets\rho_L-\delta$\;
            \tcp{Determine if minimisation has converged}
            $converged\gets \text{convergenceTest}(parameterHistory,objectiveHistory)$\;
        }
        \Return $\rho_L,\,\rho_G,\,\phi'(0),\,\phi''(0)$\;
        \textbf{End Function}
    \end{algorithm}
    \vspace{2em}
    \end{minipage}
    
\end{widetext}

\begin{widetext}
        
    \RestyleAlgo{ruled}
    \begin{minipage}{.9\linewidth}
        \vspace{2em}
    \begin{algorithm}[H]
        \caption{ConstructBinodals}
        \label{alg:two}
        \tcp{Initial parameters and phase space parameters comprising known solution}
        \KwData{$\text{Pe}_{\rm init} > 0,\,\beta\in(0,1],\,\rho_L^{\rm init}\in[0,1],\,\rho_G^{\rm init}\in[0,\rho_L^{\rm init}),\,\phi'_{\rm init}(0)<0,\,\phi''_{\rm init}(0)<0$}
        \tcp{Stepping parameters:}
        \KwData{$\text{Pe}_{\rm min}, d\text{Pe}$}
        \tcp{Output: arrays/lists of coexistence densities/values of $\text{Pe}$}
        \KwResult{$PeList,\, LiquidBinodal,\, GasBinodal$}
        
        \textbf{Function} \textsc{ConstructBinodals}($\text{Pe}_{\rm min}$,  $d\text{Pe}$, $\text{Pe}_{\rm init}$, $\beta$,  $\rho_L^{\rm init}$,  $\rho_G^{\rm init}$, $\phi'_{\rm init}(0)$, $\phi''_{\rm init}(0)$)\\
        $\text{Pe},\rho_L,\rho_G,\phi'(0),\phi''(0)\gets \text{Pe}_{\rm init},\rho_L^{\rm init},\rho_G^{\rm init},\phi'_{\rm init}(0),\phi''_{\rm init}(0)$\;
        $LiquidBinodal\gets\{\rho_L\}$\;
        $GasBinodal\gets\{\rho_G\}$\;
        $PeList\gets\{\text{Pe}\}$\;

        \While{${\rm Pe}  >  {\rm Pe}_{\rm min}$}{
            \tcc{Increment $\text{Pe}$ a small distance from calculated solution}
            ${\rm Pe}\gets {\rm Pe} - d{\rm Pe}\;$\\
            \tcc{Solve for perturbed $\text{Pe}$}
            $\rho_L,\rho_G,\phi'(0),\phi''(0)\gets \text{SelfConsistentShootingSolve}(\text{Pe},\, \beta, \,\rho_L,\, \rho_G,\, \phi'(0),\, \phi''(0))$\;
            \tcc{Store result}
            $LiquidBinodal.\text{prepend}(\rho_L)$\;
            $GasBinodal.\text{prepend}(\rho_G)$\;
            $Pelist.\text{prepend}(\text{Pe})$\;
        }
        \Return $PeList,\, LiquidBinodal,\, GasBinodal$\;    
        \textbf{End Function}  
    \end{algorithm}
    \vspace{2em}
    \end{minipage}
    
\end{widetext}
\clearpage
%

\section{Additional \texorpdfstring{$N$}{N}-body simulation illustrating fluctuating phase behaviour and pseudo-criticality}

%
\label{phase}
Here we provide additional particle simulations of the model, using the SDEs described in the main text, complementing the simulations and description offered in the main text.
%

\subsection{Additional macro-phase separation simulation}

\begin{figure}[!htp]
    \centering
    \includegraphics[width=0.99\textwidth]{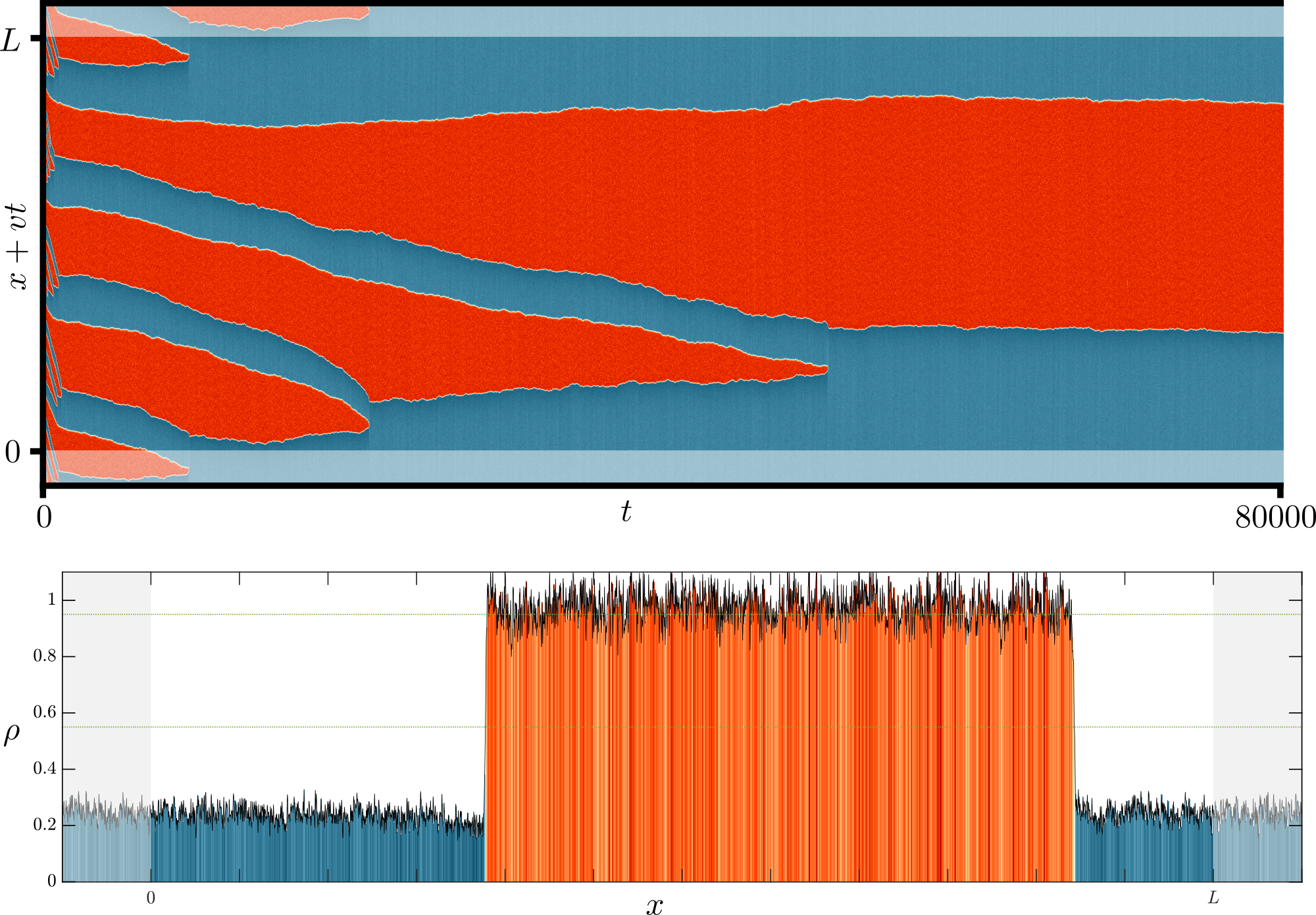}
    \caption{\label{macro_fluct_eject} {\bf Hetereogeneous coarsening in travelling macro-phase separation with fluctuations.} A density kymograph of an initially homogeneous system with $\beta=1/2$ coarsening over time in a frame where it leads to a stationary state. Examples shown use $\text{Pe}=10$, $L=128\pi$, $N=320,000$, with other parameters described in the main text appendix.}
 \end{figure}

In Fig.~\ref{macro_fluct_eject} we illustrate the coarsening phenomenon on a longer length scale than in the main text for completeness, demonstrating that the effect persists on the longest length scales we consider in simulation, but where all distinct velocities are harder to see.
%

\subsection{Additional micro-phase separation simulation}

\begin{figure}[!htp]
    \centering
    \includegraphics[width=0.99\textwidth]{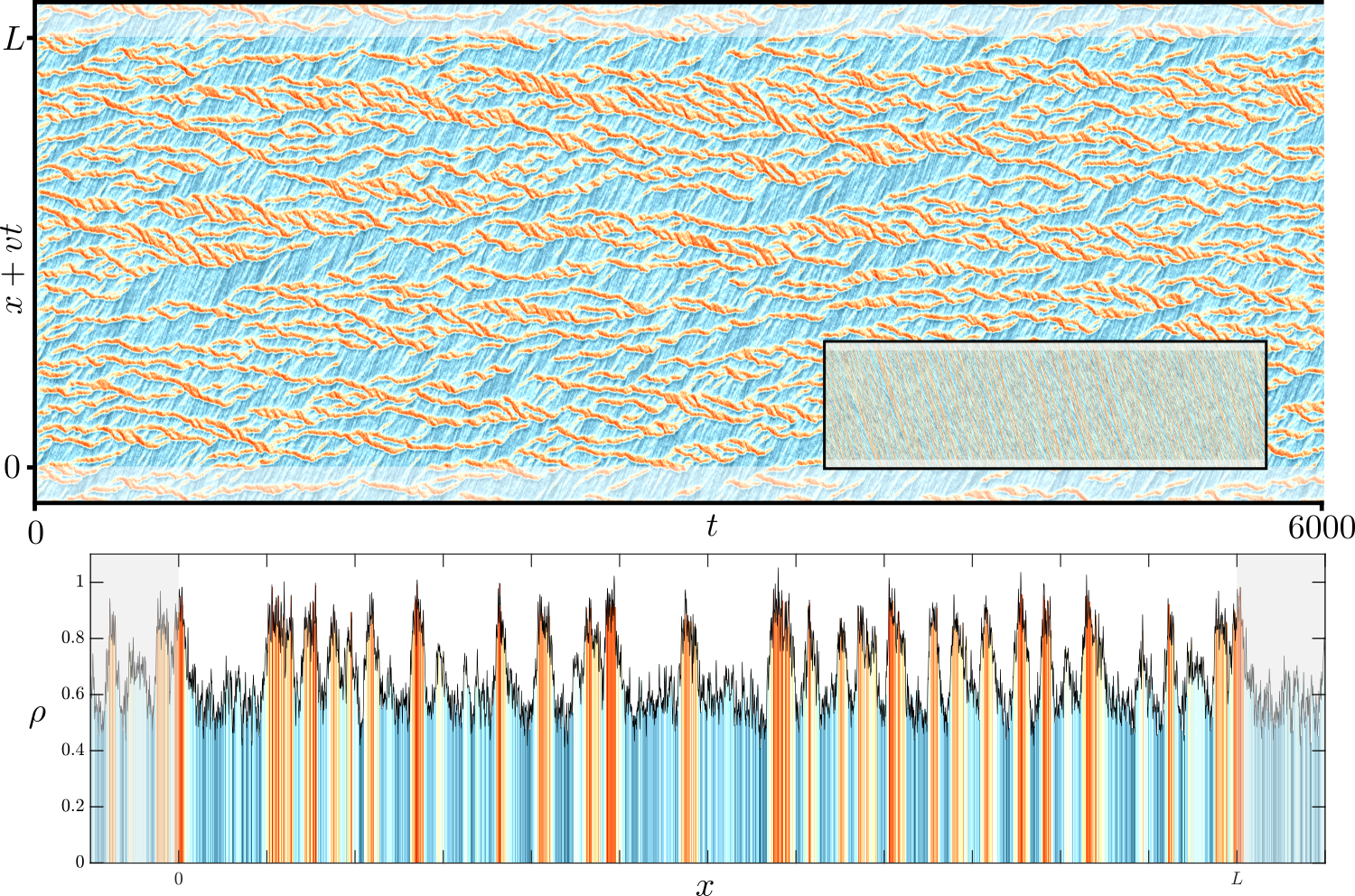}
    \caption{\label{micro_fluct} {\bf Travelling micro-phase separation with fluctuations.} The large top panel shows a density kymograph of the system in a frame where it appears stationary on average (non-transformed kymograph shown inset).  The large bottom panel shows the density profile at $t=6000$ with colours corresponding to the density kymograph. Example shown uses $\beta=1/2$, $\text{Pe}=5$, $L=128\pi$, $N=320,000$, with other parameters described in the main text appendix.}
 \end{figure}

%
Fig.~\ref{micro_fluct} shows the fluctuating behaviour of the system where there are no binodal solutions with greater resolution than can be incorporated into the main text.
%


\subsection{Additional meso-phase separation simulation}


Here we provide additional simulations of the system in the region where one binodal density is stable and one is unstable.
\par
\begin{figure}[!htp]
    \centering
    \includegraphics[width=0.99\textwidth]{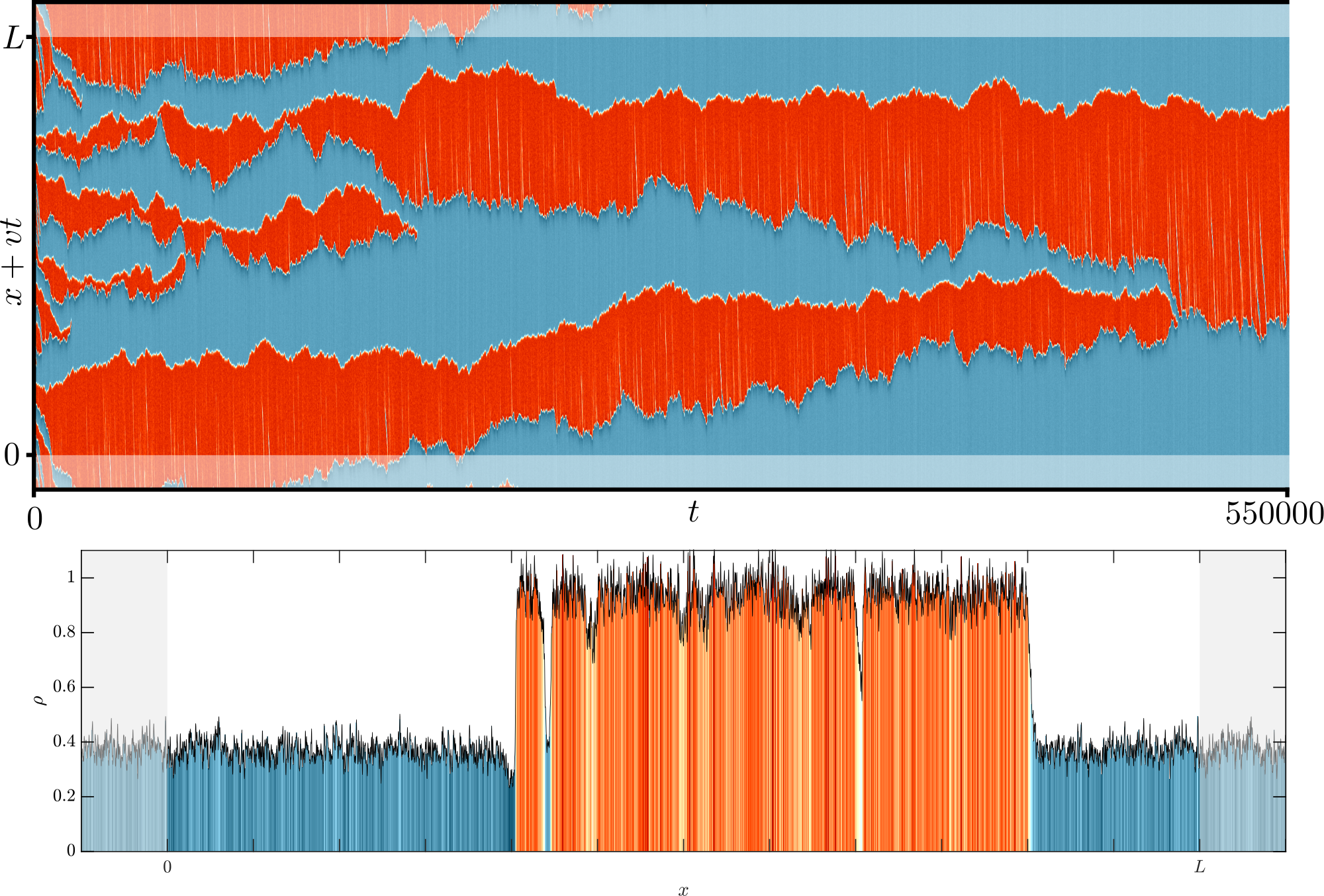}
    \caption{\label{meso_fluct0} {\bf Travelling meso-phase separation with fluctuations.} The large top panel shows a density kymograph of the system in a frame where the liquid phase envelope appears stationary on average initialised in a homogeneous state.  The bottom panel shows the density profile at $t=550000$ with colours corresponding to the density kymograph. Example shown uses $\beta=1/2$, $\text{Pe}=7.5$, $L=128\pi$, $N=320,000$, with other parameters described in the main text appendix.}
 \end{figure}

In Fig.~\ref{meso_fluct0} we show coarsening behaviour of the system where the upper binodal lies just inside the spinodal with control parameters equal to the top example in Fig.~9 in the main text. This demonstrates that the system does converge on a single liquid phase for such parameters and illustrates the tension between the coarsening mechanism and uncontrolled noise that occurs when coexistence densities lie within the spinodals. In this case the emergent length scale of the system is comparable to the size of the system.
\par
\begin{figure}[!htp]
    \centering
    \includegraphics[width=0.99\textwidth]{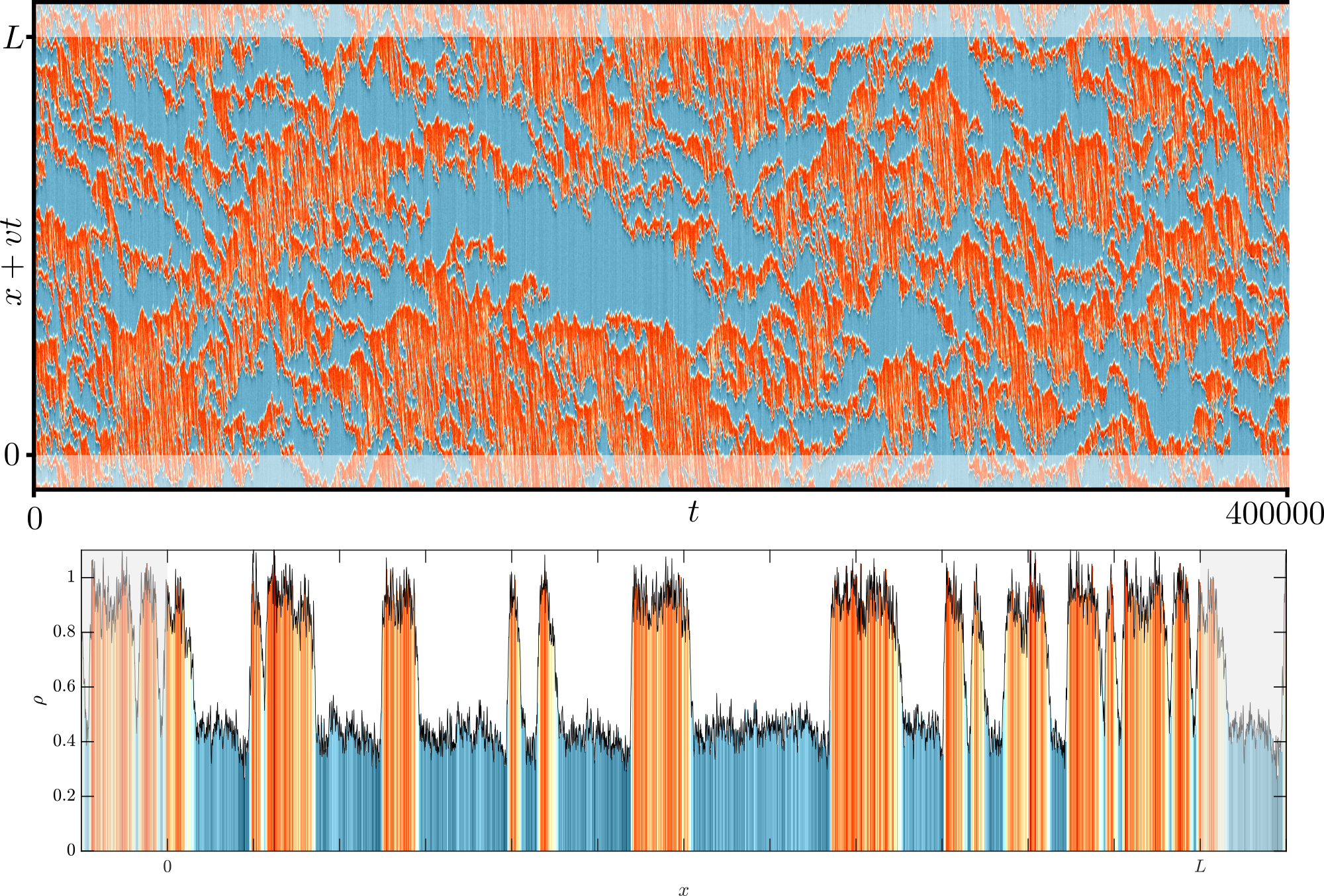}
    \caption{\label{meso_fluct1} {\bf Travelling meso-phase separation with fluctuations.} The large top panel shows a density kymograph of the system in a frame where the liquid phase envelope appears stationary on average initialised in a homogeneous state.  The bottom panel shows the density profile at $t=400000$ with colours corresponding to the density kymograph. Example shown uses $\beta=1/2$, $\text{Pe}=6.75$, $L=128\pi$, $N=320,000$, with other parameters described in the main text appendix.}
 \end{figure}

In Fig.~\ref{meso_fluct1} we show pseudo-critical behaviours at value of $\text{Pe}$ below that shown in the main text. The fluctuating domain boundaries in this case appear on shorter length scales than those shown in Fig.~9 in the main text in accordance with the prediction based on the length scale heuristic.
\par
\begin{figure}[!htp]
    \centering
    \includegraphics[width=0.99\textwidth]{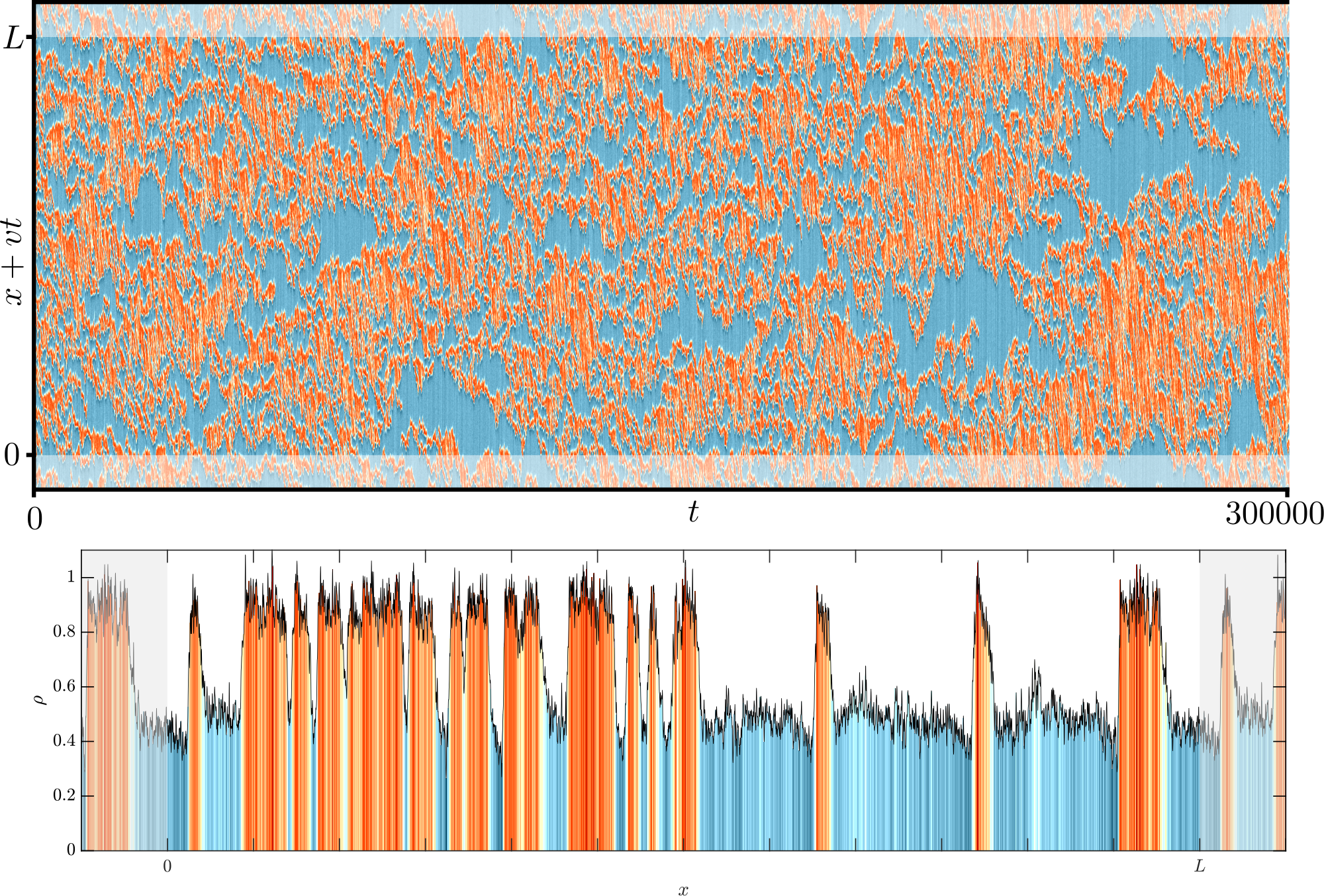}
    \caption{\label{meso_fluct2} {\bf Travelling meso-phase separation with fluctuations.} The large top panel shows a density kymograph of the system in a frame where the liquid phase envelope appears stationary on average initialised in a homogeneous state.  The bottom panel shows the density profile at $t=300000$ with colours corresponding to the density kymograph. Example shown uses $\beta=1/2$, $\text{Pe}=6.5$, $L=128\pi$, $N=320,000$, with other parameters described in the main text appendix.}
 \end{figure}

In Fig.~\ref{meso_fluct2} we show pseudo-critical behaviours at yet a lower value of $\text{Pe}$ where the system exhibits a yet shorter length scale.

